\documentclass[12pt]{article}

\pdfoutput=1

\newcommand{\beq}{\begin{equation}}
\newcommand{\eeq}{\end{equation}}
\newcommand{\beqr}{\begin{eqnarray}}
\newcommand{\eeqr}{\end{eqnarray}}
\newcommand{\beqrn}{\begin{eqnarray*}}
\newcommand{\eeqrn}{\end{eqnarray*}}
\newcommand{\beqn}{\begin{equation*}}
\newcommand{\eeqn}{\end{equation*}}
\newcommand{\bei}{\begin{itemize}}
\newcommand{\beii}{\begin{itemize} \item}
\newcommand{\eei}{\end{itemize}}
\newcommand{\ben}{\begin{enumerate}}
\newcommand{\een}{\end{enumerate}}
\newcommand{\bes}{\begin{small}}
\newcommand{\ees}{\end{small}}
\newcommand{\bec}{\begin{center}}
\newcommand{\eec}{\end{center}}

\newcommand{\tht}{\theta}
\newcommand{\eps}{\epsilon}
\newcommand{\Ex}{\mathop{\bf E\/}}
\newcommand{\xvec}{\mathbf{x}}

\usepackage{amsmath, amsthm, amssymb}
\usepackage{graphics}
\usepackage{graphicx}
\usepackage{epstopdf}

\usepackage{rotating}
\usepackage{fullpage}
\usepackage{cite}

\begin{document}

\bibliographystyle{../../plos}

%\title{When Do Feedforward Microcircuits Produce Beyond-Pairwise Correlations?}
\title{When Do Microcircuits Produce Beyond-Pairwise Correlations?}
\author{Andrea K. Barreiro$^{1,4,\ast}$, Julijana Gjorgjieva$^{3,5}$, Fred Rieke$^2$, and Eric Shea-Brown$^1$ \\  
\small{$^{1}$ Department of Applied Mathematics, University of Washington} \\ 
\small{$^{2}$Department of Physiology and Biophysics, University of Washington}  \\ 
\small{$^{3}$ Department of Applied Mathematics and Theoretical Physics, University of Cambridge}\\ 
\small{$^{4}$ present affiliation: Department of Mathematics, Southern Methodist University}\\
\small{$^{5}$ present affiliation: Center for Brain Science, Harvard University}\\
\small{$^{\ast}$ corresponding author}}
\date{}
\maketitle

\begin{abstract}
Describing the collective activity of neural populations is a daunting task:  the number of possible patterns grows exponentially with the number of cells, resulting in practically unlimited complexity.  Recent empirical studies, however, suggest a vast simplification in how multi-neuron spiking occurs: the activity patterns of some circuits are nearly completely captured by pairwise interactions among neurons.  Why are such pairwise models so successful in some instances, but insufficient in others?  Here, we study the emergence of higher-order interactions in simple circuits with different architectures and inputs.  We quantify the impact of higher-order interactions by comparing the responses of mechanistic circuit models vs. ``null" descriptions in which all higher-than-pairwise correlations have been accounted for by lower order statistics, known as pairwise maximum entropy models.  

We find that bimodal input signals produce larger deviations from pairwise predictions than unimodal inputs for circuits with local and global connectivity.  Moreover, recurrent coupling can accentuate these deviations, if coupling strengths are neither too weak nor too strong.  A circuit model based on intracellular recordings from ON parasol retinal ganglion cells shows that a broad range of light signals induce unimodal inputs to spike generators, and that coupling strengths produce weak effects on higher-order interactions.  This provides a novel explanation for the success of pairwise models in this system. Overall, our findings identify circuit-level mechanisms that produce and fail to produce higher-order spiking statistics in neural ensembles. 
\end{abstract}

% FMR: some revisions to 3rd sentence above
% E: looks good

\section*{Author Summary}
Neural populations can, in principle, produce an enormous number of distinct multi-cell patterns --- a number so large that the frequency of these patterns could never be measured experimentally.  Remarkably, the activity of many circuits is well captured by simpler probability models that rely only on the activity of single neurons and neuron pairs.  These pairwise models remove higher-order interactions among groups of more than two cells in a principled way.  Pairwise models succeed even in cases where circuit architecture and input signals seem likely to create a more complex set of outputs.  We develop a general approach to understanding which network architectures and input signals will lead such models to succeed, and which will lead them to fail.  

As a specific application, we consider the remarkable empirical success of pairwise models in capturing the activity of a class of \textit{retinal ganglion cells} --- the output cells of the retina. Our theory provides a direct explanation for these findings based on the filtering and spike generation properties of ON parasol retinal circuitry, in which collective activity arises through common feedforward inputs to spiking cells together with relatively weak gap junction coupling. Specifically, filtering of light input upstream of ON parasol cells shapes inputs to these cells in such a way that when they are processed by parasol cells, the output spiking patterns are closely fit by a pairwise model.

\section*{Introduction}
Information in neural circuits is often encoded in the activity of large, highly interconnected neural populations.  The combinatoric explosion of possible responses of such circuits poses major conceptual, experimental, and computational challenges.  How much of this potential complexity is realized?  What do statistical regularities in population responses tell us about circuit architecture?  Can simple circuit models with limited interactions among cells capture the relevant information content?  These questions are central to our understanding of neural coding and decoding.

Two developments have advanced studies of synchronous activity in recent years.  First, new experimental techniques provide access to responses from the large groups of neurons necessary to adequately sample synchronous activity patterns~\cite{Baudry2006}.  Second, maximum entropy approaches from statistical physics have provided a powerful approach to distinguish genuinely higher-order synchrony (interactions) from that explainable by pairwise statistical interactions among neurons~\cite{amari01,Martignonetal00,ssbb03}.  These approaches have produced diverse findings. 
In some instances, activity of neural populations is extremely well described by 
pairwise interactions alone, so that pairwise maximum entropy models provide a nearly complete description \cite{Shlens06,Shlens09}.  In other cases, while pairwise models bring major improvements over independent descriptions, it is not clear that they fully capture the data~\cite{Sch+06,Tang08,yu08,Martignonetal00,Montanietal09,OhiorhenuanMPSHV10,santos10}.  Empirical studies indicate that pairwise models can fail to explain the responses of spatially localized triplets of 
cells~\cite{OhiorhenuanMPSHV10,Ganmor11}, as well as the activity of populations of $\sim$100 cells responding to natural stimuli \cite{Ganmor11}.
Overall, the range of empirical results highlights the need to understand the network and input features that control the statistical complexity of synchronous activity patterns. 

% Old text: commented 7/29
%In some instances, activity of neural populations is extremely well described by 
%pairwise correlations alone, so that pairwise maximum entropy models provide a nearly complete description \cite{Shlens06,Shlens09}.  In other cases, while pairwise models bring major improvements over independent descriptions, they fail to fully capture the 
%data~\cite{Sch+06,Tang08,yu08,Martignonetal00,Montanietal09,OhiorhenuanMPSHV10,santos10}.  The range of empirical findings highlights the need to understand the network features that control the statistical complexity of synchronous activity patterns. 

Several themes have emerged from efforts to link the correlation structure of spiking activity to circuit mechanisms using generalized~\cite{amari03,KruminS09,macke09,roudi09} and biologically-based models~\cite{Martignonetal00,rth09,bohte00}.  Two findings are particularly relevant for the present study.  
First, thresholding nonlinearities in circuits with Gaussian input signals can generate correlations that cannot be explained by pairwise statistics~\cite{amari03}; the deviations from pairwise predictions are modest at moderate population sizes \cite{macke09}, 
but may become severe as population size $N$ grows large \cite{amari03, macke11}.  Cluster sizes (i.e., the number of cells firing simultaneously), in particular, may be poorly fit by pairwise models.  For large population sizes, a widespread distribution of cluster sizes requires interactions of all orders \cite{amari03}, although for population sizes explored in experimental data third or fourth order interactions may be sufficient \cite{Montanietal09}.  
% AKB: incorporated content of following sentence into earlier
%Further, for moderate population sizes thresholded Gaussian inputs cause at most modest deviations from predictions of pairwise maximum entropy models over a wide range of input parameters~\cite{macke09}. 
Poor fitting of cluster sizes by the pairwise model is also noted in networks of recurrent integrate-and-fire units with adapting thresholds and refractory potassium currents \cite{bohte00}. Small groups of cells that perform logical operations can be shown to generate higher-order interactions by introducing noisy processes with synergistic effects \cite{ssbb03}, but it is unclear what neural mechanisms might produce similar distributions. These diverse findings point to the important role that circuit features and mechanisms --- input statistics, input/output relationships, and circuit connectivity --- may play in regulating higher-order interactions.  Nevertheless, we still lack a systematic understanding that links these features and their combinations to the success and failure of pairwise statistical models.  
% FMR some edits to first third of above paragraph
% E: looks good

Second, perturbation approaches can explain why maximum entropy models with purely pairwise interactions capture circuit behavior when the population firing rate is low (i.e. the total number of firing events from all cells in the same small time window is small)~\cite{roudi09, cocco09, tkacik10}.  The fraction of multi-information \cite{ssbb03} captured by the pairwise model, a common metric of success, is necessarily low in this regime \cite{roudi09}. In this regime, as well,  higher-order interactions cannot be introduced as an artifact of under-sampling the network \cite{tkacik10}. 
The studies which find that pairwise models are successful in capturing multivariate spiking data, however, 
include cases which either extend beyond \cite{Shlens06,Shlens09}, 
or are marginally within \cite{Sch+06}, the low spikes per time bin regime. 
Nor do perturbation results necessarily account for the success of models where pairwise connections are restricted to nearest-neighbor connections\cite{Shlens06}. Therefore, an explanation for the empirical successes of pairwise models remains incomplete. 
	%The success of pairwise models in capturing multivariate spiking data, however, extends well beyond this low firing rate regime.  The basis of this unexpected success of pairwise models remains unclear.  

Here, we aim to bridge some of the gaps between our understanding of mechanistic and statistical models.  Our strategy is to systematically characterize the ability of pairwise maximum entropy (PME) models to capture the responses of several simple circuits (see Figure \ref{fig:study_schematic}).  In each case, we search exhaustively over the entire parameter space for networks with a simple thresholding model of spike generation.  Studies of feedforward circuits reveal that the success of PME models does not always bear a simple relation to network architecture --- we find examples in which networks with local connections can deviate substantially from PME predictions, while networks with global connections can be well approximated by PME models.  A more consistent determinant of the success of PME models is the unimodal vs. bimodal profile of inputs to the circuit.  In addition to departures driven by these inputs, we show that excitatory recurrent coupling can increase departures from PME models by a further 3-5 fold.  

We apply these general results to responses of specific networks based on measured properties of primate ON parasol ganglion cells.  We find that these responses are closely approximated by PME models for a wide range of input types.  PME models are successful in this case because the temporal filtering properties of parasol cells induce unimodal synaptic inputs for a broad range of light inputs and because the measured coupling strengths are insufficient to produce large effects on higher-order interactions.  This provides insight into why the measured activity patterns in these cells are well captured by PME models~\cite{Shlens06,Shlens09}. 

% FMR: simplified above paragraph quite a bit 
% E: also did some rewriting

\section*{Results}

Our aim is to determine how network architecture and input statistics influence the success of PME models.  We start by considering networks of three cells, which both allow us to make substantial analytical progress and to visualize the results geometrically.  We then extend these results in two ways: (1) to models based on the measured properties of primate ganglion cells, one system in which PME models have been very successful~\cite{Shlens06,Shlens09}; and (2) to larger networks.  

% FMR: added above paragraph
% E: looks good

\subsection*{A geometric approach to identifying higher-order interactions among triplets of cells} \label{sec:main}

One strategy to identify higher-order interactions is to compare multi-neuron spike data against a description in which any higher-order interactions have been removed in a principled way --- that is, a description in which all higher-order correlations are completely described by lower-order statistics.  Such a description may be given by a maximum entropy model~\cite{Jaynes1957a,Jaynes1957b,amari01}, which determines how much of the potential complexity of response patterns produced by large neural populations can be captured by a given set of constraints.  The idea is to identify the most unstructured, or maximum entropy, distribution consistent with the constraints.  Comparing the predicted and measured probabilities of different responses tests whether the constraints used are sufficient to explain the network activity, or whether additional constraints need to be considered.  Such additional constraints would produce additional structure in the predicted response distribution, and hence lower the entropy.  

A common approach is to limit the constraints to a given statistical order --- for example, to consider only the first and second moments of the distributions, which are determined by the mean and pairwise interactions.  In the context of spiking neurons, we denote $\mu_i \equiv \Ex [x_i]$ as the firing rate of neuron $i$ and $\rho_{ij} \equiv \Ex [x_i x_j]$ as the joint probability that neurons $i$ and $j$ will fire.  The distribution with the largest entropy for a given $\mu_i$ and $\rho_{ij}$ is often referred to as the \textit{pairwise maximum entropy} (PME) model.  The problem is made simpler if we consider only permutation-symmetric spiking patterns, in which the firing rate and correlation do not depend on the identity of the cells; i.e. $\mu_i = \mu$, $\rho_{ij} = \rho$ for $i \not= j$.  Thus the PME problem is to identify the distribution that maximizes the response entropy given the constraints $\mu$ and $\rho$.  In this section,
we review some results that help provide a geometric, and hence visual, approach to this problem (see, for example \cite{bohte00}).

We consider a permutation-symmetric network of three cells with binary responses. We assume that the response is stationary and uncorrelated in time. From symmetry, the possible network responses are 
\begin{eqnarray*}
p_0 & = & P(0,0,0)\\
p_1 & = & P(1,0,0) = P(0,1,0) = P(0,0,1)\\
p_2 & = & P(1,1,0) = P(1,0,1) = P(0,1,1)\\
p_3 & = & P(1,1,1),
\end{eqnarray*}
where $p_i$ denotes the probability that a particular set of $i$ cells spike and the remaining $3-i$ do not. 
Possible values of $(p_0, p_1, p_2, p_3)$ are constrained by the fact that $P$ is a probability distribution, meaning 
that the sum of $p_i$ over all eight states is one.  We will rearrange these response probabilities to define a more convenient coordinate system below.  

Possible solutions to the PME problem take the form of exponential functions characterized by two parameters, $\lambda_1$ and $\lambda_2$, which serve as Lagrange multipliers for the constraints,
\begin{eqnarray}
P(x_1,x_2,x_3) & = & \frac{1}{Z} \exp [ \lambda_1(x_1 + x_2 + x_3) + \lambda_2 (x_1 x_2 + x_2 x_3 + x_1 x_3) ].
\end{eqnarray}
The factor $Z$ normalizes $P$ to be a probability distribution.  We can combine the individual probabilities of events
\begin{eqnarray*}
p_0 & = & \frac{1}{Z}\\
p_1 & = & \frac{1}{Z} \exp(\lambda_1)\\
p_2 & = & \frac{1}{Z} \exp(2\lambda_1 + \lambda_2)\\
p_3 & = & \frac{1}{Z} \exp(3\lambda_1 + 3 \lambda_2)
\end{eqnarray*}
to yield the equation 
\begin{eqnarray}
\frac{p_3}{p_0} & = & \left( \frac{p_2}{p_1} \right)^3   \label{eqn:constraint_surface}.
\end{eqnarray}
This is equivalent to the condition that the {\it strain} measure defined in \cite{ovJCN10} be zero 
(in particular,  the strain is negative whenever $p_3/p_0 - (p_2/p_1)^3 < 0$, a condition identified in~\cite{ovJCN10} as corresponding to sparsity in the neural code).  Equation \ref{eqn:constraint_surface} defines, implicitly, a two-dimensional surface in the three-dimensional space of possible probability distributions which we call the \textit{maximum entropy surface}.  The family of distributions consistent with a given $\mu$ and $\rho$ forms a line (the \textit{iso-moment line}) in this space \cite{bohte00}.  The PME fit is the intersection of this line with the surface defined by Equation \ref{eqn:constraint_surface}.

This geometrical description of the PME problem takes a particularly simple form in an alternative coordinate space:
\begin{eqnarray}
f_p & = & p_3 + p_0   \nonumber\\
f_{1p} & = & \frac{p_3}{p_3+p_0} \label{eqn:fp_def}\\
f_{1m} & = & \frac{p_2}{p_2+p_1} \nonumber.
\end{eqnarray}
This set of coordinates separates network responses based on whether they are ``pure" (all cells either spike, or do not)
or ``mixed" (only a subset of cells spike). $f_p$ is the fraction of observed responses that are pure; $f_{1p}$ is the fraction of pure responses with more cells spiking than not ($p_3$ vs.  $p_0$).  $f_{1m}$ is the fraction of mixed responses with more cells firing than not ($p_2$ vs. $p_1$).  Possible probability distributions are contained within a cube in this coordinate space: $0 \leq f_p, f_{1p}, f_{1m} \leq 1$.  
The iso-moment line for a given $\mu$ and $\rho$ is still a line in this coordinate space (see Methods), and the PME approximation is given by the intersection of this line with the maximum entropy surface.
%, as shown in Figure~\ref{fig:iso_bar}.

The convenience of this coordinate system is apparent when the maximum entropy constraint (Equation \ref{eqn:constraint_surface}) is rewritten:
\begin{eqnarray}
f_{1p} & = & \frac{f_{1m}^3}{1 - 3 f_{1m} + 3 f_{1m}^2}  \label{eqn:constraint_fp}.
\end{eqnarray}
This surface is independent of $f_p$ --- i.e. the maximum entropy surface forms a curve when projected into the $(f_{1p}, f_{1m})$-plane.  In addition, each iso-moment line lies in a constant $f_p$ plane; the distance of an observed distribution $P$ from the surface is thus easily visualized.
This geometric view of the relation between the activity of a given network and its PME fit is particularly useful in visualizing results from exhaustive searches across network parameters, as described below.

We use the Kullback-Leibler divergence, $D_{KL}(P, \tilde P)$, to quantify the accuracy of the PME approximation $\tilde P$ to a distribution $P$.  
This measure has a natural interpretation as the contribution of higher-order interactions to the response entropy $S(P)$ \cite{amari01,ssbb03},
and may in this context be written as the difference of entropies $S(\tilde{P})-S(P)$. 
In addition, $D_{KL}(P,\tilde{P})$ is 
approximately $-\log_2 L$, where $L$ is the average likelihood --- that is, the average (i.e. per observation) relative likelihood that a sequence of data drawn from the distribution $P$ was instead drawn from the model $\tilde{P}$ \cite{CT91, Shlens06}.  For example, if $D_{KL} = 1$, the average likelihood that a single sample from $P$ --- i.e. a single network response --- in fact came from $\tilde P$ is $2^{-1}$ (we use the base 2 logarithm in our definition of the Kullback-Leibler divergence, so all numerical values are in units of bits).

An alternative measure of the quality of the pairwise model comes from normalizing $D_{KL}(P,\tilde{P})$ by the corresponding distance of the distribution $P$ from an \textit{independent maximum entropy} fit $D_{KL}(P,P_1)$, where $P_1$ is the highest entropy
% Note here: use multi-information terminology maybe
% See Schneidman 03
distribution consistent with the mean firing rates of the cells (equivalently, the
independent model $P_1$ is given by the product of single-cell marginal firing probabilities) \cite{amari01}. 
Many studies \cite{Shlens06, Shlens09, Sch+06, roudi09}  use 
\begin{eqnarray}
\Delta & = & \frac{D_{ind} - D_{pair}}{D_{ind}}   \nonumber \\ 
& = & 1- \frac{D_{KL}(P,\tilde{P})}{D_{KL}(P,P_1)}, \label{e.delta}
\end{eqnarray}
where following \cite{roudi09} we define $D_{ind} \equiv D_{KL}(P,P_1)$ and $D_{pair} \equiv D_{KL}(P,\tilde{P})$. 
A value of $\Delta = 1$ ($100 \%$) indicates that the pairwise model perfectly captures the additional
information left out of the independent model, while a value of $\Delta = 0$ indicates that the pairwise model gives no improvement over the independent model.
%, and therefore measures how well the pairwise model \textit{improves} on the independent model. 
Because we are interested in cases where the pairwise model fails to capture circuit outputs, at first it seems we would wish to identify circuits which produce a low $\Delta$.  However, in the circuits we explore, we find that the lowest values of $\Delta$ are achieved for nearly independent -- and therefore ``uninteresting" --  spike patterns. Therefore, we have chosen to use $D_{KL}(P,\tilde{P})$ as our primary measure of the quality of the pairwise model, while reporting values of $\Delta$ in parallel.

% FMR: moved Delta description up in text
% E: looks good

To get an intuitive picture of $D_{KL}(P,\tilde{P})$ throughout the cube of possible distributions $P$, we view this quantity along constant-$f_p$ slices in Figure \ref{fig:cube_schematic}.  $D_{KL} (P,\tilde{P})$ increases with
distance from the constraint curve (Equation \ref{eqn:constraint_fp}); along the iso-moment line for a given $(\mu, \rho)$, $D_{KL}(P,\tilde{P})$ is convex with a minimum of zero at $P=\tilde{P}$ (detailed calculations are given in Materials and Methods).  Therefore, for any choice of $\mu$ and $\rho$, $D_{KL}(P,\tilde{P})$ increases monotonically as a function of the distance along an iso-moment line. The distance, which is easily visualized, thus gives an indication of how close $P$ comes to being a pairwise maximum entropy distribution.  The observed distribution with the maximal deviation from its pairwise maximum 
entropy approximation will occur at one of the two points where the iso-moment line reaches the boundary of the cube.  The \textit{global} maximum of $D_{KL}(P,\tilde{P})$ will, therefore, also occur on the boundary.

To assess the numerical significance of $D_{KL}(P,\tilde{P})$, we can compare it with the maximal achievable value for any symmetric distribution on three spiking cells. For three cells, this value is $1$ 
(or $1/3$ bits per neuron), achieved by the XOR operation \cite{ssbb03} --- i.e. the average probability that a single output of the XOR circuit came instead from the PME approximation is $2^{-1}$.  
This distribution, along with its position in the $(f_p,f_{1p},f_{1m})$ coordinate space, 
is illustrated in Figure \ref{fig:cube_schematic} ($f_p = 0.25$ slice) and the right column of Figure \ref{fig:iso_bar}.
We will find that distributions produced by permutation-symmetric networks fall far short of this value.

In summary, we have shown that identifying high-order interactions in the joint firing patterns of three cells is equivalent to showing that spiking probabilities lie a substantial distance from a constraint surface that is easy to visualize.  Given this geometric description of the problem, we next consider how the distance from the constraint surface depends on circuit connectivity, nonlinear properties of the individual circuit elements, and the statistics of the input signals (Figure \ref{fig:study_schematic}). 

\subsection*{When do triplet inputs produce higher-order interactions in spike outputs?}  \label{sec:N3}

%We then explored high-order correlations that are produced by mechanistic models of spike generation.  To gain insight into the surprising success of pairwise models in classes of retinal ganglion cells \cite{Shlens06, Shlens09} that are known to have small or negligible coupling \cite{tr08}, we restricted our analysis to feedforward circuits.

% FMR: above paragraph seemed out of place and felt like it disrupted flow.  
% E: agreed

We first considered a simple feedforward circuit in which three spiking cells sum and threshold their inputs.
Each cell $j$ received an independent input $I_j$ and a ``triplet'' --- or global --- input $I_c$ that is shared among all three cells.  Comparison of the total input $S_j = I_c + I_j$ with a threshold $\Theta$ determined whether or not the cell spiked in that time bin. The nonlinear threshold can produce substantial differences between input and output correlations~\cite{RochaDoironSJR07, Vilela:2009p370, brown07, me08, BarreiroST10}.  An additional parameter, $c$, identified the fraction of the total input variance $\sigma^2$ originating from the global input; that is, $c \equiv {\rm Var}[I_c]/ {\rm Var}[I_c + I_j]$.

What types of inputs might be natural to consider?  The constraint surface (Equations \ref{eqn:constraint_surface} and \ref{eqn:constraint_fp}) can give us some intuition into what types of inputs will be more or less likely to produce spiking responses that will deviate from the pairwise model. For example, suppose that $I_c$ can take on values that cluster around two separated values, $\mu_A < \mu_B$, but that it is very unlikely that $I_c$ ever takes on values in the interval between; that is, the distribution of $I_c$ is \textit{bimodal}. If $\mu_B$ is large enough to push the cells over threshold but $\mu_A$ is not, then we see that any contribution to the right-hand side of Equation \ref{eqn:constraint_surface}, $p_2/p_1$, depends only on the distribution of the independent inputs $I_j$; if either one or two cells spike, then the common input must have been drawn from the cluster of values around $\mu_A$, because otherwise all three cells would have spiked. Keeping the values of $\mu_A$ and $\mu_B$ the same, therefore, but changing the relative likelihood of drawing the common input from one cluster or the other, would change the ratio $p_3 / p_0$ without changing the ratio $p_2/p_1$.  Hence the constraint specifying those network responses exactly describable by PME models can be violated when the common input is bimodal.

%; moreover, this general effect holds in cases beyond the specific choices of $\mu_A$ and $\mu_B$ in the present example.
% A: removed E's summary

%Similarly, suppose that neither $\mu_A$ nor $\mu_B$ is enough to push the cells above threshold without some push from the independent input, and furthermore that the support of the distribution of $I_i$ has a lower bound such that if the common input is drawn from the cluster around $\mu_A$, then there is no way that $I_i$ can compensate to get cell $i$ to threshold.  Again we find that the ratio $p_2/p_1$ is decoupled from $p_3/p_0$; if any cells fire at all, then the common input must have come from the cluster around $\mu_B$. 

% FMR: removed paragraph above - it was obscure for me
% E:  put in a quick summary at end of paragraph above that one to describe it.  If we needed that paragraph to respond to reviewers, can sharpen and re-include it.

In contrast, we may instead consider a \textit{unimodal} input --- one whose distribution has a single peak or range of most likely values --- of which a Gaussian input is a natural example.  Here, the distribution of the common input $I_c$ is completely described by its mean and variance; both parameters can impact the ratio $p_3/p_0$ (by altering the likelihood that the common input alone can trigger spikes) and the ratio $p_2/p_1$.  Each value of $I_c$ is consistent with both events $p_1$ and $p_2$, with the relative likelihood of each event depending on the specific value of $I_c$; it is no longer clear how to separate the two. 

Motivated by these observations, we choose our inputs to come from either one of three types of unimodal distributions --- Gaussian, skewed, or uniform --- or a bimodal distribution. The one-sided skewed shape, in particular, is chosen to mimic qualitative features of inputs to retinal ganglion cells (e.g. \cite{tr08} and below).  We indeed find both qualitative and quantitative differences in the capacity of each class of inputs to generate deviations from the pairwise model.

\subsubsection*{Unimodal inputs fail to produce higher-order interactions in three cell feedforward circuits} \label{sec:unimodal}
We first considered unimodal inputs, which were chosen from a distribution with a single peak (or range of
most likely values).  Gaussian inputs provide a natural example.  If $I_c$ and each $I_j$ are Gaussian, then the joint distribution of ${\bf S} = (S_1, S_2, S_3)$ is multivariate normal, and therefore characterized entirely by its means and covariances.  Because the PME fit to a continuous distribution is precisely the multivariate normal that
is consistent with the first and second moments, every such input distribution on ${\bf S}$ \textit{exactly} coincides with its PME fit.  However, even with Gaussian inputs, outputs (which are now in the binary state space $\{0,1\}^3$)
will deviate from the PME fit \cite{amari03, macke09}.  As shown below, non-Gaussian unimodal inputs can produce outputs with larger deviations.  Nonetheless, these deviations in all cases are small, and PME models are quite accurate descriptions of circuits with a broad range of unimodal inputs.

In more detail, we considered a circuit of three cells with inputs $I_c$ and $I_j$ that could be Gaussian, uniform, or skewed.  For each type of input distribution, we probed the output distribution across a range of values for $c$, $\sigma$, and $\Theta$ that explored ``all" possible activity patterns.  In particular, we covered a full range of firing rates (or equivalently spikes per time bin), not limited to the low firing rate regime treated in \cite{roudi09}.  Figure \ref{fig:max_DKL_N3}A-C shows observed distributions for different marginal input statistics (left column). The central column compares all observed distributions with the PME constraint curve, projected into the $(f_{1p},f_{1m})$-plane.  

The right column of Figure \ref{fig:max_DKL_N3}A-C shows $D_{KL}(P,\tilde{P}$) as a function of $c$ and $\sigma$ for the value of $\Theta$ that maximized $D_{KL}(P, \tilde{P})$ (or one of them, if multiple such values exist).  Different scales are used to emphasize the structure of the data. For the unimodal cases shown, $D_{KL}$ peaked in regions with comparatively low input variance ($\sigma < 1$) and large relative strength of common input ($c > 0.5$).  However, $D_{KL}(P,\tilde{P})$ never reached a very high numerical value for unimodal inputs; the maximal values achieved for Gaussian, skewed, and uniform distributions are $0.00376$ ($\Delta = 0.989$), $0.0152$ ($\Delta = 0.999$), and $0.0186$ ($\Delta = 0.9428$) respectively (compare with Figure \ref{fig:iso_bar}).  Thus hundreds or thousands of samples would be required to reliably distinguish the outputs of these networks from the PME approximation.

% FMR: added last sentence above
% E:  looks good

%%%%%%%%%%%%%%%%%%%%%%%%%%%%%%%%%%%%%%%%

%%%%%%%%%%%%%%%%%%%%%%%%%%%%%%%%%%%%%%%%

Clear patterns emerged when we viewed $D_{KL}(P,\tilde{P})$ as a function of \textit{output} spiking statistics rather
than \textit{input} statistics.  Figure~\ref{fig:outputstats_N3}A-B show the same data contained in 
the center column of Figure \ref{fig:max_DKL_N3}, but now plotted with respect to the output firing rate, which is the same for all cells.  The data were segregated according to the correlation coefficient $\rho$ between the responses of cell pairs,
%\begin{equation*}
%\rho  =  \frac{{\rm{Cov}}(x_i,x_j)}{\sqrt{ {\rm{Var}}(x_i) {\rm{Var}}(x_j)}} 
% =  \frac{\Ex[x_i x_j]-\Ex[x_i] \Ex[x_j]}{\Ex[x_j^2]-\Ex[x_j]^2}, 
%\end{equation*}
with lighter shades indicating increasing correlation. For a fixed correlation, there was generally a one-to-one relationship
between firing rate and $D_{KL}(P,\tilde{P})$. For unimodal distributions (Figure \ref{fig:outputstats_N3}A-B),
$D_{KL}(P,\tilde{P})$ showed a double-peaked relationship with firing rate, with larger values attained
at low and high firing rates, and a minimum in between.  Additionally, $D_{KL}(P,\tilde{P})$ had a non-monotonic relationship with spike correlation:  it increased from zero for low values of correlation, obtained a maximum for an intermediate value, and then decreased.  These limiting behaviors agree with intuition:  a spike pattern that is completely uncorrelated can be described by an independent distribution (a special case of PME model), and one that is perfectly correlated can be completely described via (perfect) pairwise interactions alone.  

\subsubsection*{Bimodal triplet inputs can generate higher-order interactions in three cell feedforward circuits} \label{sec:bimodal_global}
Having shown that a wide range of unimodal common inputs produced spike patterns that are well-approximated 
by PME fits, 
we next examined bimodal inputs. Figure \ref{fig:max_DKL_N3}D shows
results from a simple ensemble of bimodal inputs --- Bernoulli-distributed common and independent inputs --- 
that produced moderate deviations from the pairwise approximation.
The common input was ``1" with probability $p$ and ``0" with probability $1-p$. 
The independent inputs were each chosen to be ``1" with probability $q$ and ``0" with probability $1-q$. The 
threshold of the cells was between 1 and 2, so that spiking required \textit{both} common and independent inputs to be active. The space of possible spiking distributions was explored by varying $p$ and $q$. 

This circuit produced response distributions that deviated modestly from PME fits, and these distributions preferentially lie on one side of the constraint curve (Figure \ref{fig:max_DKL_N3}D, center).  The largest values of $D_{KL}(P,\tilde{P})$ occurred where moderate correlated input is coupled with strong background input $(q > 0.5$; Figure \ref{fig:max_DKL_N3}D, right), and reached values that are five times higher than those found for a unimodal distribution (the maximal value achieved is $0.091$). The location of the distribution generating this maximum value is demonstrated in the center column of Figure \ref{fig:iso_bar}.

Both of these observations can be explained by direct calculation of the spiking probabilities. Substituting the probabilities of different events
--- $p_0  = 1-p + p(1-q)^3$, $p_1 = pq(1-q)^2$, $p_2  =  pq^2(1-q)$ and $p_3 =  pq^3$ ---
into the constraint surface specifying PME models (Equation \ref{eqn:constraint_surface}) and dividing by $q^3$, we can write
\begin{eqnarray}
\frac{p}{1-p + p(1-q)^3} & = & \frac{1}{(1-q)^3}
\end{eqnarray}
which gives us an intuition for how to violate the constraint; for a fixed $q$,
we manipulate the left-hand side by changing $p$.

Another way to view this is by observing that the right hand side of Equation \ref{eqn:constraint_surface}
can be written without reference to the probability of common input; because $P[1\, {\rm spike} \mid I_c = 0] = 0$ and $P[2\, {\rm spikes} \mid I_c = 0] = 0$, one may write
\begin{eqnarray}
\frac{p_2}{p_1} & = & \frac{P[2 \, {\rm spikes} \mid I_c = 1] P[I_c = 1]}{P[1 \, {\rm spike} \mid I_c = 1] P[I_c = 1]} \nonumber \\
& = & \frac{P[2 \, {\rm spikes} \mid I_c = 1]}{P[1 \, {\rm spike} \mid I_c = 1] }
\end{eqnarray}
which has no dependence on the statistics of the common input. So the \textit{left} hand side of the constraint equation (Eqn.  \ref{eqn:constraint_surface}) can be 
manipulated by shifting $p$, without making any changes to the right hand side.

In Figure \ref{fig:outputstats_N3}C, we again present values of $D_{KL}(P,\tilde{P})$ as a function of the firing rate and pairwise correlation elicited by the full range of possible bimodal inputs.  We see that $D_{KL}(P,\tilde{P})$ is maximized at a single, intermediate firing rate, and for correlation values near 0.6. 

We find distinctly different patterns when we view $\Delta$ (i.e. $D_{KL}$ normalized by the distance of $P$ from an independent maximum entropy fit, Equation \ref{e.delta}), for these same simulations, as a function of output spiking statistics (Figure~\ref{fig:outputstats_N3}D-F).
For unimodal distributions (Figure~\ref{fig:outputstats_N3}D-E), $\Delta$ is very close to 1, with the few exceptions at extreme firing rates. For bimodal inputs (Figure~\ref{fig:outputstats_N3}F), 
$\Delta$ may be appreciably far from 1 --- as small as 0.5 --- with the smallest numbers (suggesting a poor fit of the pairwise model) occurring for low correlation $\rho$. This highlights one interesting example where these two metrics for judging the quality of the pairwise model, $D_{KL}(P,\tilde{P})$ and $\Delta$, yield contrasting results.

\subsubsection*{An analytical explanation for unimodal vs. bimodal effects}

We next develop analytical support for the distinct impact of unimodal vs. bimodal inputs on fits by the PME model.  Specifically, we calculate $D_{KL}(P,\tilde{P})$ for small deviations from the PME constraint surface of Equation \ref{eqn:constraint_surface}.  We summarize the results of this calculation here; details are in Materials and Methods.  We considered narrow distributions of common input $I_c$, with a small parameter ${\bf c}$ indicating their variance.  By approximating the distribution of network outputs by a Taylor series in ${\bf c}$, we found that the leading order behavior of $D_{KL}(P,\tilde{P})$ depended on ${\bf c^3}$ for unimodal distributions --- i.e the low order terms in ${\bf c}$ dropped out (for \textit{symmetric} distributions, such as Gaussian, the growth was even smaller: ${\bf c^4}$).  For bimodal distributions, on the other hand, the leading order term of $D_{KL}(P, \tilde{P})$ grew like ${\bf c^2}$.  
The key point is that, as the strength of common input signals increased, circuits with bimodal inputs diverged from the PME fit much more rapidly than circuits with unimodal inputs. 
 
 %Moreover, a similar analysis can be done for $c$ near 1, yielding similar results where $D_{KL}(P,\tilde{P})$ depended on ${\bf  (1-c)^4}$ or  ${\bf  (1-c)^3}$ for unimodal distributions and  ${\bf  (1-c)^2}$ for bimodal distributions.  Thus, the same conclusion that circuits with bimodal inputs diverge more rapidly from PME fits holds for strong common inputs as well.   
 % AKB: commented out
  
\subsection*{When do pairwise inputs produce higher-order interactions in spike outputs?} \label{sec:anatomical} 
In the previous sections, we considered permutation-symmetric distributions generated by a single, global, common input. Another class of permutation-symmetric distributions can be generated when common inputs are shared \textit{pairwise} --- i.e. by two cells but not three at once. We now show that significant departures from the pairwise maximum entropy model (PME) can be generated with pairwise bimodal inputs.  This provides a specific example in which network architecture and output statistics are not in simple correspondence.

Our circuit setup included three cells, each of which received and summed two inputs and spiked if this sum exceeded a threshold.
We denote the inputs $I_{12}$, $I_{23}$, $I_{13}$ so that cell 2 received $I_{12}$ and $I_{23}$, and so forth, as illustrated in Figure \ref{fig:study_schematic}.  Each input was chosen from a binary distribution
with parameters $m$ and $r$ so that $P[I_{ij} = m] = r$ and $P[I_{ij} = 0] = 1-r$.
Without loss of generality, we chose $m = 1$ and chose the threshold such that $1 < \Theta < 2$. Therefore, both
pairwise inputs to a cell must be active in order for a cell to fire.
It is not possible in this circuit for precisely \textit{two} cells to fire; for two cells to fire (say cell 1 and cell 2), 
both inputs to each cell must be active. However, this implies that both
inputs to cell 3 ($I_{13}$ and $I_{23}$) are active as well. If two cells fire, then the third must fire as well; that is, $p_2 = 0$.

The remaining probabilities are easily computed by itemizing and computing the
probabilities for each event and are as follows: $p_3 = r^3$, $p_1 =  r^2(1-r)$, and $p_0 = 3r(1-r)^2 + (1-r)^3$.
This distribution has a unique PME fit consistent with both the first and second moments.  However, the PME fit can be far from the actual distribution; we found that $D_{KL}(P,\tilde{P})$ depended on the input rate $r$ and could exceed $0.5$.  Thus observation of the network response (single draw from $P$) would, on average, have a likelihood ratio of 0.7 of coming from $\tilde{P}$ versus $P$, and observation of 15 network responses would have a likelihood ratio of less than $0.01$. 

We will revisit the contrast between global and pairwise common inputs below; we also refer to the pairwise case as {\it local} inputs.

\subsection*{Recurrent coupling produces modest effects on higher-order interactions}
Neural circuits are rarely purely feedforward, and recurrent connectivity can play an important role in shaping the resulting activity. Therefore, we next modify our thresholding model to incorporate the effects of recurrent coupling among the spiking cells.  We take all-to-all coupling among our $N=3$ cells, and assume that this coupling is rapid compared with the timescale over which inputs arrive at the cells, so that coupling has its full impact on the spike events that are recorded in a single realization of the model.  We limit our study to excitatory interactions, anticipating an application below to the effects of gap junctions known to mediate recurrent coupling between retinal ganglion cells.

In more detail, our coupling model may be described as follows: first, inputs arrive at each cell as for the cases without coupling.  If the inputs elicit any spikes, there is a second stage in which the input to each neuron receiving a connection from a spiking cell is increased by an amount $g$.  This represents a rapid depolarizing current, assumed for simplicity to add linearly to the input currents.  If the second stage results in additional spikes, the process is repeated:  recipient cells receive an additional current $g$, and their summed inputs are again thresholded.  The sequence terminates when no new spikes occurred on a given stage; e.g., for $N=3$, there are a maximum of three stages.  The spike pattern that is recorded on a given trial is the total number of spikes fired across all of the stages.

For the simplest case of globally structured inputs and all-to-all coupling, the probability of each spike pattern (or count) can be written down analytically, by evaluating the probabilities along the tree of possible events that unfold through the stages just described.  For other cases, these trees are more complex, and we instead evaluate probabilities via Monte-Carlo sampling (unimodal inputs), or through enumeration of a finite number of possible input states (bimodal inputs).

We begin by describing our results for globally structured, unimodal inputs.  We range over parameters $c$, $\sigma$, and $\Theta$ as for the uncoupled cases above, and additionally vary $g$ over a wide range of values, from $g=0$ to $g=2.4$ (comparable to the maximum threshold value).  Our basic finding is that coupling has only modest effects on $D_{KL}$ values, when compared with the theoretical range of values that could be obtained.  Specifically, the largest value found over all parameters tested was $D_{KL} = 0.0099$ for the case of Gaussian inputs, 0.108 for uniform, and 0.0599 for skewed.  All of these values are roughly $3-5 \times$ values obtained for the uncoupled cases, and, for Gaussian and skewed inputs, remain much smaller than values found for bimodal inputs.  The largest value achieved over any unimodal input and for any coupling strength ($0.108$) is comparable to 
the largest value found with bimodal inputs without coupling. 

Figure~\ref{f.DKL_vs_coup}A shows the data for Gaussian inputs as a scatter plot of $D_{KL}$ values vs. coupling strengths $g$, showing that intermediate values of coupling --- for which the size of coupling terms is about the same as the standard deviation of the inputs ---  have the greatest ability to drive departures from the PME model.  Moreover, when coupling strength was varied while keeping other parameters fixed, $D_{KL}$ was also generally maximized at intermediate values of the coupling.
In Figure~\ref{f.DKL_vs_coup}B, we show $D_{KL}(P,\tilde{P})$ vs. coupling strength $g$, for representative values of the circuit parameters $c$, $\sigma$ and $\Theta$.  For each case of global Gaussian, skewed, uniform, and bimodal inputs (thin lines), we see a peak in $D_{KL}(P,\tilde{P})$ at medium values of $g$.

% as  is found where $g \approx 1.2$. The center panel shows the result from a circuit forced with uniform inputs; the maximal value of  is found where $g \approx 0.9$. 

We next introduce coupling to circuits with pairwise, locally structured unimodal inputs (as in the preceding section). 
Here, we range over parameters $\sigma$ and $\Theta$ ($c$ is fixed at 0.5), and additionally vary $g$ over the same wide range of values above. 
For local inputs, the maximum value of $D_{KL}$ found could be increased modestly by including coupling (from $0.0013$ to $0.0086$ with Gaussian inputs,
from $0.0150$ to $0.0157$ with skewed inputs, and from $0.0233$ to $0.0547$ with uniform inputs), but remains substantially below the values found for global inputs with coupling. As in circuits with global inputs, $D_{KL}$ was generally maximized at intermediate values of the coupling; this is illustrated for uniform inputs in the third panel of \ref{f.DKL_vs_coup}B (note that global and local inputs are identical for Gaussian inputs).
An exception was observed for skewed inputs, illustrated in the top panel of \ref{f.DKL_vs_coup}B, where $D_{KL}$ shows a maximum at $g=0$.

Finally, we introduce coupling to circuits with bimodal inputs. We note that because the inputs can only take on a finite number of configurations, with all other parameters fixed, the output distribution of the network can only change at a few discrete values of $g$. Moreover, large values of $g$ induce ``all-or-none" firing behavior, which is perfectly matched by pairwise maximum entropy.
For global inputs, we find modest increases in $D_{KL}$ which are 
maximized at an intermediate value of the coupling; for pairwise inputs, the output distribution attains its maximum level of higher order interactions at $g = 0$ and decreases only when $g$ is sufficiently strong enough to create an ``all-or-none" firing pattern (bottom panel, Fig.~\ref{f.DKL_vs_coup}B).

% AKB: commented list below
%\bei
%\item Bimodal inputs plus coupling?
%\item Local coupling
%\item $N>3$
%\item Revisit these results -- at least one showing maximal effect of $g$ when it's about 1 standard deviation of unimodal inputs -- below, in RGC section 
%\eei

%\noindent{\it Mid-course summary:} 
In summary, both input statistics and connectivity shape the development of higher-order interactions in our basic circuit model.  Other parameters being equal, bimodal inputs can generate a larger $D_{KL}(P,\tilde{P})$ than unimodal inputs.  For a particular choice of input marginals, global inputs can generate greater deviations than purely pairwise inputs (with the exception of one case, that locally structured bimodal inputs). However, the goodness of the PME fit alone does not distinguish between global and pairwise anatomical projections.  Adding fast recurrent excitation 
%sharply 
increases the accessible magnitude of higher-order interactions;
%, if the threshold $\Theta$ is allowed to vary;
for a fixed $\Theta$, the range of accessible $D_{KL}$ increases --- and then decreases --- with the magnitude of recurrent excitation.
With these principles in hand, we now study a more biologically realistic network model. 

%%%%%%%%%%%%%%%%%%%%%%%%%%%%%%%%%%%%%%%%

\subsection*{An experimentally constrained model for correlated firing in retinal ganglion cells} \label{s:RGC}
PME approaches have been effective in capturing the activity of small retinal ganglion cell (RGC) populations~\cite{Sch+06,Shlens06,Shlens09}.  This success does not have an obvious anatomical correlate --- i.e. there are multiple opportunities in the retinal circuitry for interactions among three or more ganglion cells.  Why do these apparently fail to generate higher-order interactions in output spiking data?  To answer this question, we explored the properties of circuits composed of cells with input statistics, recurrent connectivity and spike-generating mechanisms based directly on experiment.  We based our model on ON parasol RGCs, one of the RGC types for which PME approaches have been applied extensively~\cite{Shlens06,Shlens09}.  We first describe the RGC model, then apply this to feedforward circuits, and finally consider recurrent connections.

% FMR: added last sentence - also reorganized sections to go with this sentence
% E:  looks good

\subsubsection*{RGC model}
We modeled a single ON parasol RGC in two stages (for details see Materials and Methods).  First, we characterized the light-dependent excitatory and inhibitory synaptic inputs to cell $k$ $(g^{\text{exc}}_k(t), g^{\text{inh}}_k(t))$ in response to randomly fluctuating light inputs $s(t)$ via a linear-nonlinear model, e.g.:
\begin{eqnarray}
g^{\text{exc}}_k(t) & = & N^{\text{exc}}[L^{\text{exc}} \ast s_k(t) + \eta^{\text{exc}}_k], \label{eq:g_exc}
\end{eqnarray}
where $N^{\text{exc}}$ is a static nonlinearity, $L^{\text{exc}}$ is a linear filter, and $\eta^{\text{exc}}_k$ is an effective input noise that captures variability in the response to repetitions of the same time-varying stimulus.  These parameters were determined from fits to experimental data collected under conditions similar to those in which PME models have been tested empirically.  The modeled excitatory and inhibitory conductances captured many of the statistical features of the real conductances, particularly the correlation time and skewness.   

% FMR: added last sentence
% E:  looks good

Second, we used Equation \ref{eq:g_exc} and an equivalent expression for $g^{\text{inh}}_k(t)$ as inputs to an integrate-and-fire model incorporating a nonlinear voltage and history-dependent term to account for refractory interactions between spikes~\cite{badel07}. The voltage evolution equation was of the form
\begin{eqnarray}
\frac{dV}{dt} & = &  F(V, t-t_{last}) + \frac{I_{input}(t)}{C},   \label{eqn:fI_curve_intext}
\end{eqnarray}
where $F(V,t-t_{last})$ was allowed to depend on the time of the last spike $t_{last}$ (more details in Methods).
We fit the parameters of this model to a dynamic clamp experiment~\cite{sharpe1993,Murphy2006} in which currents corresponding to $g^{\text{exc}}(t)$ and $g^{\text{inh}}(t)$ were injected into a cell and the resulting voltage response measured.  Excitatory and inhibitory synaptic currents injected during one time step were determined by scaling the conductances by driving forces based on the measured voltage in the previous time step.   
Recurrent connections were implemented by adding an input current proportional to the voltage difference between the two coupled cells.  

% FMR: clarified dynamic clamp above a little
% E:  looks good

The prescription above provided a flexible model that we used to study the responses of small RGC networks to a wide range of light inputs and circuit connectivities.  Specifically, we simulated RGC responses to light stimuli that were (1) constant, (2) time-varying and spatially uniform, and (3) varying in both space and time.  Correlations between cell inputs arose from shared stimuli, from shared noise originating in the retinal circuitry~\cite{tr08}, or from recurrent connections~\cite{Dacey1992,tr08}.  Shared stimuli were described by correlations among $s_k$.  Shared noise arose via correlations in $\eta_k$ as described in Materials and Methods.  The recurrent connections were chosen to be consistent with observed gap-junctional coupling between ON parasol cells.  We also investigated how stimulus filtering by  $L^{\text{exc}}$ and $L^{\text{inh}}$ influenced network statistics.

For each network model, we determined whether the accuracy of a PME fit to the outputs was predictable based on the structure of the input distributions, using the results developed above for the idealized circuit model.  We focused on excitatory conductances because they exhibit stronger correlations than inhibitory conductances in ON parasol RGCs \cite{tr08}.  To compare our results with empirical studies, constant light and spatially and temporally fluctuating checkerboard stimuli were used as in \cite{Shlens06, Shlens09}. 

% FMR: reorganized above two paragraphs
% E:  looks good

\subsubsection*{Feedforward RGC circuits}
We start by considering networks without recurrent connectivity and without temporal modulations in the light input.  Thus we set $s_k(t)=0$ for $k=1,2,3$, so that the cells received only Gaussian correlated noise $\eta^{\text{exc}}_k$ and $\eta^{\text{inh}}_k$ and constant excitatory and inhibitory conductances.  Time-dependent conductances were generated and used as inputs to a simulation of three model RGCs.  Simulation length was sufficient to ensure significance of all reported deviations from PME fits (see Materials and Methods).  Under these conditions the excitatory conductances were unimodal and broadly Gaussian.  As expected from earlier results on threshold models, the spiking distributions were well-modeled by a PME fit, as shown in Figure \ref{fig:fullfield_composite}A; $D_{KL}(P,\tilde{P})$ is $2.90 \times 10^{-5}$ bits. This agrees with the very good fits found experimentally in \cite{Shlens06} under constant light stimulation.

For full-field simulations, each cell received the same stimulus, $s_k(t)=s(t)$, where $s(t)$ refreshes every few milliseconds with an independently chosen value from one of several marginal distributions. The shared stimulus produced strong pairwise correlation between conductances of neighboring cells. However, results from our threshold model (Fig.~\ref{fig:max_DKL_N3}) suggest that this is not the overall determining factor in whether or not spiking outputs will be well-modeled by a PME fit; rather, the shape of the marginal distribution of inputs (here, conductances) should be more important than the strength of pairwise correlations. 

% FMR: removed "excitatory" above - aren't inhibitory conductances also correlated?
% E:  looks good

We first examined the effects of different marginal statistics of light stimuli, standard deviation of full-field flicker, and refresh rate on the marginal distributions of excitatory conductances.  For a short refresh rate ($8$ ms) and small flicker variance ($1/6$ or $1/4$ of baseline light intensity), temporal averaging via the filter $L^{\text{exc}}$ and the approximately linear form of $N^{\text{exc}}$ over these light intensities produced a unimodal, modestly skewed distribution of excitatory conductances, regardless of whether the flicker is drawn from a Gaussian or binary distribution (see Figure \ref{fig:fullfield_composite}B-C, center panels). For a slower refresh rate ($100$ ms) and large flicker variance ($1/3$ or $1/2$ of baseline light intensity), excitatory conductances had multi-modal and skewed features, again regardless of whether the flicker is drawn from a Gaussian or binary distribution (Figure \ref{fig:fullfield_composite}D). Other parameters being equal, binary light input produced more skewed conductances.  While some conductance distributions had multiple local maxima, these were never well-separated, with the envelope of the distribution still resembling a skewed distribution. 

As expected from our studies with the simple thresholding model of spike generation, the largely unimodal shape of input distributions was reflected in the ability of PME fits to accurately capture spiking distributions.  $D_{KL}(P,\tilde{P})$ values computed from the observed distributions were small, never exceeding $0.0067$; these are numbers comparable to what is achievable by skewed inputs in our simple thresholding circuit model.  To test the sensitivity of this conclusion to the finite sampling in our simulations, we performed an analysis in which the data was divided into 20 subsets and the maximum entropy analysis was performed individually on each subset.  The resulting KL-distance remained small, never exceeding $0.0089$.  In summary, even high-contrast, bimodal, highly spatially correlated stimulus variations do not produce a large departure from the PME fit.

When we examined all of the spiking distributions produced in this sequence of simulations, we found a common pattern in the way in which the PME fit deviated from observed distributions.  Single spiking events were over-predicted by PME fits, whereas double spiking events were under-predicted.  We note that this is the same situation observed in our simple threshold model with bimodal global inputs (see Figure 3 and Materials and Methods), and corresponds to the case of negative strain identified by Ohiorhenuan et al. \cite{ovJCN10}.  This finding is extremely robust; upon perturbing the distributions by estimated standard errors, as described in Materials and Methods, only 22 out of 840 perturbed distributions showed a positive strain while the remainder had negative strain.

% FMR: added sentence above about negative strain - is this right? AKB: YES
% E:  looks good

Overall, high-pass filtering --- a consequence of the differentiating linear filter in Equation \ref{eqn:linear_filter} and illustrated in Figure \ref{fig:fullfield_composite}D --- was responsible for significantly reducing the bimodality of the input stimuli.  
%%%%%%%%%%%%%%%%%%%
%  (compare Figure \ref{fig:fullfield_composite}C, left panel (input, blue histogram) --- which illustrates marginal statistics with the same shape but lower variance --- vs. Figure \ref{fig:fullfield_composite}D, center panel (output, conductance histogram)).  
% AKB --- I don't really under stand the placement of this phrase here. This seems to address the point in the previous paragraph.
In Figure  \ref{fig:fullfield_composite}E, we produce a filter without the biphasic, high-pass shape of Equation \ref{eqn:linear_filter} (i.e., without the negative dip at longer time lags) by simply rectifying the experimentally determined filter. 
This filter produced conductance distributions that more completely reflect the bimodal shape of binary light inputs (Figure \ref{fig:fullfield_composite}E, center panel).  The resulting simulation produces six-times greater $D_{KL}(P,\tilde{P})$  (Figure \ref{fig:fullfield_composite}E, right panel).  This raises the intriguing suggestion that greater $D_{KL}(P,\tilde{P})$ may occur for other cell types primarily characterized via monophasic filters (such as midget cells), or for different light stimuli (e.g., at different mean levels) for which the retinal circuit acts to primarily integrate rather than differentiate over time.

% FMR: edited last sentence
% E:  looks good

In Figure \ref{fig:rectified_vs_orig}A, we examine this effect over all full-field stimulus conditions by plotting deviations from the pairwise model, in terms of $D_{KL}(P,\tilde{P})$, in simulations with a rectified filter against the same quantity with a non-rectified filter.  An increase in $D_{KL}(P,\tilde{P})$ was observed across stimulus conditions, with a markedly higher effect for longer refresh rates. Large effects were accompanied by a striking increase in the bi- or multi-modality of excitatory conductances (see Figure \ref{fig:rectified_vs_orig}B, illustrating binary stimulus at $100$ ms refresh rate and standard deviation of $1/4$), 
while small effects were accompanied by small or no discernible change in the marginal distribution of excitatory synaptic conductances (see Figure \ref{fig:rectified_vs_orig}C, illustrating binary stimulus at $8$ ms refresh rate and standard deviation of $1/2$).  This consistent change could not be attributed to changes in lower order statistics; there was no consistent relationship between the change in pairwise model performance and either firing rate or pairwise correlations (data not shown). 

Moving beyond full field light stimuli, we asked whether pairwise maximum entropy models capture RGC responses to stimuli with varying spatial scales.  We fixed stimulus dynamics to match the two cases that yielded the highest $D_{KL}(P,\tilde{P})$ under the full field protocol: for both Gaussian and binary stimuli, 
this was an $8$ ms refresh rate and $\sigma = 1/2$.  The stimulus was generated as a random checkerboard with squares of variable size; each square in the checkerboard, or \textit{stixel}, was drawn independently from the appropriate marginal distribution and updated at the corresponding refresh rate.  The conductance input to each RGC was then given by convolving the light stimulus with its receptive field, where the stimulus is positioned with a fixed rotation and translation relative to the receptive fields. This position was drawn randomly at the beginning of each simulation and held constant throughout (see Figure \ref{fig:stixel}D,G for examples, and Materials and Methods for further details).    

The RGC spike patterns remained very well described by PME models for the full range of spatial scales.  Figure \ref{fig:stixel}A shows this by plotting $D_{KL}(P,\tilde{P})$ vs. stixel size.  Values of $D_{KL}(P,\tilde{P})$ increased with spatial scale, sharply rising beyond  $128 \, \rm{\mu m}$, where a stixel is approximately the same size as a receptive field center. The points at $512 \, \rm{\mu m}$ are from the corresponding {full field} simulations, illustrating that introducing spatial scale via stixels produces even closer fits by PME models.  

%remains low across simulations, with a maximum value achieved of $0.063$. 

Values reported in Figure \ref{fig:stixel}A are \textit{averages} of  $D_{KL}(P,\tilde{P})$ produced by 5 random stimulus positions.  At stixel sizes of $128 \,  \rm{\mu m}$ and $256 \,  \rm{\mu m}$, the resulting spiking distributions differed significantly from position to position; in Figure \ref{fig:stixel}B, we show the probabilities of the distinct singlet (e.g. $P(1,0,0)$) and doublet (e.g. $P(1,1,0)$) spiking events produced at $256 \,  \rm{\mu m}$.
Each stimulus position creates a ``cloud" of dots (identified by color); large dots show the average over 20 sub-simulations. Each sub-simulation is identified by a small dot of the same color; because the simulations are very well-resolved, most of these are contained within the large dots (and hence not visible here). Heterogeneity across stimulus positioning is indicated by the distinct positioning of differently colored dots.
At smaller spatial scales, the process of averaging stimuli over the receptive fields results in spiking distributions that are largely unchanged as stimulus position changes, as shown in Figure \ref{fig:stixel}C, where singlet and doublet spiking probabilities are plotted for $60 \,  \rm{\mu m}$ stixels.  Thus, filtered light inputs are largely homogeneous from cell to cell, as each receptive field samples a similar number of independent, statistically identical inputs: Figure \ref{fig:stixel}D shows the projection of input stixels onto cell receptive fields from an example with $60 \,  \rm{\mu m}$ stixels. The resulting excitatory conductances (Figure \ref{fig:stixel}E) and spiking patterns (Figure \ref{fig:stixel}F) are very close to cell-symmetric.

By contrast, spiking patterns showed significant heterogeneity from cell to cell,when the stixel size was large; this arises because each cell in the population may be located differently with respect to stixel boundaries, and therefore receive a distinct pattern of input activity. Figure \ref{fig:stixel}G shows the projection of input stixels onto cell receptive fields from one example, with $256 \,  \rm{\mu m}$ stixels. The difference in statistical properties of inputs is reflected in the marginal distributions of excitatory conductances that each cell receives, shown in Figure \ref{fig:stixel}H. It is also apparent in the observed output spiking patterns, where one particular doublet spiking pattern (here, $110$) is significantly more likely to occur than the others (see Figure \ref{fig:stixel}I). However, PME models gave excellent fits to data regardless of heterogeneity in RGC responses, as illustrated for this particular example in Figure \ref{fig:stixel}I; over all 20 sub-simulations (as above), and over all individual stixel positions, we found a maximal $D_{KL}(P,\tilde{P})$ value of $0.00811$. 

\subsubsection*{Recurrent connectivity in the RGC circuit}
We next considered the role of recurrence in shaping higher order interactions by incorporating gap junction coupling into our simulations.  We did this separately for each full field stimulus condition:  Gaussian and binary marginal distributions, with variances $1/16$, $1/12$, $1/8$, $1/6$, $1/4$, $1/3$, or $1/2$ of baseline light intensity, and refreshed at either 8, 40, or 100 ms intervals.  In each case, we added gap junction coupling with strengths from 1 to 16 times an experimentally measured value \cite{tr08}, and compared the resulting $D_{KL}$ with that obtained without recurrent coupling.  Results are shown in Figure \ref{fig:recur_vs_orig}.  

At the measured coupling strength ($g^{\text{gap}}=1.1$ nS) itself, the fit of the pairwise model barely changed (Figure \ref{fig:recur_vs_orig}B). At twice the measured coupling strength ($g^{\text{gap}}=2.2$ nS), recurrent coupling had increased higher order interactions, as measured by larger values of $D_{KL}$ for all tested stimulus conditions. Higher order interactions could be further increased, particularly for long refresh rates (100 ms), by increasing the coupling strength to 4 or 8 times its baseline level ($g^{\text{gap}}=8.8$ nS; see Figure \ref{fig:recur_vs_orig}C,\,D). Consistent with our intuition that very strong coupling leads to ``all-or-none" spiking patterns, $D_{KL}(P,\tilde{P})$ decreased as $g^{\text{gap}}$ increased further, often to a level below what was seen in the absence of coupling (Figure \ref{fig:recur_vs_orig}F). 

Again, we considered the possibility that firing rates could explain the increase; however, firing rates actually decreased with the addition of coupling.  Our findings are consistent with our results for the threshold model (Figure \ref{f.DKL_vs_coup}), in that the impact of coupling is maximized at intermediate values of the coupling strength. Moreover, the impact of recurrent coupling on the maximal values of $D_{KL}$ evoked by visual stimuli is small overall, and almost negligible for experimentally measured coupling strengths. 

%F:  need some more details above - what marginals and other stimulus parameters?  Generally could flesh out a little. 
% E:  I think what is now there should be close to enough detail.

\subsection*{Scaling of higher-order interactions with population size} \label{s:largeN}
The results above identify the input statistics --- particular unimodality vs bimodality --- as a key determinant of the success of PME models.  How robust is this conclusion to network size?  The permutation-symmetric architectures we have considered can be scaled up to more than three cells in several natural ways; for example, we can consider $N$ cells with a global common input. The pairwise (local) input structure can also be scaled up to consider $N$ cells on a ring, with each pair of adjacent cells receiving a common, pairwise input (see Figure \ref{fig:study_schematic}). 

%We next used the methods described in the previous section to study networks with these architectures and sizes up to $N=16$.

% FMR: added a few sentence to start of this section.  Also commented out sentence above - ref to "previous section" was not clear and it seemed redundant with rest of paragraph
% E:  looks good

We first considered a sequence of models in which a set of $N$ threshold spiking units received global input $I_c$ (with mean $0$ and variance $\sigma^2 c$) and an independent input $I_j$ (with mean $0$ and variance $\sigma^2 (1-c)$).  As for the three cell networks above, the output of each cell was determined by summing and thresholding these inputs.  The probability distribution of network outputs was computed as described in the Methods and then fit with a pairwise maximum entropy distribution.  As also for the three cell networks, we explored a range of $\sigma$, $c$, and $\Theta$ and recorded the maximum value of $D_{KL}(P,\tilde{P}) $ between the observed distribution $P$ and its PME fit $\tilde{P}$.  Figure \ref{fig:max_DKL_Nmedium} shows this $D_{KL}/N$ (i.e. entropy per cell \cite{macke09}) for Gaussian, uniform, skewed, and bimodal input distributions. 

We found that the maximum $D_{KL}(P,\tilde{P})$ increased roughly linearly with $N$ for bimodal inputs, and superlinearly for unimodal inputs.  The relative ordering found at $N = 3$ --- that the maximal achievable $D_{KL}(P,\tilde{P})$ is lowest for Gaussian inputs, followed by skewed, uniform, and bimodal inputs consecutively --- remained the same.  
The sidebar of Figure \ref{fig:max_DKL_Nmedium} shows that the probability distributions produced by these inputs qualitatively agree with this trend:  departures from PME were more visually pronounced for global bimodal inputs (top histogram) than for global unimodal inputs (third histogram from top).  At $N=16$, the value $D_{KL}/N \approx 0.1$ for bimodal global inputs corresponds to a likelihood ratio of 0.33 that a single draw from $P$ (single network output) in fact came from the PME fit $\tilde P$ versus $P$; a likelihood $< 0.01$ is reached for 4 draws. 

We next considered pairwise inputs for $N>3$ cells by adopting a ring structure with nearest-neighbor common inputs (illustrated in Figure \ref {fig:study_schematic}).  For unimodal inputs, we computed $D_{KL}(P,\tilde{P})$ while varying $\sigma$ and $\Theta$; for bimodal inputs, we varied the probability of each Bernoulli input.  Figure \ref{fig:max_DKL_Nmedium} shows the maximal 
$D_{KL}(P,\tilde{P})$ per neuron.  Overall, circuits with bimodal pairwise inputs showed appreciable values that are  about half of that found for bimodal global inputs. The relatively large deviation at $N=3$ receded, replaced by deviations that were similar to those seen for global, unimodal inputs. For pairwise unimodal inputs, values of $D_{KL}(P,\tilde{P})$ remained very small.
%, until $N \gtrapprox 10$.  

Finally, we extended our recurrent model to $N>3$ cells. As for the three cell networks, we explored a range of $\sigma$, $c$, and $\Theta$ for unimodal inputs, or $s$, $p$, and $\Theta$ for bimodal inputs. In addition, the coupling strength, $g$, was varied for each type of input. Coupling was either all-to-all (global) or applied pairwise in a ring structure (local). Figure \ref{fig:max_DKL_Recur} shows the maximal 
$D_{KL}(P,\tilde{P})$ per neuron, for each type of input and network architecture,
up to population size $N=8$. Recurrent coupling increases the available range of higher order interactions in most settings from the level achieved with purely feed forward connections. However, the maximal value of  $D_{KL}(P,\tilde{P})$ per cell is steady or decreasing with $N$, suggesting that recurrence has less impact as population size grows.

To summarize, the greater impact of bimodal vs. unimodal input statistics on maximal values of $D_{KL}(P,\tilde{P})$ that can be obtained in a given circuit persists from $N=3$ up to $N=16$ (tested up to $N=8$ in the recurrent case).  Moreover, agreeing with intuition, global inputs of a given type can generate greater deviations than purely pairwise inputs (with the exception of one case, $N=3$).  Overall (again excepting this one case), for the circuit parameters producing maximal deviations, it becomes easier to statistically distinguish between spiking distributions and their PME fits as $N$ increases in feedforward networks. In contrast, the maximal attainable
$D_{KL}(P,\tilde{P})$ per cell appears to decrease with $N$ in many recurrent networks.

\section*{Discussion}

We use simple mechanistic models to identify which combinations of network architectures and input signals produce spike patterns with higher-order interactions and which do not.  Deviations in circuit outputs from pairwise maximum entropy (PME) predictions were much smaller than the maximum theoretically attainable values for a general spiking pattern.  Moreover, output statistics were not simply related to network architecture.  Nonetheless, several simple principles emerge that determine the strength of higher-order interactions.  First, bimodal input distributions produced stronger higher-order interactions than unimodal distributions.  Second, networks with shared inputs among all cells produced greater higher-order interactions than those with pairwise inputs (except for the case of three cell networks receiving bimodal input).  Third, recurrent excitatory or gap junction coupling could produce a further, moderate increase higher-order correlations; the effect was greatest for coupling of intermediate strength.  Our overall results held for networks with nonlinear integrate-and-fire units based on measured properties of retinal ganglion cells.  Together with the facts that ON parasol cell filtering suppresses bimodality in light input, and that coupling among ON parasol cells is relatively weak, our findings provide an explanation for why their population activity is well captured by PME models.  

% FMR: simplified above paragraph and updated not to focus on feedforward
% E: looks good

\subsection*{Comparison with empirical studies}
How do our maximum entropy fits compare with empirical studies? 
In terms of $D_{KL}(P,\tilde{P}$) --- equivalently, the logarithm of the average relative likelihood that a sequence of data drawn from $P$ was instead drawn from the model $\tilde{P}$ --- numbers 
obtained from our RGC models are very similar to those obtained by experiments on retinal ganglion cells \cite{Shlens06,Shlens09}. We find that $D_{KL}(P,\tilde{P}) = 2.90 \times 10^{-5}$ bits under constant light conditions, compared to an experimental value of
$0.0008$ \cite{Shlens06} (inferred from a reported likelihood ratio of $0.99944$).  Under full-field, time varying light conditions, as well as spatiotemporally varying stixel simulations, we find average log-likelihood ratios of up to one order of magnitude larger -- bounded above by $0.007$.  We can view this as a model of the checkerboard experiments of  \cite{Shlens06}, for which close fits by PME distribution were also observed (likelihood numbers were not reported).  Similarly, the values of $\Delta$ that are produced by our RGC model are close to those found by \cite{Shlens06, Sch+06} under comparable stimulus conditions. 
We obtain $\Delta = 99.5 \%$ (for cell group size $N=3$) under constant illumination, which is near the range reported by \cite{Shlens06} ($97.8 - 99.2 \%$, $N=3-7$).  For full-field stimuli we find a range of numbers from $95.7-99.3 \%$ ($N=3$).  

The simple threshold models that we have developed, meanwhile, give us a roadmap for how circuits could be driven in such a way as to lower $\Delta$.  Figure \ref{fig:outputstats_N3}D-F show $\Delta$ plotted as a function of firing rate for the data presented in Figure \ref{fig:outputstats_N3}A-C:  circuits of $N=3$ cells receiving global common inputs.  We observe that $\Delta \approx 1$ for Gaussian and skewed inputs over a broad range of firing rates and pairwise correlation coefficients, but that values of $\Delta$ can be depressed by 10-15\% in the presence of a bimodal common input.  Indeed, Shlens et al. \cite{Shlens06} showed that adding global
bimodal inputs to a purely pairwise model can lead to a comparable departure in $\Delta$.  Our results are consistent with this finding, and explicitly demonstrate that the bimodality of the inputs --- as well as their global projection --- are characteristics that lead to this departure. 

While meaningful in an experimental study with non-neglible pairwise correlations, we caution that using $\Delta$ as a metric can be problematic when an idealized circuit is explored over its full range of parameters, because it may flag ``uninteresting" cases in which cells are nearly independent, and a pairwise model adds little additional value.  Specifically, 
%if $\Delta$ is close to 1, then $D_{pair}$ must be close to zero. If $D_{pair}$ is close to zero, 
%then $\Delta$ \textit{may} be close to 1, unless $D_{ind}$ is comparable to $D_{pair}$.
%For example, 
if $D_{ind}$ is small (i.e., the true distribution is well-approximated by the independent distribution $P_1$),
then $\Delta$ may be appreciably far from 1 although $D_{pair}$ is small. 
Thus, a poor pairwise maximum entropy fit, as measured by
$\Delta$ (that is, $\Delta < 1$) is not necessarily indicative of a poor performance in 
$D_{KL}(P,\tilde{P})$. For example, in the bimodal common input case (Figure \ref{fig:outputstats_N3}F), 
the very lowest values of $\Delta$ are achieved for low correlation $\rho$; in essence, when the independent model already
does a good job of representing the output distribution.  As suspected this performance is not
reflected in Figure \ref{fig:outputstats_N3}C, where low correlation gives low $D_{KL}(P,\tilde{P})$. In summary, 
$\Delta$ can be as low as $0.5$ for distributions that are barely perceptibly different when measured by the Kullback-Leibler divergence.

Figure \ref{fig:outputstats_N12}A-B extend our observations to a circuit of $N=12$ cells forced by Gaussian and skewed inputs respectively.  We find that small $\Delta$ occurs for $\sigma \ll 1$, coincident with low population firing rates.
Here $P$ is nearly independent (because it is dominated by 
$0$ spiking events, the distribution is well-modeled by independent non-spiking neurons) and the improvement of the pairwise model over the independent model is negligible. 
The low population firing rate regime ($N \nu < 1$, where $\nu$ is the single cell firing rate per time bin and $N$ is the population size) is also the regime where $1-\Delta$ is linear in $N-2$ \cite{roudi09}. If a nontrivial
deviation of $\Delta$ (from 1) 
is observed for $N \nu < 1$ (such as, in Figure \ref{fig:outputstats_N3}, $N=3$), then $1-\Delta$ must continue to grow as $N$ grows; 
equivalently, $\Delta$ must decrease.
The growth of $1-\Delta$ with $N$ for particular points in this region is illustrated in Figure \ref{fig:outputstats_N12}C. 

\subsection*{Using correlation structure to infer anatomical structure}

We address two questions about the relationship between the architecture of feedforward circuits and the statistical structure of the spike patterns that they produce, based on our comparisons between global-input and pairwise nearest-neighbor network architectures in the Results.

First, if a circuit produces spike patterns that deviate substantially from pairwise maximum-entropy (PME) predictions, can we conclude that it has beyond-pairwise anatomical projections --- that is, common inputs received by more than two cells?  
For a small group of $N=3$ cells, the answer is no: we find that, among all cases we study, the largest deviation from PME predictions occurs for {\it purely pairwise} (binary) inputs, so that departures from PME models do not imply departures from pairwise nearest-neighbor network architectures.  For larger $N$, the answer is a qualified yes:  for marginal statistics of a given type, we show in Figure \ref{fig:max_DKL_Nmedium} that the greatest deviations from PME models do correspond to global common (as opposed to purely pairwise) inputs.  However, without knowing input marginals, values of $D_{KL}$ are still not predictive of anatomy:  for example, if $N=16$, then roughly the same values of $D_{KL}(P,\tilde{P})$ follow from global inputs with uniform marginals as for pairwise nearest-neighbor inputs with binary marginals.

Second, if a circuit produces spike patterns that are well-described by PME models, does this imply that it has a pairwise architecture?  Again the answer is no, as the success of PME models depends on $N$ and marginal statistics. For $N>3$ and {\it known} input marginals, better fits by PME imply pairwise nearest-neighbor connectivity; otherwise, such inferences cannot be made.  

% FMR: changed second sentence above
% E:  looks good

\subsection*{Scope and open questions}

Our first set of findings are for a set of circuit models with a simple thresholding nonlinearity at each cell.  These models were chosen to be simple enough to allow analytical insights and a complete parametric study.  

While our  retinal ganglion cell model demonstrates that these findings, based on a simple threshold model, do carry over to describe the spiking statistics of a more realistic spiking model (here, a time-dependent, nonlinear integrate-and-fire system), there are many aspects of circuits left unexplored by our study.  
Prominent among these is heterogeneity.  Only a few of our simulations produce heterogeneous inputs to model RGCs, and all of our studies apply to cells with identical response properties.  This is in contrast to studies such as~\cite{Sch+06} which examine correlation structures among multiple cell types. For larger networks, feedforward connections with variable spatial profiles also occur, between the extremes of ``nearest neighbor" and global input connections examined here.  It is also possible that more complex input statistics could lead to greater higher-order interactions \cite{bethge08}. Finally, Figure \ref{fig:max_DKL_Nmedium} indicates that some trends in $D_{KL}(P,\tilde{P})$ vs. N appear to become nonlinear for $N \ \gtrapprox 10$; for larger networks, our qualitative findings could change.   
 
Our study also leaves largely open the role of different retinal filters in generating higher-order interactions. 
We have found that the specific filtering properties of ON parasol cells suppress bimodality in light inputs, suggesting that
other classes of retinal ganglion cells may produce more robust higher-order interactions (compare Figures \ref{fig:fullfield_composite}D, E).
This predicts a specific
mechanism for the development of higher-order interactions in preparations that include multiple classes of ganglion cells \cite{Sch+06}.  Finally, we considered circuits with a single step of inputs and simple excitatory or gap junction coupling; a plethora of other network features could also lead to higher-order interactions, including multilayer feedforward structures, together with lateral and feedback coupling.  We speculate that, in particular, such mechanisms could contribute to the higher-order interactions found in cortex~\cite{oizumi10, Tang08, Montanietal09, OhiorhenuanMPSHV10}.  

A final outstanding area of research is to link tractable network mechanisms for higher-order interactions with their impact (or lack of impact) on information encoded in neural populations~\cite{Montanietal09,oizumi10}.
A simple starting point is to consider rate-based population codes in which each stimulus produces a different ``tuned" average spike count (see, e.g., Ch. 3 of~\cite{Day+01}).  One can then ask whether spike responses can be more easily decoded to estimate stimuli for the full population response (i.e., $P$) to each stimulus or for its pairwise approximation ($\tilde P$).  In our preliminary tests where higher-order correlations were created by inputs with bimodal distributions, we found examples where decoding of $P$ vs. $\tilde P$ differed substantially.  However, a more complete study would be required before general conclusions about trends and magnitudes of the effect could be made; such a study would include complementary approach in which the full spike responses $P$ are themselves decoded via a ``mismatched" decoder based on the pairwise model $\tilde P$~\cite{oizumi10}.
Overall, we hope that the present paper, as one of the first that connects circuit mechanisms to higher-order statistics of spike patterns, will contribute to future research that takes these next steps.

%%%%%%%%%%%%%%%%%%%%%%%%%%%%%%%%%%%%%%%%
%%%%%%%%%%%%%%%%%%%%%%%%%%%%%%%%%%%%%%%%
%%%%%%%%%%%%%%%%%%%%%%%%%%%%%%%%%%%%%%%%
%%%%%%%%%%%%%%%%%%%%%%%%%%%%%%%%%%%%%%%%

\newpage

\section*{Materials and Methods}

\subsection*{$D_{KL}$ and distance from PME surface} \label{sec:dkl_distance}
We can see that the curve along which $\mu$ and $\rho$ are constant --- the \textit{iso-moment line} --- remains a line in
the $(f_p, f_{1p}, f_{1m})$ coordinate space by inverting the equations $\mu = p_1 + 2 p_2 + p_3$, $\rho = p_2 + p_3$,
and $1 = p_0 + 3 p_1 + 3 p_2 + p_3$.
The first coordinate, $f_p \equiv p_0 + p_3 = 3(\rho-\mu)-1$, is constant for a fixed $(\mu, \rho)$.  Similarly, $f_{1m}$ and $f_{1p}$ can be written as linear functions of the remaining free parameter.

To see that $D_{KL}(P,\tilde{P})$ is convex along an iso-moment line, we consider $D_{KL}(P,\tilde{P})$ as $P$
varies so as to remain on an iso-moment line. Letting $\tilde{f}_{1m}$ and $\tilde{f}_{1p}$ be the coordinates of the PME fit, and defining $dm = f_{1m} - \tilde{f}_{1m}$ and
$dp = f_{1p} - \tilde{f}_{1p}$, we find
\begin{eqnarray*}
dp & = & \frac{(f_p - 1)/3}{\sqrt{f_p^2 + (1-f_p)^2/9}} \, dx \\
dm & = & \frac{f_p}{\sqrt{f_p^2 + (1-f_p)^2/9}} \, dx
\end{eqnarray*}
where $dx$ is an increment of distance along the iso-moment line.
Inverting Equations \ref{eqn:fp_def}
and substituting the results into the definition of $D_{KL}$, we can write
\begin{eqnarray}
D_{KL}(P,\tilde{P}) & = & f_p (1-\tilde{f}_{1p}) \left(1-\frac{dp}{1-\tilde{f}_{1p}} \right) \log \left(1-\frac{dp}{1-\tilde{f}_{1p}} \right) \nonumber \\
& + & (1-f_p)(1-\tilde{f}_{1m}) \left(1-\frac{dm}{1-\tilde{f}_{1m}} \right) \log \left(1-\frac{dm}{1-\tilde{f}_{1m}} \right) \nonumber \\
& + & (1-f_p) \tilde{f}_{1m} \left( 1 + \frac{dm}{\tilde{f}_{1m}} \right) \log \left( 1 + \frac{dm}{\tilde{f}_{1m}} \right) \nonumber\\
& + & f_p \tilde{f}_{1p} \left( 1 + \frac{dp}{\tilde{f}_{1p}} \right) \log \left( 1 + \frac{dp}{\tilde{f}_{1p}} \right).   \label{eqn:dkl_fcoordinates}
\end{eqnarray}
This is a convex function of $dx$; we can see this by observing that each of the four terms is a 
function of the form 
\begin{eqnarray*}
F(dx) & = & \alpha \left( 1+\frac{dx}{\beta} \right) \log \left(1 + \frac{dx}{\beta} \right)
\end{eqnarray*}
(in the first term, for example, we have $\alpha = f_p (1-\tilde{f}_{1p})$ and $\beta = -(1-\tilde{f}_{1p})$).
This can be readily shown to be convex by taking the second derivative with respect to $dx$ and
verifying that it is positive:
\begin{eqnarray*}
F''(dx) & = & \frac{\alpha}{\beta^2 \left(1+\frac{dx}{\beta} \right)},
\end{eqnarray*}
where we can verify that $\alpha > 0$ and $|dx| < \beta$.
The sum of convex functions is likewise convex. Because $D_{KL}(P,Q)$ is non-negative for any distributions $P$ and $Q$, $D_{KL}(P,\tilde{P})$ achieves its unique minimum along 
an iso-moment line at $P=\tilde{P}$, and it must monotonically increase as a function of $|dx|$.

To get an intuitive picture of $D_{KL}(P,\tilde{P})$ as it varies in the $(f_p, f_{1p}, f_{1m})$-coordinate space, we view 
this quantity along constant-$f_p$ slices (Figure \ref{fig:cube_schematic}). $D_{KL} (P,\tilde{P})$ increases with
distance from the constraint curve. Generally, the range of this distance peaks at $f_p = 0.25$.

To further quantify the relationship between distance and $D_{KL}$, we approximate the logarithms in Equation \ref{eqn:dkl_fcoordinates}
for small arguments and find that $D_{KL}$ increases quadratically with distance for small arguments:
\begin{eqnarray}
D_{KL}(P,\tilde{P}) & \approx & (dx)^2 C(f_p, \tilde{f}_{1m}) + O(dx^3)   \label{eqn:dkl_approx}
\end{eqnarray}
where 
\begin{eqnarray*}
C(f_p,\tilde{f}_{1m}) & = & \frac{f_p (1-f_p)^2/9}{f_p^2 + (1-f_p)^2/9} \frac{(1-3 \tilde{f}_{1m} + 3 \tilde{f}_{1m}^2)^2}{\tilde{f}_{1m}^3 (1-\tilde{f}_{1m})^3} \nonumber\\
& +&  \frac{f_p^2(1-f_p)}{f_p^2 + (1-f_p)^2/9} \frac{1}{\tilde{f}_{1m}(1-\tilde{f}_{1m})}
\end{eqnarray*}     

\subsection*{Numerical sampling of 3 cell network} \label{sec:quadrature}
For general circuit set-ups, it may be necessary to probe the output distribution by sampling.
In the case of global input, however, it is more computationally efficient and accurate to compute the output spiking probability distribution using quadrature. To be concrete,
a set of $N=3$ threshold spiking units is forced by a common input
$I_c$ (drawn from a probability distribution $P_C(y)$) and an independent input $I_j$ (drawn from a probability distribution $P_I(y)$). The output of each cell $x_j$ is determined by summing and thresholding these inputs:
\begin{eqnarray*}
x_j = H(I_j + I_c - \Theta)
\end{eqnarray*}
Conditioned on $I_c$, the probability of each spike is given by:
\begin{eqnarray*}
{\bf Prob}[x_j = 1 \mid I_c =  a] &=& {\bf Prob}[I_j + a - \Theta >0]\\
& = & {\bf Prob}[I_j > \Theta - a]\\
& = & \int_{\Theta - a}^{\infty} P_I(y) \, dy
\end{eqnarray*}
Similarly, we have the conditioned probability that $x_j = 0$:
\begin{eqnarray*}
{\bf Prob}[x_j = 0 \mid I_c =  a] &=& {\bf Prob} [I_j + a - \Theta < 0]\\
& = & {\bf Prob}[I_j < \Theta - a]\\
& = & \int_{-\infty}^{\Theta - a} P_I(y) \, dy
\end{eqnarray*}
Because these are conditionally independent, 
the probability of any spiking event $(x_1, x_2, x_3) = (A_1, A_2, A_3)$  is
given by the integral of the product of the conditioned probabilities against the
density of the common input.
\begin{equation}
{\bf Prob} [x_1=A_1,x_2=A_2, x_3=A_3]  = \int_{-\infty}^{\infty} dy \, P_C(y) \prod_{j=1}^3 {\bf Prob}[x_j=A_j \mid I_c=y] \label{eqn:for_quad}
\end{equation}
The integrand in the previous equation is numerically evaluated via an adaptive quadrature routine, 
such at Matlab's {\tt quad}. This is easily generalized to an arbitrary number of cells $N$.

Unimodal inputs $I_j, I_c$ were chosen from different marginals with mean $0$ and variance $\sigma^2$. For Gaussian input, $P(x) \propto e^{-x^2/2\sigma^2}$; for uniform inputs, $P(x) \propto 1$ for $|x|<\sqrt{3 \sigma^2}$ and $0$ otherwise. For skewed input, $P(x) \propto (x+\mu)e^{-(x+\mu)^2/2a}$, for $x > -\mu$, where the parameter $a$ sets the variance $2a(1 - \frac{\pi}{4} )$ and shifting by $\mu = \sqrt{\frac{a \pi}{2}}$ ensures that the mean of $P(x)$ is zero.

When sampling was necessary in thresholding models, for $N \le 14$ the number of samples was chosen so that the mean bin occupancy of each unique circuit output was $> 100$ (and usually $>1000$). 
For $N=16$ --- the most poorly resolved case --- the mean bin occupancy was lower $(~40)$. 

\subsection*{Bimodal triplet inputs always generate distributions with negative strain} \label{sec:bimodal_details}
Another information theoretic quantity that relates to the ability of a distribution to be characterized by a pairwise model is the strain:
\begin{eqnarray*} 
\gamma & = & \frac{1}{8} \log \left( \frac{P(1,1,1) P(1,0,0) P(0,1,0) P(0,0,1)}{P(0,0,0) P(1,1,0) P(1,0,1) P(0,1,1)} \right).
\end{eqnarray*}
Indeed, this quantity must be zero for any distribution that satisfies the PME constraint (Equation \ref{eqn:constraint_surface}). 
Negative values of strain occur to the  
\textit{left} side of the PME constraint curve in the $(f_{1p},f_{1m})$-plane, whereas positive values occur to the \textit{right}.

For a circuit forced by common binary inputs, the simplicity of our setup allows us to show why observed distributions occur to the left side of the PME constraint curve.
We approach this by showing that given the $f_{1m}$ coordinate of an observed distribution, the $f_{1p}$ coordinate is less than
the PME fit would predict. A point on the constraint surface corresponding to a particular value of $f_{1m}$ may be written
\begin{eqnarray*}
\tilde{f}_{1p} & = & \frac{f_{1m}^3}{1 - 3 f_{1m} + 3 f_{1m}^2}  \\
& = & \frac{q^3}{q^3 + (1-q)^3}
\end{eqnarray*}
whereas
\begin{eqnarray*}
f_{1p} & = & \frac{pq^3}{pq^3 + (1-p) + p(1-q)^3}
\end{eqnarray*}
and
\begin{eqnarray*}
\frac{1}{f_{1p}} & = & \frac{1-p + p(q^3 + (1-q)^3)}{pq^3}\\
& = & \frac{1-p}{pq^3} + \frac{1}{\tilde{f}_{1p}}
\end{eqnarray*}
This makes it clear that $\frac{1}{f_{1p}} \ge \frac{1}{\tilde{f}_{1p}}$ with equality if and only if $p=1$. Therefore
\begin{eqnarray}
f_{1p} < \tilde{f}_{1p}  \label{eqn:underpredict}
\end{eqnarray} 
unless $p=1$, in which case they coincide.

According to Equation \ref{eqn:underpredict}, $p_0$ is \textit{under-predicted} by the PME model,
whereas $p_3$ is \textit{over-predicted}; this is precisely the condition of ``sparse coding" \cite{OhiorhenuanMPSHV10,ovJCN10}.

%Alternately, we can say that $f_{1m} > \tilde{f}_{1m}$; because $f_{1m} = \frac{p_2}{p_2+p_1}$, this implies that 
%$p_2$ is under-predicted by the PME model,
%whereas $p_1$ is over-predicted. 
 
\subsection*{An analytical explanation for unimodal vs. bimodal effects} \label{sec:unimodal_vs_bimodal}

We consider an analytical argument to support the 
our numerical results that bimodal inputs generate
larger deviations from PME model fits than unimodal inputs. As a metric, we consider
$D_{KL}(P,\tilde{P})$ --- where $P$ and $\tilde{P}$ are again the true and model distributions respectively ---
when we perturb an independent spiking distribution by adding a common, global input of variance ${\bf c}$.
To simplify notation, the small parameter in the calculation will be denoted $\epsilon = \sqrt{{\bf c}}$.

We begin by observing that when $\tilde{P}$ is a maximum entropy distribution; that is, its logarithm is given 
as a sum over functionals whose averages over $P$ must also be satisfied by $\tilde{P}$, or
\begin{eqnarray*}
\log(\tilde{p}(\xvec)) & = & \sum_{\mu} \lambda_{\mu} f_{\mu}(\xvec) - \log Z, \qquad \sum_{\xvec} \tilde{p}(\xvec)  f_{\mu}(\xvec) = \sum_{\xvec} p(\xvec)  f_{\mu}(\xvec);
\end{eqnarray*}
the KL-distance may be written as a difference of entropies:
\begin{eqnarray*}
D_{KL}(P,\tilde{P}) & \equiv & \sum_{\xvec} p(\vec x) \log \left( \frac{p(\xvec)}{\tilde{p}(\xvec)} \right)\\
& = & \sum_{\xvec} p(\xvec) \log( p(\xvec) ) - \sum_{\xvec} p(\xvec) \log( \tilde{p}(\xvec) )\\
& = & - S(P) - \sum_{\xvec} p(\xvec) \left( -\log Z + \sum_{\mu} \lambda_{\mu} f_{\mu}(\xvec) \right)\\
& = & - S(P)  + \log Z -  \sum_{\mu} \lambda_{\mu} \sum_{\xvec} p(\xvec)  f_{\mu}(\xvec)\\
& = & -S(P) + \log Z -  \sum_{\mu} \lambda_{\mu} \sum_{\xvec} \tilde{p}(\xvec)  f_{\mu}(\xvec)\\
& = & -S(P) - \sum_{\xvec} \tilde{p}(\xvec) \log( \tilde{p}(\xvec) )\\
& = & -S(P) + S(\tilde{P}).
\end{eqnarray*}
Here, the entropy of a probability distribution $P$ is given
\begin{eqnarray}
S(P) & = & -p_0 \log(p_0) - 3 p_1 \log(p_1) - 3 p_2 \log(p_2) - p_3 \log(p_3)  \label{eqn:entropy}
\end{eqnarray}
if we use the fact that the distributions are permutation-symmetric (i.e. $p_1 \equiv P(1,0,0)=P(0,1,0)=P(0,0,1)$).
We take the logarithms in the definitions of entropy $S$ and KL-divergence $D_{KL}$ to be base 2, so that any numerical values 
of these quantities are in units of bits.

We now compute $S(P)$ and $S(\tilde{P})$ by deriving a series expansion for each set of event probabilities.
We can compute the true distribution $P$ using the expressions derived in 
Equation \ref{eqn:for_quad}; to recap,
let the common input have probability density $p(c)$, and the independent input to each cell have density $p_s(x)$.  Let $\theta$ be the threshold for generating a spike (i.e., a ``1" response).  For each cell, a spike is generated if $x + c > \tht$, i.e., with probability 
	$$ d(c) = \int_{\tht-c}^\infty p_s(x) dx \; \; . $$
Given $c$, this is conditionally independent for each cell.  We can therefore write our probabilities by integrating over $c$ as follows:
	\beqr
	p_0 &=& \int_{- \infty}^{\infty} p(c) (1-d(c))^3 \, dc   \label{eqn:p0} \nonumber \\
	p_1 &=& \int_{- \infty}^{\infty} p(c)  d(c)(1-d(c))^2 \, dc \label{eqn:p1} \\
	p_2 &=& \int_{- \infty}^{\infty} p(c)  d(c)^2(1-d(c)) \, dc \label{eqn:p2} \nonumber \\
	p_3 &=& \int_{- \infty}^{\infty} p(c)  d(c)^3 \, dc \label{eqn:p3} \nonumber
	\eeqr
We develop a perturbation argument in the limit of very weak common input.  That is, $p(c)$ is close to a delta function centered at $c=0$.  Take $p(c)$ to be a scaled function
\begin{eqnarray}
p(c) & = & \frac{1}{\epsilon} f \left( \frac{c}{\epsilon} \right)   \label{eqn:pdef}
\end{eqnarray}
We place no constraints on $f(c)$, other than that it must be normalized ($\Ex[1] = 1$) and that its moments must be finite (so $\Ex[c]$, $\Ex[c^2]$, and so forth must exist, where
$\Ex[g(c)] \equiv \int_{-\infty}^{\infty} g(c) f(c) \, dc$). 

For the moment, assume that the function $f(c)$ has a single maximum at $c = 0$. To evaluate the integrals above, we Taylor-expand $d(c)$ around $c=0$.  Anticipating a sixth-order term to survive, we keep all terms up to this order. This gives, for small $y$, 
$$d(y) \approx d(0) + \sum_{k=1}^6 a_k y^k + O(y^7)$$
where $a_1 = p_s(\tht)$ (the other coefficients $a_2$-$a_6$ can be given similarly in terms of the 
independent input distribution at $\tht$). 
Substituting this into the expressions for $p_0$, etc., above, with $p(c)$ given as in Equation \ref{eqn:pdef}, 
gives us each event as a series in $\epsilon$; for example,
\begin{eqnarray*}
p_3 & = & d_0^3 + \left( 3 a_1 d_0^2 \Ex[c] \right) \epsilon + \left( (3 a_1^2 d_0 + 3 a_2 d_0^2)  \Ex[c^2] \right) \epsilon^2 + ...
\end{eqnarray*}
The entropy $S(P)$ is now given by using these series expansions in Equation \ref{eqn:entropy}.

We note that our derivation does not rely on the fact that the distribution of common input is peaked at
$c=0$ in particular. For example, we could have a common input centered around $\mu$. 
The common input distribution function would be of the form
\begin{eqnarray*}
p(c) & = & \frac{1}{\eps} f \left( \frac{c-\mu}{\eps} \right)
\end{eqnarray*}
Changing $\eps$ regulates the variance, but doesn't change the mean or the peak (assuming,
without loss of generality, that the peak
of $f$ occurs at zero). The peak of $p(c)$ now occurs at $\mu$, and the appropriate Taylor
expansion of $d(y)$ is
$$d(y) \approx d(\mu)  + \sum_{k=1}^6 b_k (y-\mu)^k + O(y^7),$$ 
where the coefficients $b_k$ now depend on the local behavior of $d$ around $\mu$. 
The expectations that appear in the expansion of $p_3$, and so forth, are now centered moments taken around $\mu$; 
the calculations are otherwise identical.
In other words, perturbation expansion requires the \textit{variance} of the common input to be small,
but not the mean. 

For bimodal inputs, we consider a common input with a probability distribution of the following form:
\begin{eqnarray*}
p(c) & = & (1-\eps^2) \frac{1}{\eps} f \left( \frac{x}{\eps} \right)  + \eps^2 \frac{1}{\eps} f \left(\frac{x-1}{\eps} \right)
\end{eqnarray*}
so that most of the probability distribution is peaked at zero, but there is a second peak of higher
order (here taken at $c=1$, without loss of generality). Again, we approximate the integrals given in Equations \ref{eqn:p1}, and therefore the entropy $S(P)$, by Taylor expanding $d$;
\begin{eqnarray*}
d(c) & \approx & d(0) +  \sum_{k=1}^6 a_k c^k + O(c^7); \quad (c \approx 0)\\
& \approx & d(1)  + \sum_{k=1}^6 b_k (c-1)^k+ O((c-1)^7); \quad (c \approx 1)
\end{eqnarray*}
around the two peaks $0$ and $1$ respectively.
For each integral we have the same contributions from the unimodal case, 
multiplied by $(1-\eps^2)$, as well as 
the corresponding contributions from the second peak multiplied by $\eps^2$ (these weightings are chosen so
that the common input has variance of order $\eps^2$, as in the unimodal case). This makes clear at what order
every term enters. 

We now construct an expansion for the PME model $\tilde{P}$:
\begin{eqnarray*}
\tilde{P} (x_1, x_2, x_3) & = & \frac{1}{Z} \exp \left( \lambda_1 (x_1 + x_2 + x_3) + \lambda_2 ( x_1 x_2 + x_2 x_3 + x_1 x_3 ) \right)
\end{eqnarray*}
We approach this problem by describing $\lambda_1$ and $\lambda_2$ as a series in $\epsilon$. We match
coefficients by forcing the first and second moments of $\tilde{P}$ to match those of $P$ --- as they must.
Specifically, take
\begin{eqnarray*}
\lambda_1 & = & \tilde{\lambda} + \sum_{k=1}^{6} \epsilon^k u_k + O(\epsilon^7) \\
\lambda_2 & = & \sum_{k=1}^{6} \epsilon^k v_k + O(\epsilon^7) 
\end{eqnarray*}
where $\lambda_1 = \tilde{\lambda}$, $\lambda_2 = 0$ are the corresponding parameters from the independent case. 
The events $\tilde{p}_0$, $\tilde{p}_1$, $\tilde{p}_2$ and $\tilde{p}_3$ can be written as a series in $\epsilon$.
We then require that the mean and centered second moments of $\tilde{P}$ match those of $P$; that is
\begin{eqnarray*}
p_1 + 2 p_2 + p_3 & = & \tilde{p}_1 + 2 \tilde{p}_2 + \tilde{p}_3\\
p_2 + p_3 - (p_1 + 2 p_2 + p_3)^2 & = & \tilde{p}_2 + \tilde{p}_3 - (\tilde{p}_1 + 2 \tilde{p}_2 + \tilde{p}_3)^2.
 \end{eqnarray*}
At each order $k$, this yields a system of two linear equations in $u_k$ and $v_k$; we solve, 
inductively, up to the desired order;
we now have $\tilde{P}$, and therefore $S(\tilde{P})$, as a series in $\epsilon$.

Finally, we combine the two series to find that in the \textit{unimodal} case, 
\begin{eqnarray}
D_{KL}(P,\tilde{P}) & = & S(\tilde{P})-S(P)   \nonumber \\
& = & \epsilon^6 \left[  \frac{a_1^6 (2 \Ex[c]^3 - 3 \Ex[c] \Ex[c^2] + \Ex[c^3] )^2}{2 (1-d_0)^3 d_0^3} \right]+ O(\epsilon^7)
\end{eqnarray}
If the first two odd moments of the distribution are zero (something we can expect for ``symmetric" distributions, such as a Gaussian), then this sixth-order term is zero as well.

For the \textit{bimodal} case
\begin{eqnarray*}
D_{KL}(P,\tilde{P}) & = & S(\tilde{P})-S(P) \nonumber \\
& = & \epsilon^4 \left[ \frac{(d_1 - d_0)^6}{2 (1-d_0)^3 d_0^3} \right] + O(\epsilon^5)
\end{eqnarray*}
This last term depends on the distance $d_1 - d_0$, in other words, how much more likely
the independent input is to push the cell over threshold when common input is ``ON".   We can also view this as
depending on the ratio $\frac{d_1 - d_0}{1 - d_0}$, which gives the fraction of previously 
non-spiking cells that now spike as a result of the common input.
 
{\it The main point here, of course, is that $D_{KL}(P,\tilde{P})$ is of order $\eps^4$ rather than $\eps^6$.}  So, as the strength of a common binary vs. unimodal input increases, spiking distributions depart from the PME more rapidly.

\subsection*{Experimentally-based model of a RGC circuit} \label{sec:RGC_details}
We model the response of a individual RGC using data collected from a representative primate ON parasol cell, following methods in~\cite{Murphy2006,tr08}.  Similar response properties were observed in recordings from 16 other cells.
To measure the relationship between light stimuli and synaptic conductances, the retina was exposed to a full-field, white noise stimulus.
%\begin{table}
%\begin{tabular}{cccc}\hline
%Function type&Parameter&Excitatory&Inhibitory\\  \hline
%Filter $L$& $P$      & $-8 \times 10^4$ pA/s & $-1.8 \times 10^5$ pA/s \\
%	  & $n$      & 3.6 			   & 3.0				   \\  
% 	 & $\tau$ &12 ms			   & 16 ms 		  	   \\
%	 & $T$     &105 ms			   & 120 ms 		  	   \\\hline
%\\
%Nonlinearity $N$& $A$          & $-5 \times 10^{-5}$ & $ 1 \times 10^4$  \\
%	        & $B$          &$0.42 $			   & $	0.37$		   \\  
%	        & $C$          &$-57$			   & $250$		  	   \\ \hline
%\\
%Noise term $\eta$& mean & $30$ & $-1200$ \\
%			& std & $500$ & $780$ \\
%		 	& $\tau_{\eta}$ & $22$ ms & $33$ ms \\ \hline
%\end{tabular}
%\caption{Simulation parameters. \label{tab:sim_params} }
%\end{table}
The cell was voltage clamped at the excitatory (or inhibitory) reversal potential $V_E = 0$ mV ($V_I = -60$ mV),
and the inhibitory (or excitatory) currents were measured in response to the stimulus. These currents were then turned into equivalent conductances by dividing by the driving force of $\pm 60$ mV; in other words 
\begin{align}
I^{\text{exc}} & =  g^{\text{exc}}(V-V_E); \qquad V-V_E=-60 \; \rm{mV} \nonumber \\
I^{\text{inh}} & =  g^{\text{inh}}(V-V_I); \qquad V-V_I=60 \; \rm{mV} \nonumber
\end{align}
The time-dependent conductances $g^{\text{exc}}$ and $g^{\text{inh}}$ were now injected into the same cell using a dynamic clamp 
(i.e., input current was varied rapidly to maintain the correct relationship between the conductance and the
membrane voltage) and the voltage was measured at a resolution of $0.1$ ms. 

To model the relationship between the light stimulus and synaptic conductances, the current measurements $I^{\text{exc}}$ and $I^{\text{inh}}$ were
fit to a linear-nonlinear model:
\begin{eqnarray*}
I^{\text{exc}}(t) & = & N^{\text{exc}}[L^{\text{exc}} \ast s(t) + \eta^{\text{exc}}], \label{eq:currents}  \\
I^{\text{inh}}(t) & = & N^{\text{inh}}[L^{\text{inh}} \ast s(t) + \eta^{\text{inh}}]
\end{eqnarray*}
where $s$ is the stimulus, $L^{\text{exc}}$ ($L^{\text{inh}}$) is a linear filter, $N^{\text{exc}}$ ($N^{\text{inh}}$) is a nonlinear function, and $\eta^{\text{exc}}$ ($\eta^{\text{inh}}$) is a noise term. The linear filter was fit by the function
\begin{eqnarray}
L^{\text{exc}}(t) & = & P_{\text{exc}}(t/\tau_{\text{exc}})^{n_{\text{exc}}} \exp(-t/\tau_{\text{exc}}) \sin(2\pi t/T_{\text{exc}})  \label{eqn:linear_filter}
\end{eqnarray}
and the nonlinear filter by  the polynomial
\begin{equation*}
N^{\text{exc}} =  A_{\text{exc}} x^2 + B_{\text{exc}} x + C_{\text{exc}};
\end{equation*}
$L^{\text{inh}}$ and $N^{\text{inh}}$ were fit using the same parametrization.  
The noise terms $\eta^{\text{exc}}_k$, $\eta^{\text{inh}}_k$ were fit to reproduce the statistical characteristics of the residuals from this fitting. We simulated the noise terms $\eta^{\text{exc}}$ and $\eta^{\text{inh}}$ using Ornstein-Uhlenbeck processes with the appropriate parameters; these were entirely characterized by the 
mean, standard deviation, and time constant of autocorrelation $\tau_{\eta,\text{exc}}$ ($\tau_{\eta,\text{inh}}$), as well as pairwise correlation coefficients for noise terms entering neighboring cells. The noise correlation coefficients were estimated from the dual recordings of \cite{tr08}.

Linear filter parameters computed were $P_{\text{exc}} = -8 \times 10^4$ pA/s, $n_{\text{exc}}=3.6$, $\tau_{\text{exc}} = 12$ ms, $T_{\text{exc}}=105$ ms, and $P_{\text{inh}} = -1.8 \times 10^5$ pA/s, $n_{\text{inh}}=3.0$, $\tau_{\text{inh}} = 16$ ms, $T_{\text{exc}}=120$ ms. Nonlinearity parameters were $A_{\text{exc}} = -5 \times 10^{-5}$, $B_{\text{exc}} = 0.42$, $C_{\text{exc}} = -57$, and $A_{\text{inh}} = 1 \times 10^4$, $B_{\text{inh}}=0.37$, $C_{\text{inh}} =250$. Noise parameters were measured to be ${\rm mean}(\eta^{\text{exc}}_k) = 30$, ${\rm std}(\eta^{\text{exc}}_k) = 500$, $\tau_{\eta,\text{exc}} = 22$ ms, and 
${\rm mean}(\eta^{\text{inh}}_k) = -1200$, ${\rm std} (\eta^{\text{inh}}_k) = 780$, $\tau_{\eta,\text{inh}} = 33$ ms. In addition, excitatory (inhibitory) noise to different cells $\eta^{\text{exc}}_k$, $\eta^{\text{exc}}_j$ ($\eta^{\text{inh}}_k$, $\eta^{\text{inh}}_j$) had a correlation coefficient of $0.3$ ($0.15$).

To obtain the rectified filter shown in Figure \ref{fig:fullfield_composite}E, we simply took the absolute value of the filter function; 
i.e. $L^{\text{exc,R}}(t) = \left| L^{\text{exc}}(t) \right|$.

\subsubsection*{Model fitting} 

We create a model of the cell as a nonlinear integrate-and-fire model using the method of Badel et al. \cite{badel07},
in which the membrane voltage is assumed to respond as 
\begin{eqnarray}
\frac{dV}{dt} & = &  F(V, t-t_{last}) + \frac{I_{input}(t)}{C}   \label{eqn:fI_curve}
\end{eqnarray}
where $C$ is the cell capacitance, $t_{last}$ is the time of the last spike before time $t$, 
and $I_{input}(t)$ is a time-dependent input current. We use the current-clamp data,
which yields cell voltage in response to the input current $I_{input}(t) = g^{\text{exc}}(t)(V-V_E) + g^{\text{inh}}(V-V_I)$, to fit 
a function $F(V,t)$. When voltage data is segregated according to the (binned) time since the last spike,
the $I-V$ curve is well fit by a function of the form
\begin{eqnarray}
F(V) & = & \frac{1}{\tau_m} \left(E_L - V + \Delta_T e^{(V-V_T)/\Delta_T} \right) \label{eqn:fI_shape}
\end{eqnarray}
The membrane time constant $\tau_m$, resting potential ($E_L$),  spike width $\Delta_T$
and knee of the exponential curve $V_T$
are parameterized as a function of $t-t_{last}$. 
%The fitted parameters are illustrated in Figure~\ref{fig:Badel_fits}.
Our model neuron comprises Equations (\ref{eqn:fI_curve}, \ref{eqn:fI_shape}) for $V<V_{threshold}$; 
a spike was detected when $V$ reached $V_{threshold}=-30$ mV,
with a voltage reset to $V_{reset}=-55$ mV after 2 ms. In addition, the cell was unable to spike for an absolute refractory period of $\tau_{abs} = 3$ ms.

The capacitance was inferred from the voltage trace data by finding, at a voltage value where the
voltage/membrane current relationship is approximately Ohmic, the value of $C$ that
minimizes error in the relation Equation \ref{eqn:fI_curve} \cite{badel07}. The estimated value was $C=28$ pF.

\subsubsection*{Recurrent model}
Gap junction coupling was introduced as an additional current on the right-hand side of Equation \ref{eqn:fI_curve}: 
\begin{eqnarray*} 
\frac{I_{gap,j}}{C} & = & -\frac{g^{\text{gap}}}{C}\sum_{k \not=j}(V_j-V_k)
\end{eqnarray*}
The coupling strength $g^{\text{gap}}$ was held constant during a simulation. When coupling was present (i.e. when $g^{\text{gap}} \not= 0$), $g^{\text{gap}}$ was varied from a baseline level ($1.1$ nS)\cite{tr08} to 16 times this value ($17.6$ nS) between simulations.

For simulations that include electrotonic coupling, the spike trajectory was modeled with greater care. A typical (averaged) spike waveform was extracted from voltage traces of a primate ON parasol cell. The spike waveform was used to replace 1 ms of the membrane voltage trajectory during and after a spike; at the end of the 1 ms, the voltage was released at approximately $-58$ mV.  The cell was unable to spike for an absolute refractory period of $\tau_{abs}=3$ ms. A relative refractory period was induced by introducing a declining threshold for the period of 3-6 ms following a spike, after which $V_{th}$ limits to -30 mV. 

\subsubsection*{Cell receptive field}
We defined each cell's stimulus as the linear convolution of an image with its receptive field. The receptive fields include an ``on" 
center and an ``off" surround, as in \cite{kalmar02}:
\begin{eqnarray*}
s_j(\vec x) = \exp \left( -\frac{1}{2}(\vec x - \vec x_j) {\bf Q} (\vec x - \vec x_j) \right) -  k \exp \left( -\frac{1}{2} r (\vec x - \vec x_j) {\bf Q} r (\vec x - \vec x_j) \right)
\end{eqnarray*}
where the parameters $k$ and $1/r$ give the relative strength and size of the surround. 
${\bf Q}$ specifies the shape of the center and was chosen to have a 1 standard deviation (SD) radius of $50 \,  \rm{\mu m}$ and to be perfectly spherical. The receptive field locations $\vec x_1$, $\vec x_2$, and $\vec x_3$ were chosen so that the 1 SD outlines of the receptive field centers 
will tile the plane 
(i.e. they just touch). Other parameters used were $k = 0.3$, $r = 0.675$.

\subsubsection*{Numerical methods and convergence testing}
For each simulation, Equations \ref{eqn:fI_curve}, \ref{eqn:fI_shape} were integrated using the Euler method for $> 10^5$ ms with a time step of
$0.1$ ms. For each stimulus condition, 20 simulations (or sub-simulations) were run, for a total integration time of $> 20 \times 10^5$ ms.
The synaptic noise terms, $\eta^{\text{exc}}_k$ and $\eta^{\text{inh}}_k$, as well as the light input, were generated
independently for each sub-simulation. To discretize spiking outputs, 5 ms bins were used.

These 20 sub-simulations were used to estimate
standard errors in both the probability distribution over spiking events and $D_{KL}(P,\tilde{P})$. 
For example, in the constant light case, we generated the following distribution on spiking events:
$P(0,0,0) = 0.816 \pm 0.004$, $P(0,0,1) = 0.0457 \pm 0.0015$; $P(0,1,0) = 0.0448 \pm 0.0015$, $P(1,0,0) = 0.0459 \pm 0.0017$,
$P(0,1,1) = 0.00554 \pm 0.00054$, $P(1,0,1) = 0.00545 \pm 0.00051$, $P(1,1,0) = 0.00545 \pm 0.00056$, and $P(1,1,1) = 0.00116 \pm 0.00020$.  
Numbers reported in the Results are, unless specified otherwise, produced by collating the data from the $20$ simulations. 

To test our finding that the observed distributions were well-modeled by the PME fit, we also performed the PME analysis on each of the 20 simulations for each stimulus condition. While in 
general $D_{KL}(P,\tilde{P})$ can be quite sensitive to perturbations in $P$, the numbers remained small under this analysis.
To confirm that our results for $D_{KL}(P,\tilde{P})$ are sufficiently resolved to remove bias from sampling, we performed an analysis in which we collect the 20 simulations in subgroups of 1, 2, 4, 5, 10, and 20, and plot the mean $D_{KL}$ with estimated standard errors.
As expected (e.g. \cite{paninski03}), bias decreases as the length of subgroup increases and asymptotes at
--- or before --- the full simulation length. Results are shown in Figure \ref{fig:suppl} for the 
RGC simulations under full-field stimulation, as well as two 
representative cases with ``stixel" stimuli. 

For the thresholding model, sampling was used for pairwise input cases in 
Figure \ref{fig:max_DKL_Nmedium}, and all data presented in Figure \ref{fig:max_DKL_Recur}.
In all but three simulations, the sampling frequency was chosen so that the mean bin occupancy
(that is, the average frequency of each unique spiking pattern) 
was $> 100$ (and usually $> 1000$).  The exceptions are for the cases where the network size $N=16$ and 
unimodal pairwise inputs are used, for which the mean bin occupancy was $ \approx 40$. 
We show the bias for $N=16$, pairwise Gaussian inputs in Figure \ref{fig:suppl}A, which confirms that the bias is close to its asymptotic value of zero.

To provide a cross-validation test, we divided our data into halves (which we denote $P_1$ and $P_2$, each including data from 10 subsimulations) and 
performed the PME analysis on one half (say $P_1$) to yield a model $\tilde{P}_1$. 
We then computed 
$D_{KL}(P_2,\tilde{P}_1)$ and  $D_{KL}(P_2,P_1)$ (as in \cite{yu2011}), which we refer to 
the \textit{cross-validated} and \textit{empirical} likelihood respectively. 
The former tests whether the PME fit is robust to over-fitting; 
the latter tests how well-resolved our ``true" distribution is in the first place. 
In Figure S2, we plot these numbers vs. the numbers reported in the paper. 
Most cross-validated likelihoods fall on or near the identity line; most empirical likelihoods are close to zero (and importantly, significantly smaller than either $D_{KL}(P,\tilde{P})$ or $D_{KL}(P_2,\tilde{P}_1)$, indicating that $D_{KL}(P,\tilde{P})$ is accurately resolved). 
We conclude that the deviations that we observe when these conditions are met 
can not be accounted for by the differences in testing and training data.
 
Finally, to test the robustness of our finding that the strain $\gamma < 0$ for
the full-field RGC simulations, 
we perturbed our spiking event distributions randomly, 
with perturbations weighted by the estimated standard errors. This was repeated 20 times for each stimulus condition. Out of the resulting 840 perturbations 
(20 each for 42 stimulus conditions), $\gamma > 0$ in only 22 trials.

\section*{Financial Disclosure}  

This research was supported by NSF grant DMS-0817649 and by a Career Award at the Scientific 
Interface from the Burroughs-Wellcome Fund (ESB), by the Howard Hughes Medical Institute and by NIH grant EY-11850 (FMR),
and a Cambridge Overseas Research Studentship (JG).  The funders had no role in study design, data collection and analysis, 
decision to publish, or preparation of the manuscript.  
\newpage

%\bibliography{../../RGC_3rd_order_corr_v4}

\newpage
\setcounter{figure}{0}
\begin{figure}[htbp]
\begin{center}
\includegraphics[width=5in]{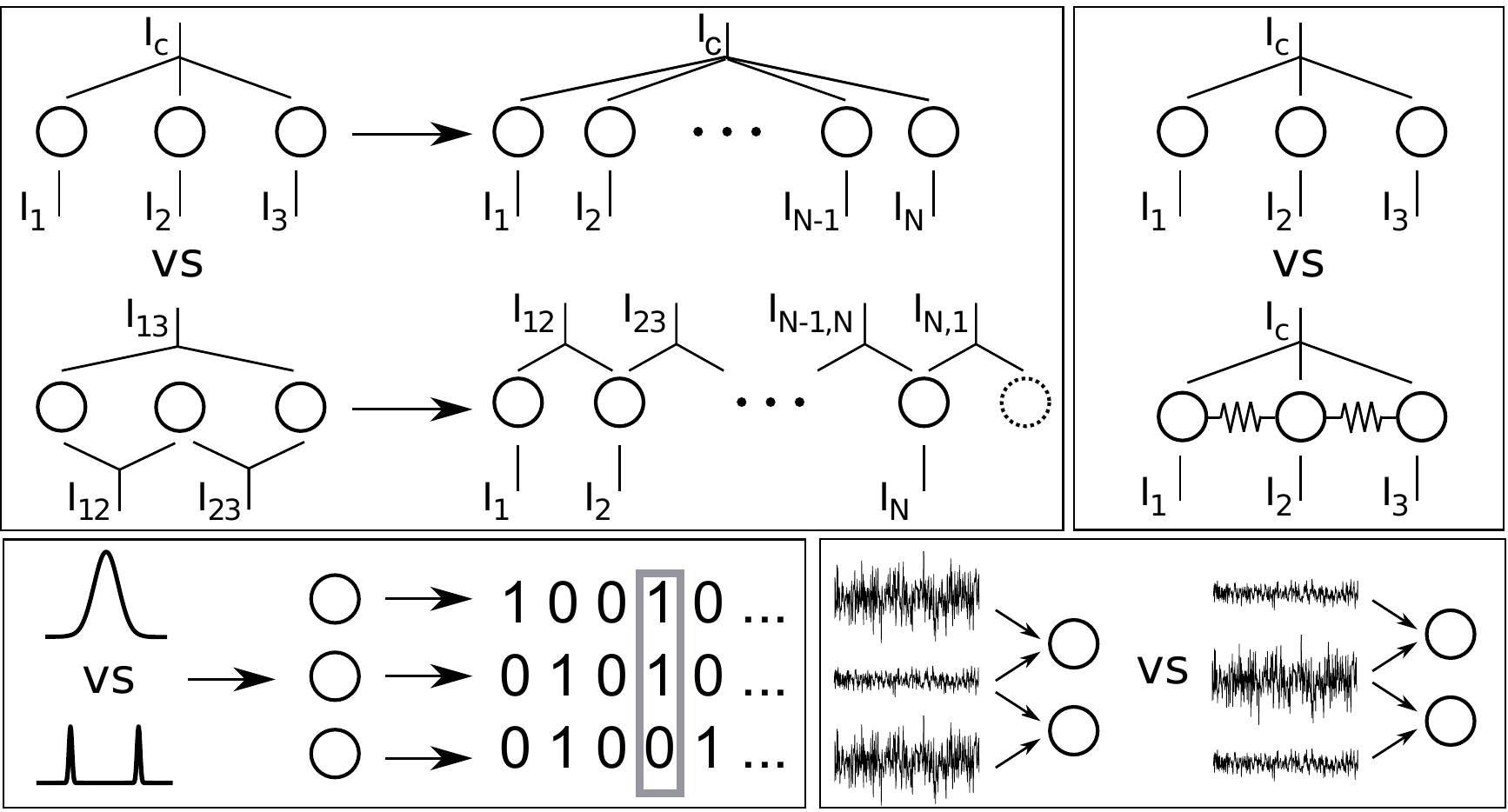}
\caption{Schematic showing different axes on which we explore microcircuits in this study. (Top left) 
Network architecture: global vs. pairwise inputs and scaling up system size $N$. 
(Top right)  Presence vs. absence of reciprocal coupling.
(Bottom left) Input statistics: unimodal vs. bimodal marginal statistics.
(Bottom right) varying strength of common input.}  \label{fig:study_schematic}
\end{center}
\end{figure}

\begin{figure}[htbp]
\begin{center}
\includegraphics[width=5in]{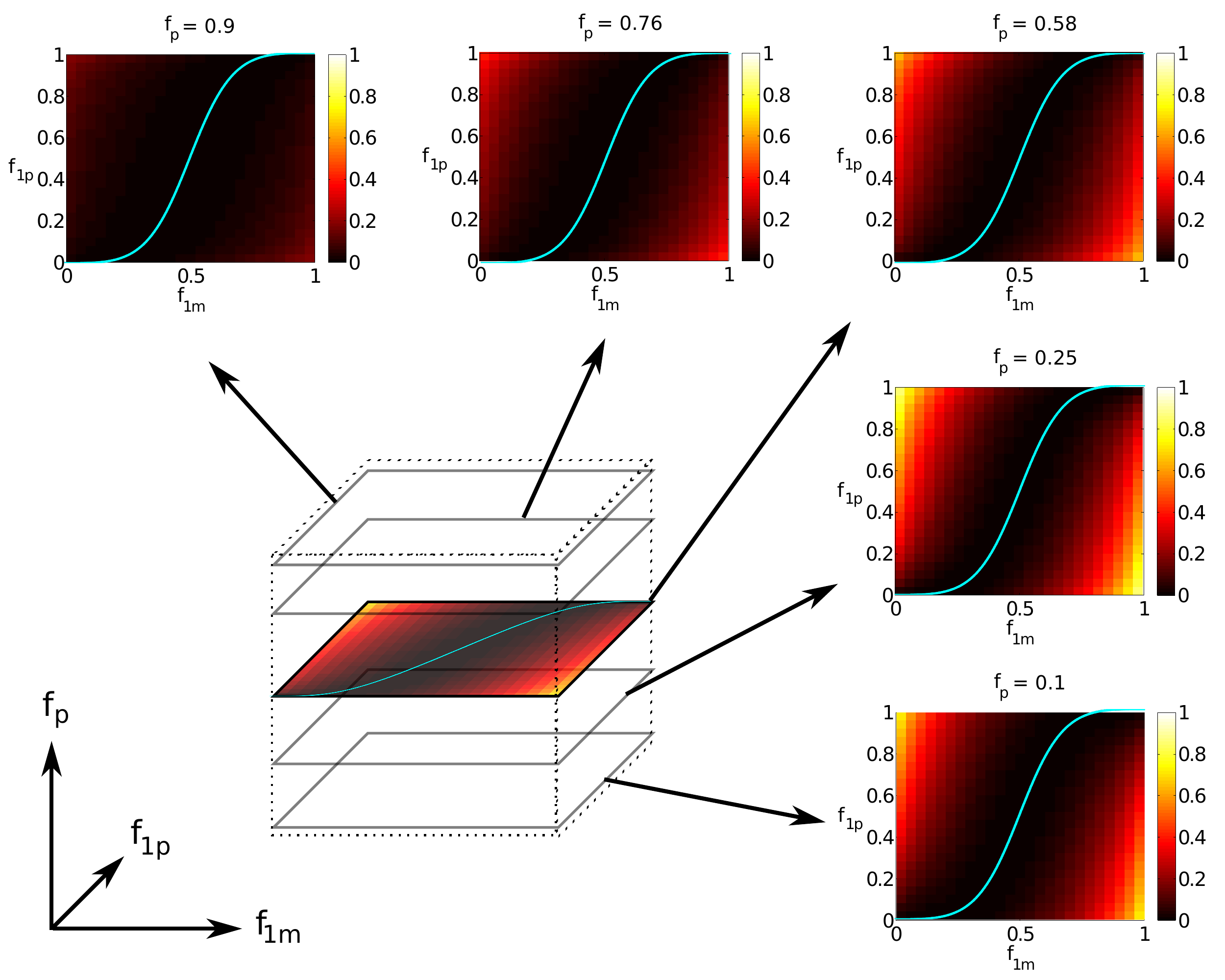}
\caption{Geometrical organization of $D_{KL}(P,\tilde{P})$ within the space of three-cell spiking distributions.  Outer plots:  Slices of $D_{KL}(P,\tilde{P})$ along surfaces $f_p = constant$. Center: schematic of the $(f_p, f_{1p}, f_{1m})$ coordinate space. Counterclockwise from lower right:
$f_p = 0.1, 0.25, 0.58, 0.76, 0.9$. $f_p = 0.25$ contains the maximal attainable $D_{KL}(P,\tilde{P})$ over all admissible $P$. $f_p = 0.616$ contains the maximal attainable $D_{KL}$ from pairwise bimodal inputs (see Results).}  \label{fig:cube_schematic}
\end{center}
\end{figure}

\begin{figure}[htbp]
\begin{center}
\includegraphics[width=5in]{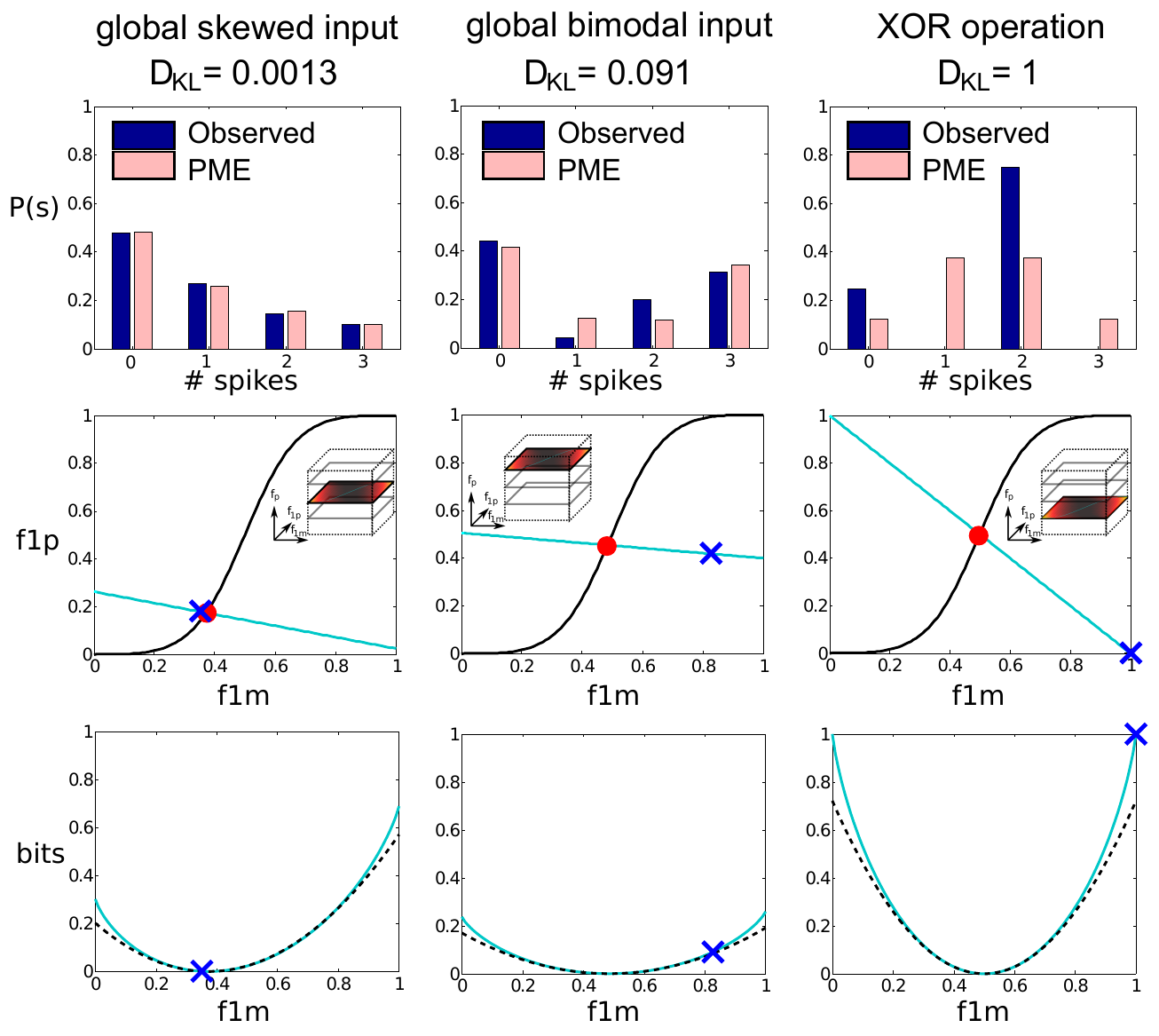}
\caption{Examples of spiking distributions with small, intermediate, and large deviations from PME models (as quantified by the KL- divergence, $D_{KL}(P,\tilde{P})$, between the observed distribution $P$ and its PME fit $\tilde{P}$). (Top row) Bar plot contrasting three distributions with their pairwise maximum entropy (PME) fits.  The probability shown is the total probability of all events with a certain number of spikes. 
From left: global skewed input, global binary input, XOR operator. $D_{KL}(P,\tilde{P})$ is, from left, $0.0013$ (skewed), $0.091$ (bimodal), and $1$ (XOR). (Middle row) The same distributions (crosses) projected into the $(f_{1m},f_{1p}$)-plane and their corresponding PME fits (circles). The cyan line is the \textit{iso-moment line} of all distributions with the same first and second moments, and the black curve is the PME constraint surface. (Bottom row) $D_{KL}(P,\tilde{P})$ along the iso-moment line (cyan solid) and the quadratic approximation derived in the Methods.
%in Equation \ref{eqn:dkl_approx} 
(black dashed). }  \label{fig:iso_bar}
\end{center}
\end{figure}

\begin{figure}[p!]
\begin{center}
\includegraphics[height=5.0in]{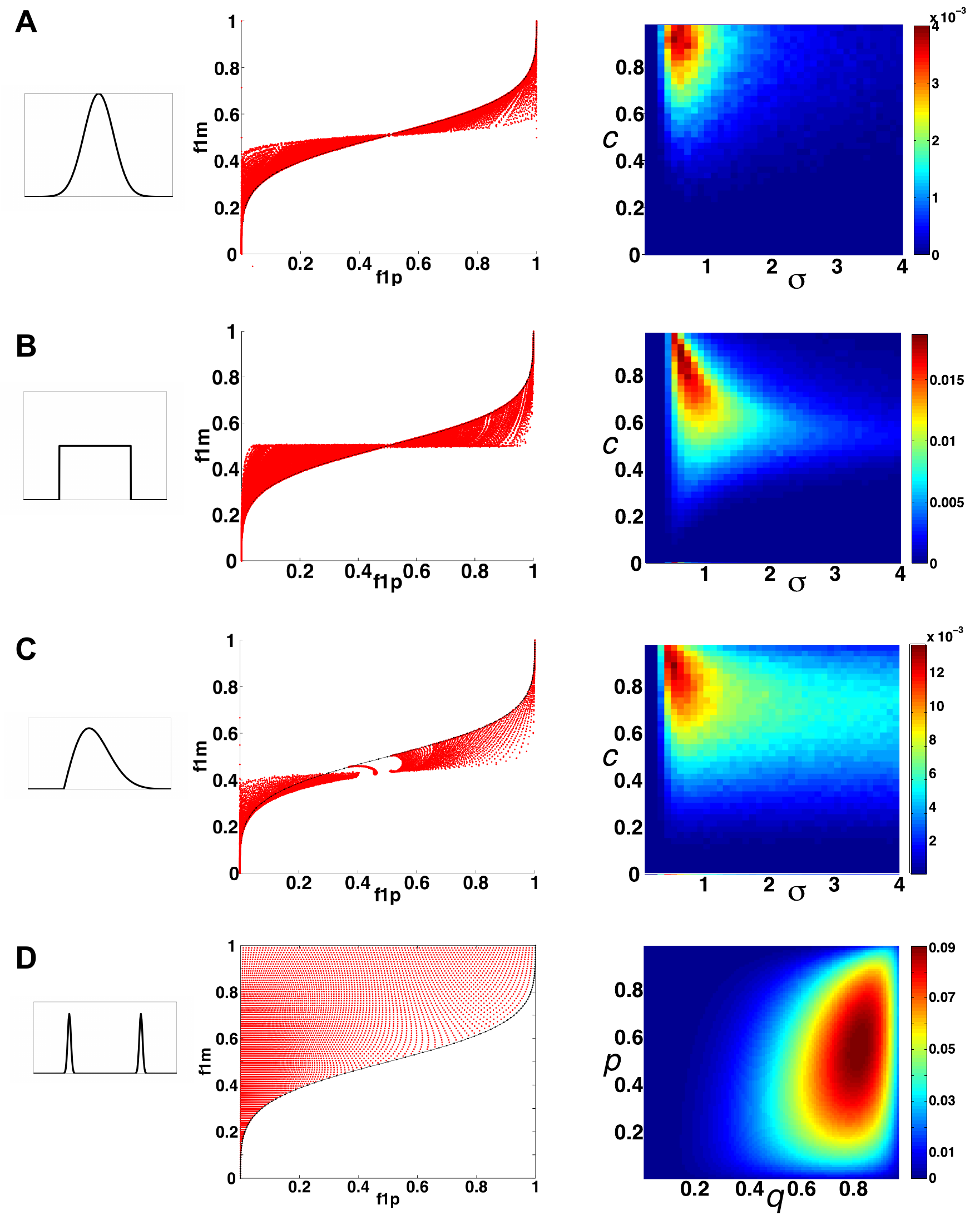}
\caption{Deviation from PME fit for circuits receiving independent and (global) common input. Results shown for $N=3$; (A) Gaussian, (B) uniform, (C) skewed, and (D) bimodal. 
For each choice of marginal input statistics, possible input parameters are varied over a broad range as described in the Results;
over $c \in [0,1]$, $\sigma \in [0,4]$, and $\Theta \in [-1,3]$ (unimodal inputs), over $p \in [0,1]$ and $q \in [0,1]$ (bimodal inputs). (Left column) Schematic of input distributions. (Center column) Projection of all distributions onto the $(f_{1p},f_{1m})$-plane. (Right column) For the value of $\Theta$ for which the maximal value is achieved, a slice of $D_{KL}(P,\tilde{P})$ (unimodal inputs); contour plot of $D_{KL}(P,\tilde{P})$ where $1 < \Theta < 2$ (bimodal inputs).} \label{fig:max_DKL_N3}
\end{center}
\end{figure}

\begin{figure}[t!]
\begin{center}
\includegraphics[width=5.0in]{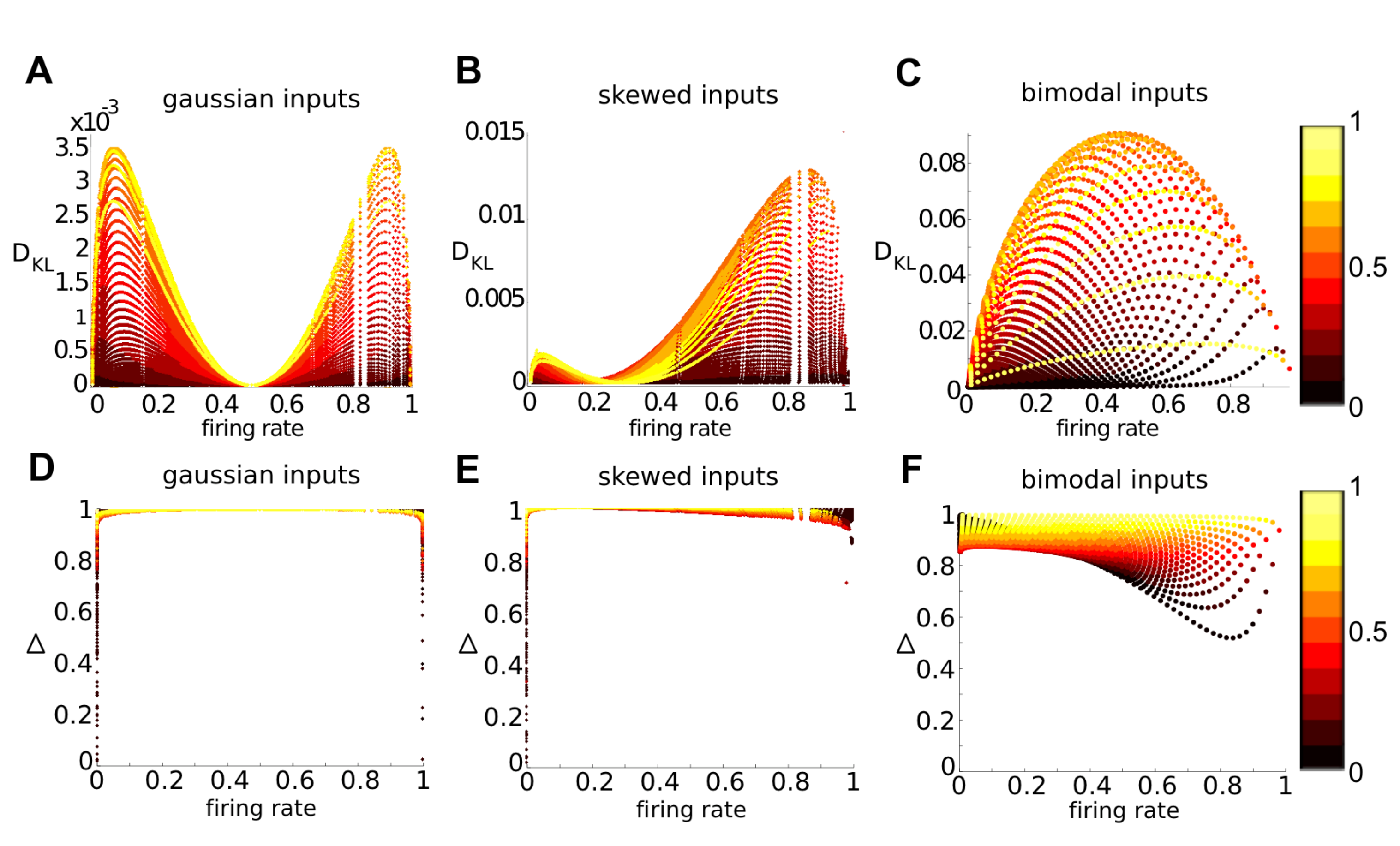}
\caption{Relationship between measures of higher-order interactions and other output firing statistics. 
(A-C) $D_{KL}(P,\tilde{P})$ versus firing rate $\Ex[x_j]$, for data
obtained from $N=3$ thresholding cells. For each choice of marginal input statistics, 
possible input parameters were varied over a broad range as described in the Results;
over $c \in [0,1]$, $\sigma \in [0,4]$, and $\Theta \in [-1,3]$ for unimodal inputs, and over $p \in [0,1]$ and $q \in [0,1]$ for bimodal inputs.
In each panel, data is organized by $\rho$,
from dark ($\rho \in (0,0.1)$) to light ($\rho \in (0.9,1)$);
(A) Gaussian inputs, (B) skewed inputs, (C) bimodal inputs.
(D-F) The fraction of multi-information captured by the PME model, $\Delta$, 
%(defined in Discussion) 
versus firing rate $\Ex[x_1]$, for these same distributions. 
Data is organized by correlation coefficient $\rho$,
as in panels (A-C);
D) Gaussian, E) skewed, F) bimodal. } \label{fig:outputstats_N3}
\end{center}
\end{figure}

\begin{figure}
\includegraphics[width=3in]{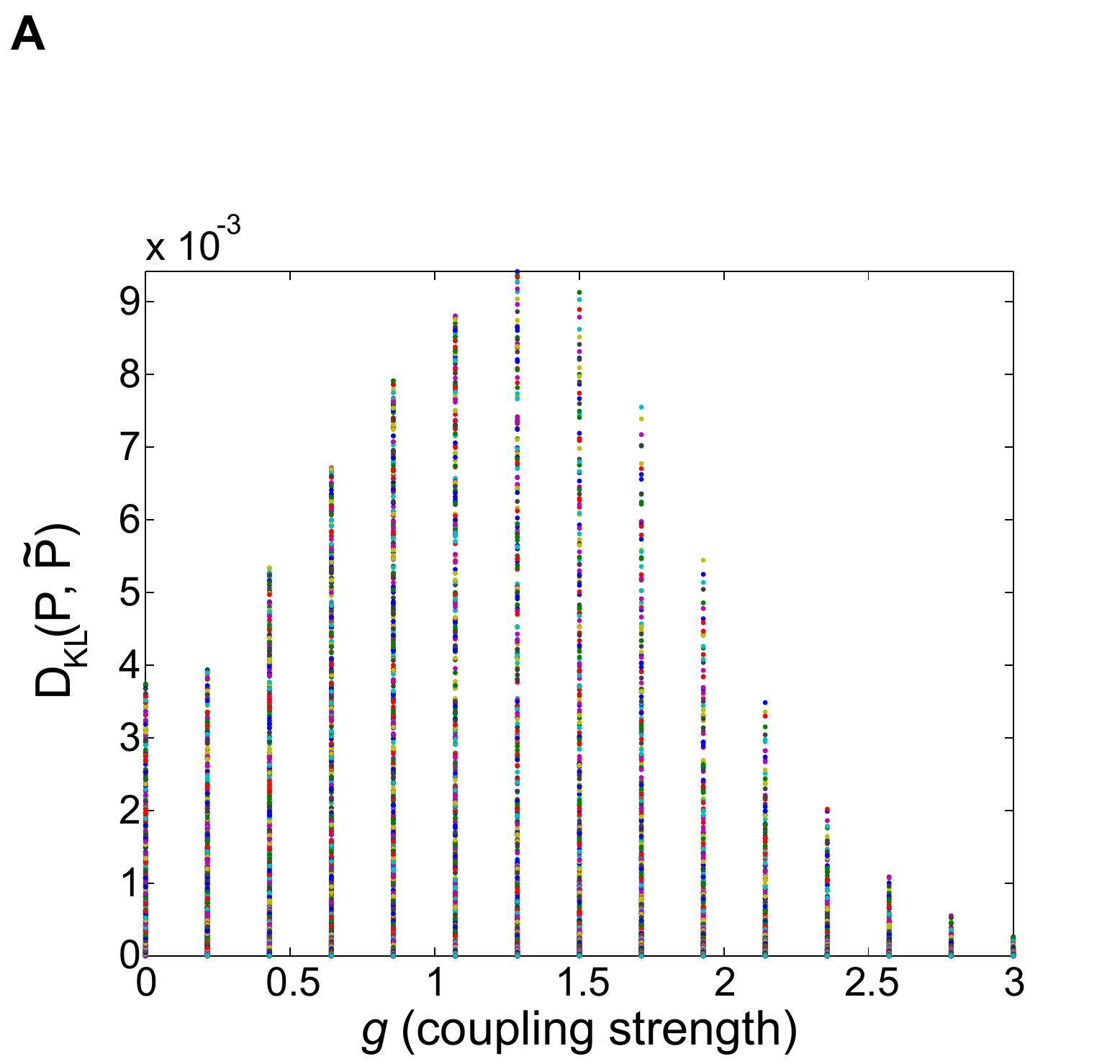}
\includegraphics[width=3in]{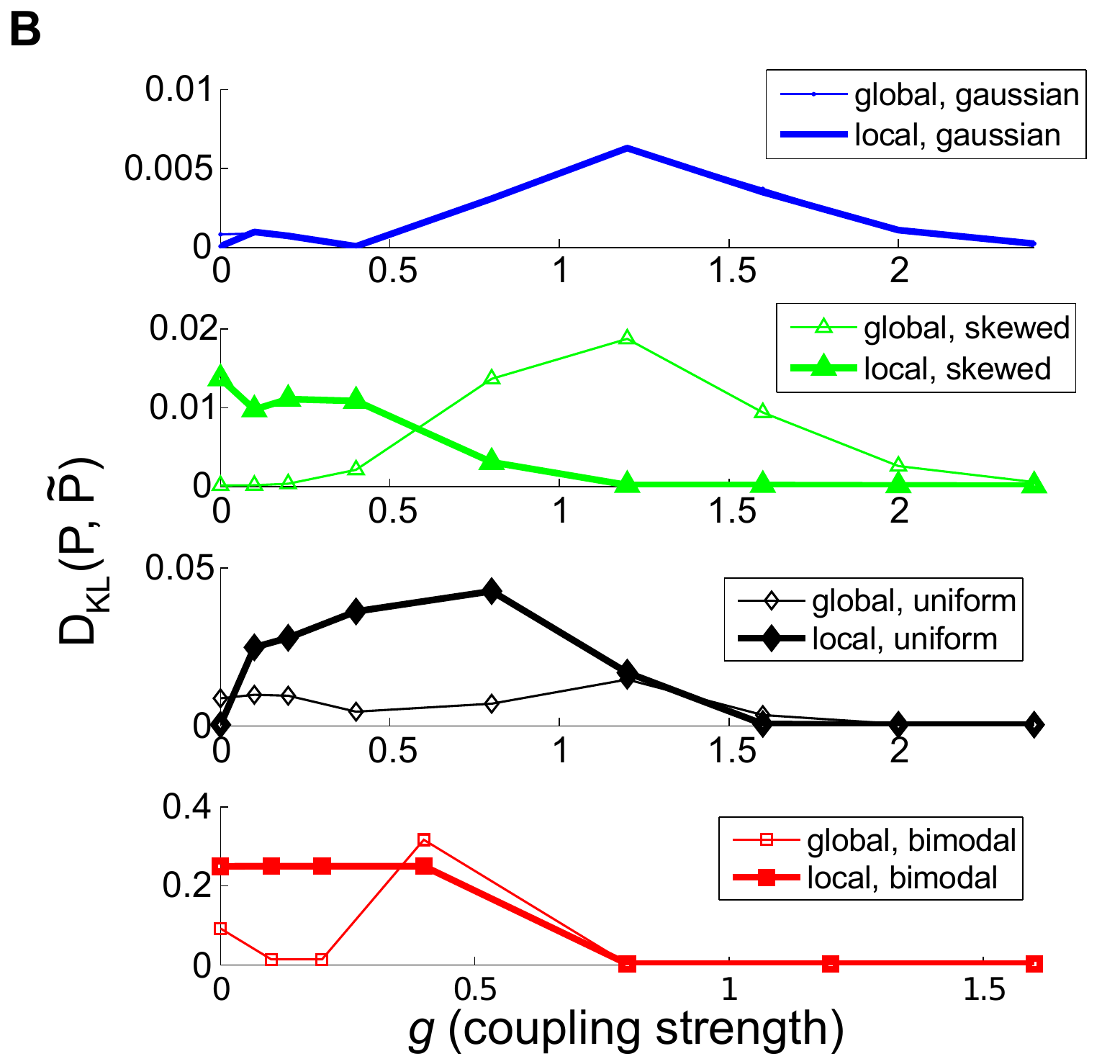}
\caption{Effect of recurrent coupling on PME fits of circuit outputs for the thresholding model.  Left panel: scatter plot of $D_{KL}$ achieved with Gaussian inputs vs. coupling strength $g$, for $N=3$ cells.  Right panel:  another view of $D_{KL}(P,\tilde{P})$ vs. coupling strength $g$, for $N=3$ cells.  In the top three panels, for unimodal inputs, $c$, $\sigma$, and $\Theta$ were fixed at a single representative value, while $g$ was varied. The top panel shows the results from circuits with Gaussian inputs (global and pairwise inputs are equivalent) and the next two for skewed and uniform inputs with either global (thin lines) or local (heavy lines) inputs.  For all unimodal inputs,  $(c,\sigma, \Theta)$ were $(0.5,1,1)$.
Analogous results from circuits with either global (thin line) or local (heavy line) bimodal inputs are shown in the bottom panel. Here, $p$, $q$, and $\Theta$ were fixed at a single representative value, while $g$ was varied. The parameters $(p,q,\Theta)$ were $(0.81, 0.5, 1.5)$ and $(0.5,0,1.5)$ for global and pairwise inputs respectively.}
\label{f.DKL_vs_coup}
\end{figure}

%%%%%%%%%%%%%%%%%%%%%%%%%%%%%%%%%%%%%%%%

%---------------------------------------------------------
%  FIGURE 6, now?
%

\begin{figure}[p!]
\begin{center}
\includegraphics[width=5in]{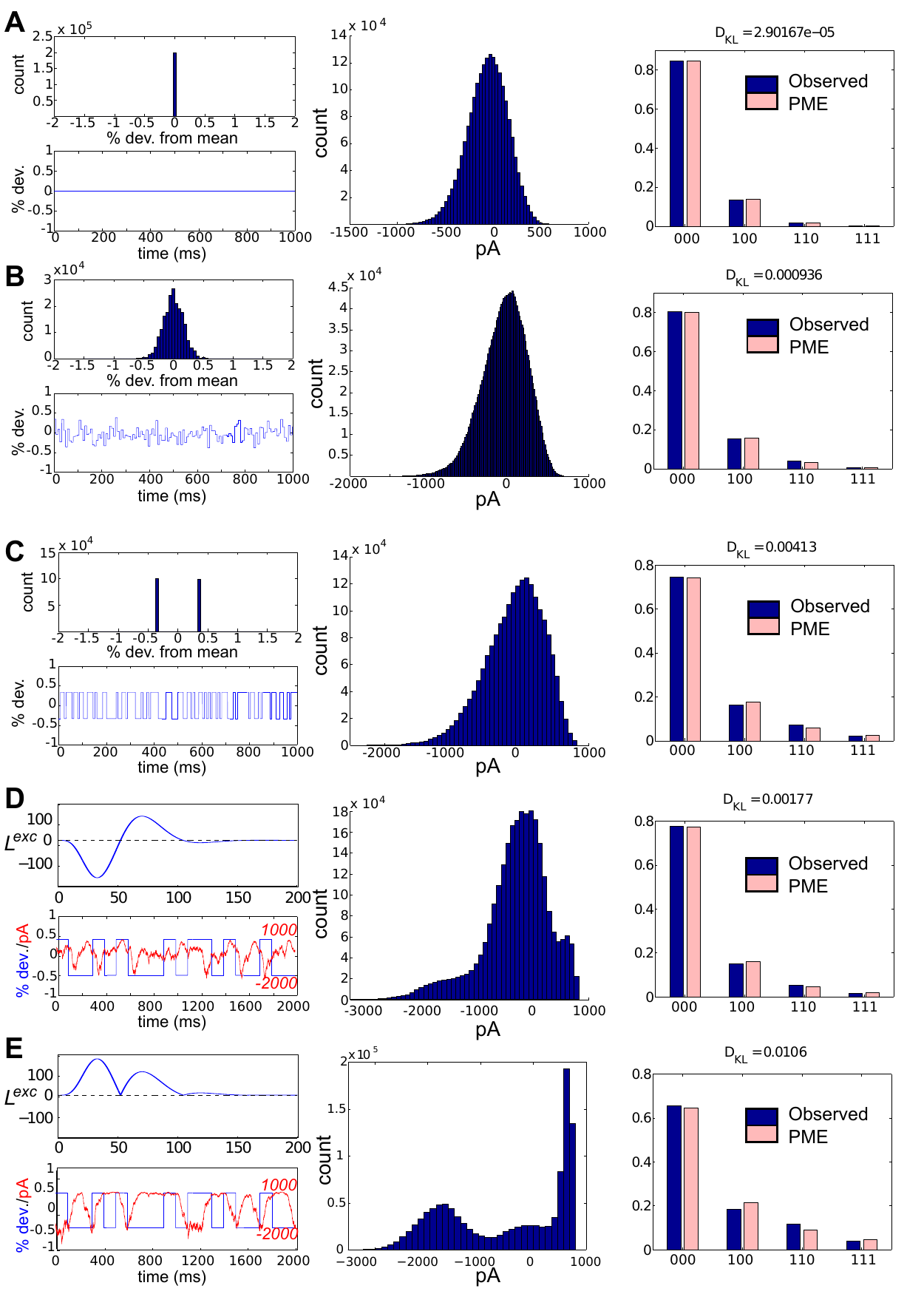}
%\caption{Composite of results for RGC simulations with constant light and full field flicker. In each row, we have (left) a histogram and time series of stimulus, (center) 
%a histogram of excitatory conductances and (right) the resulting distribution on spiking
%patterns. (A) Gaussian noise only. (B) Gaussian input, standard deviation $1/6$, refresh rate $8$ ms. (C) Binary input, standard deviation $1/3$, refresh rate $8$ ms.  (D) Binary input, standard deviation $1/2$, refresh rate $100$ ms. The normalized excitatory conductance (red dashed) is superimposed on the stimulus (blue solid) to illustrate that the LN model that processes light input acts as a (time-shifted) high pass filter.} 
\label{fig:fullfield_composite_pic_only}
\end{center}
\end{figure}

\setcounter{figure}{6}
\begin{figure}[p!]
\begin{center}
\caption{Composite of results for RGC simulations with constant light and full field flicker. In rows (A-C), we have (left) a histogram and time series of stimulus, (center) 
a histogram of excitatory conductances (multiplied by driving voltage $-60$ mV, and therefore in units of pA)  and (right) the resulting distribution on spiking
patterns. (A) Gaussian noise only. (B) Gaussian input, standard deviation $1/6$, refresh rate $8$ ms. (C) Binary input, standard deviation $1/3$, refresh rate $8$ ms.  (D-E) Binary input, standard deviation $1/3$, refresh rate $100$ ms. In the top left panels, the excitatory filter $L^{\text{exc}}(t)$ is shown instead of a stimulus histogram; in the bottom left panels, the normalized excitatory conductance (in pA --- red dashed line) is superimposed on the stimulus (blue solid). (D) Filter fit from data, parameters given in Methods. Both the filter and conductance trace illustrate that the LN model that processes light input acts as a (time-shifted) high pass filter. (E) As in (D), but with rectified filter, producing a bimodal distribution of conductances.} 
\label{fig:fullfield_composite}
\end{center}
\end{figure}

 \begin{figure}[p!]
\begin{center}
\includegraphics[width=\textwidth]{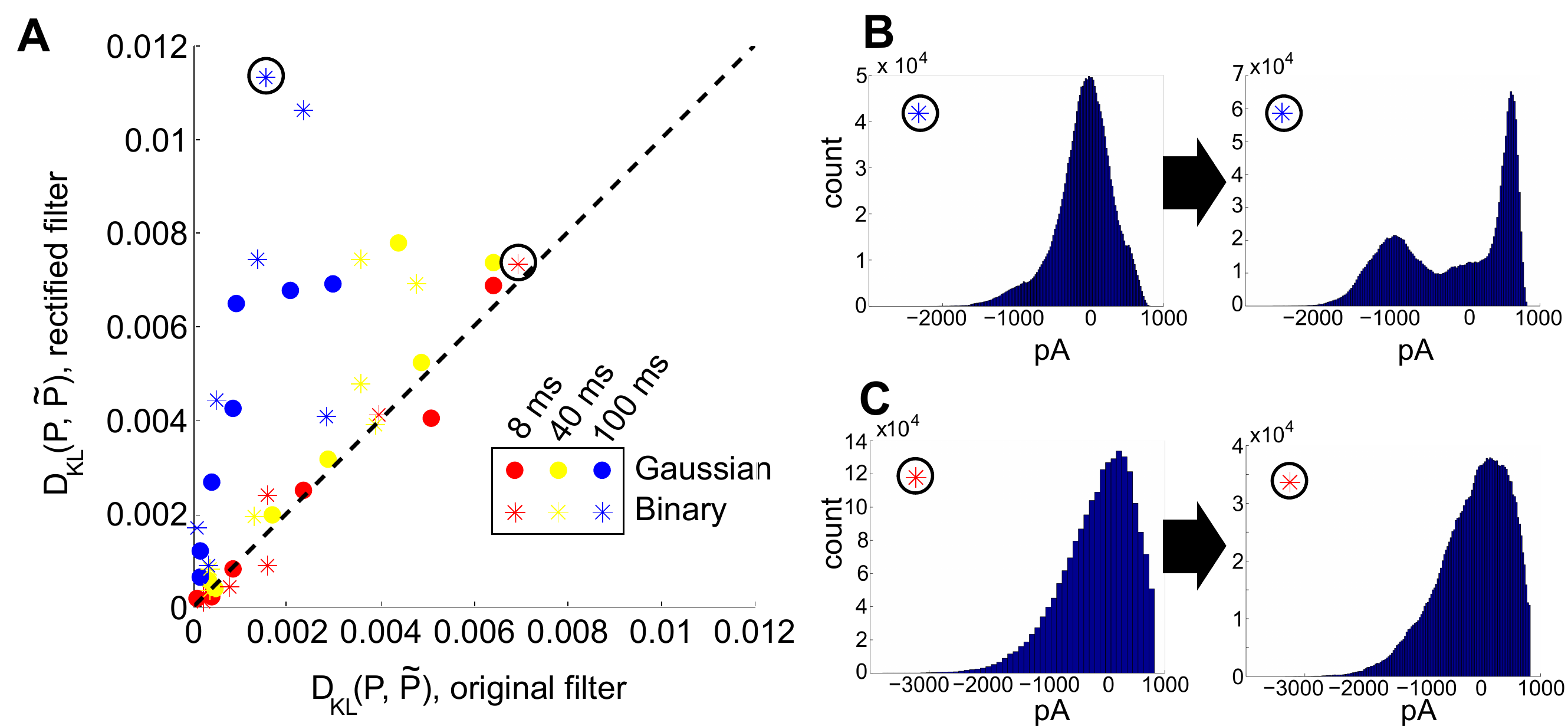}
\caption{Comparison of RGC simulations computed with the original ON parasol filter, vs. simulations incorporating a rectified filter. (A) $D_{KL}(P, \tilde{P})$ for original vs. rectified filter. Data is organized by stimulus refresh rate (8, 40, and 100 ms) and marginal statistics (Gaussian vs. binary). (B-C) Histograms of excitatory conductances (multiplied by driving voltage $-60$ mV, and therefore in units of pA) for representative simulations, under original (left) and rectified (right) filters. (B) Binary stimulus with standard deviation of $1/4$ and refresh rate 100 ms. The histogram shows a significant shift towards a bimodal structure, with a corresponding increase in $D_{KL}(P, \tilde{P})$.  (C) Binary stimulus with standard deviation of $1/2$ and refresh rate 8 ms. The histogram shows little qualitative change under the change in filter, and little change in $D_{KL}(P, \tilde{P})$.} \label{fig:rectified_vs_orig}
\end{center}
\end{figure}

%---------------------------------------------------------
%  FIGURE 8
%
% NOTE (12/12/11):
% FIGURES D,E,F and G,H,I have been switched, so that 
%     60 um is shown before 256 um (matches order in text)
%
\begin{figure}[p!]
\begin{center}
\includegraphics[width=0.3\textwidth]{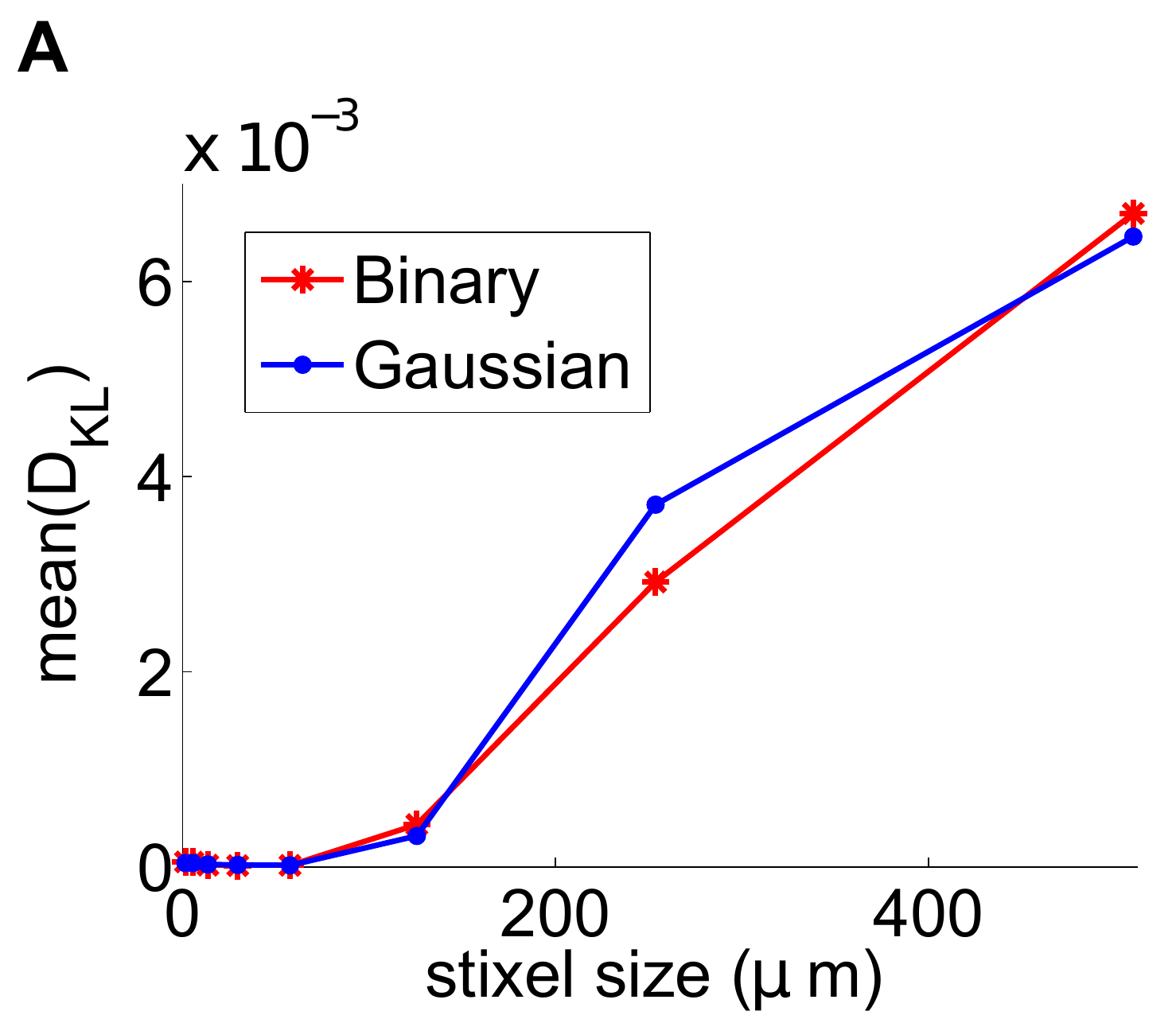}
\includegraphics[width=0.3\textwidth]{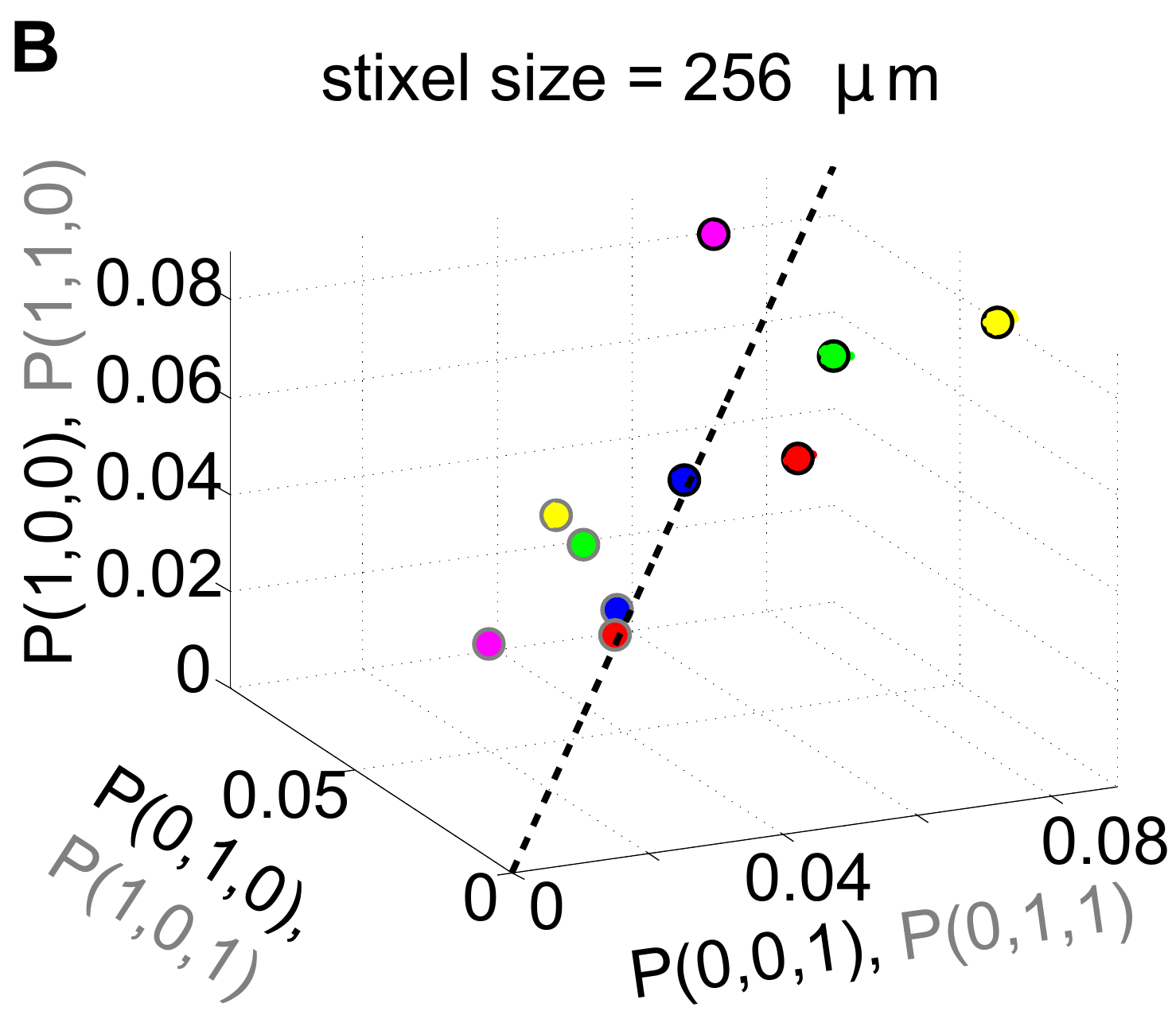}
\includegraphics[width=0.3\textwidth]{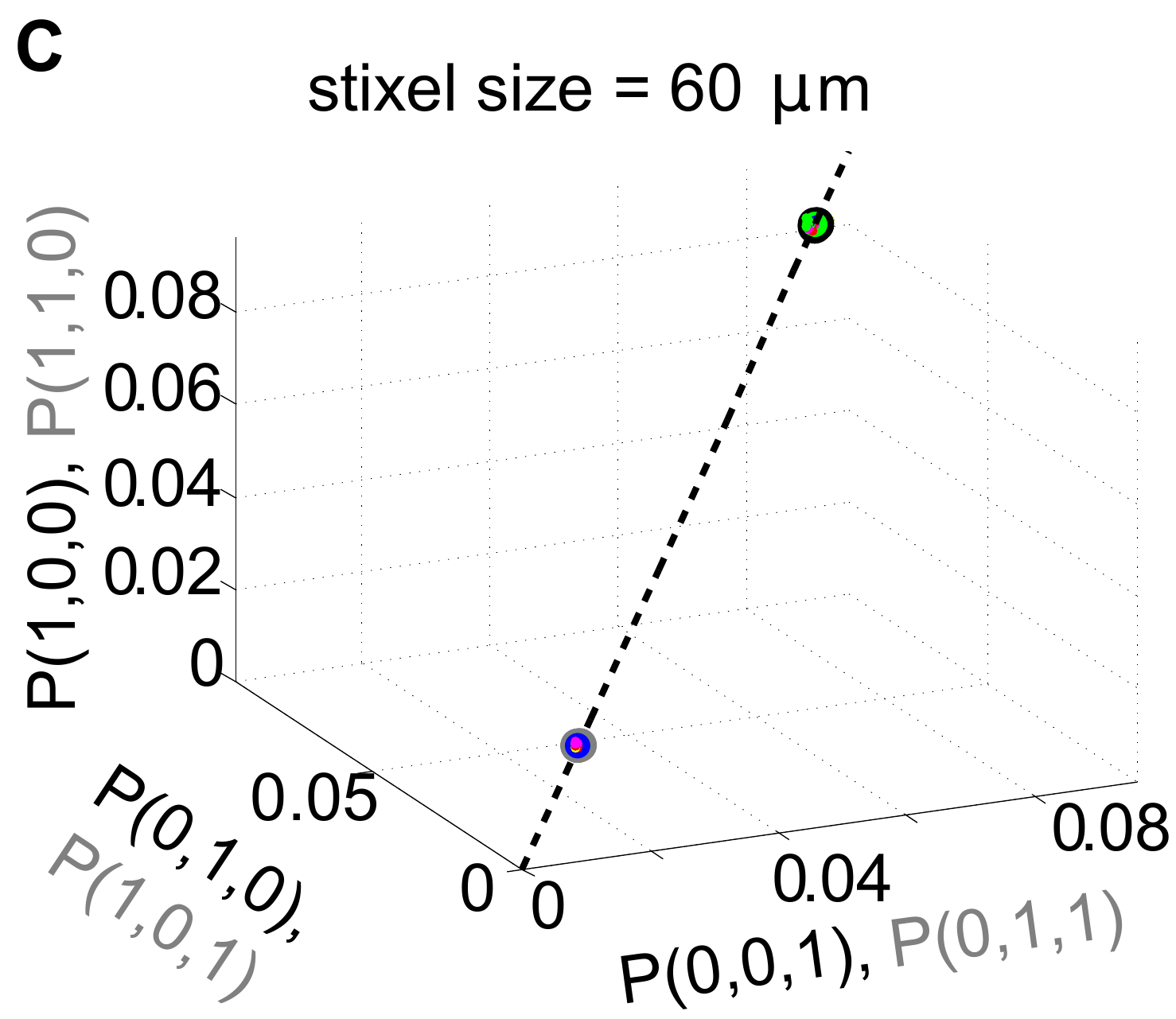}\\
\includegraphics[width=0.3\textwidth]{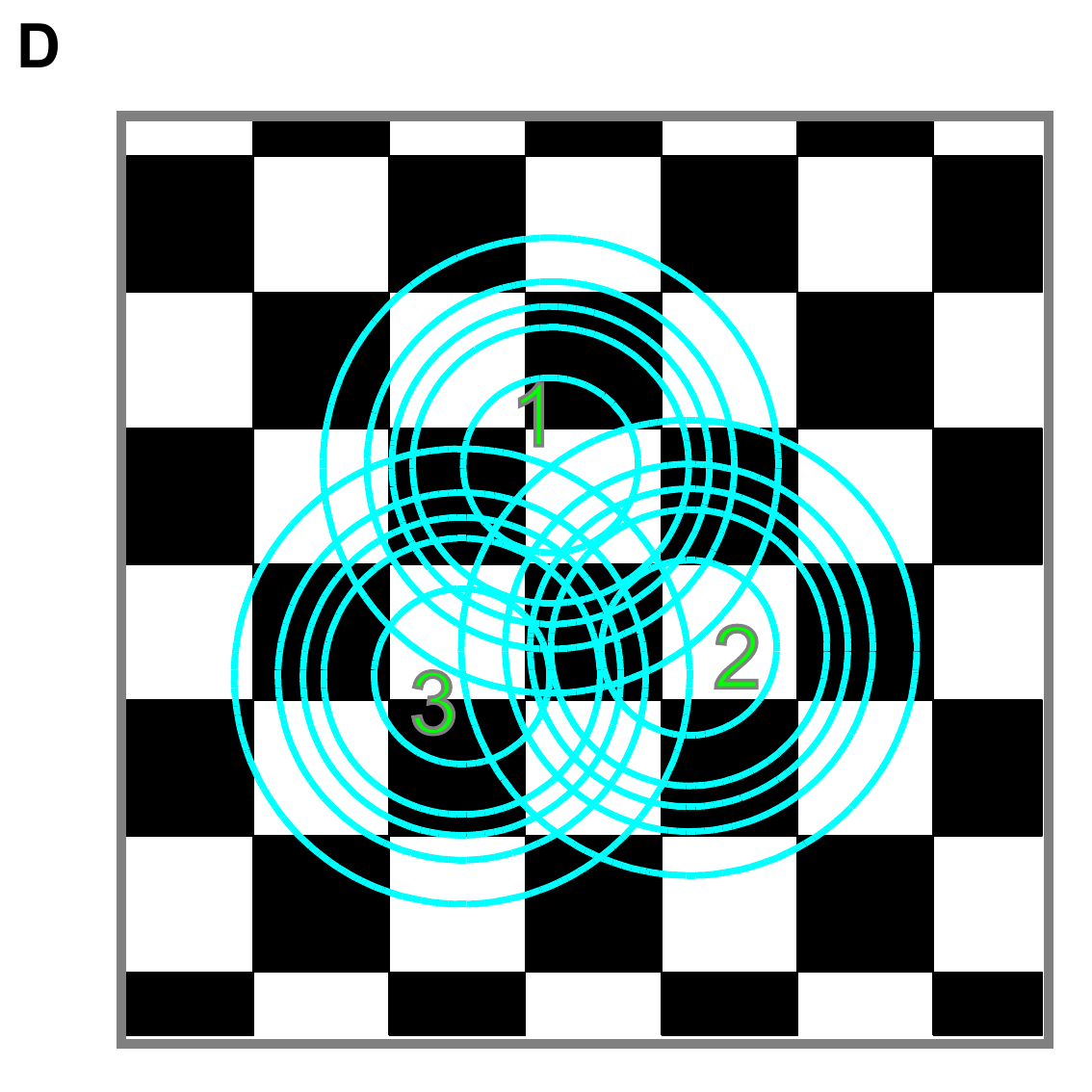}
\includegraphics[width=0.3\textwidth]{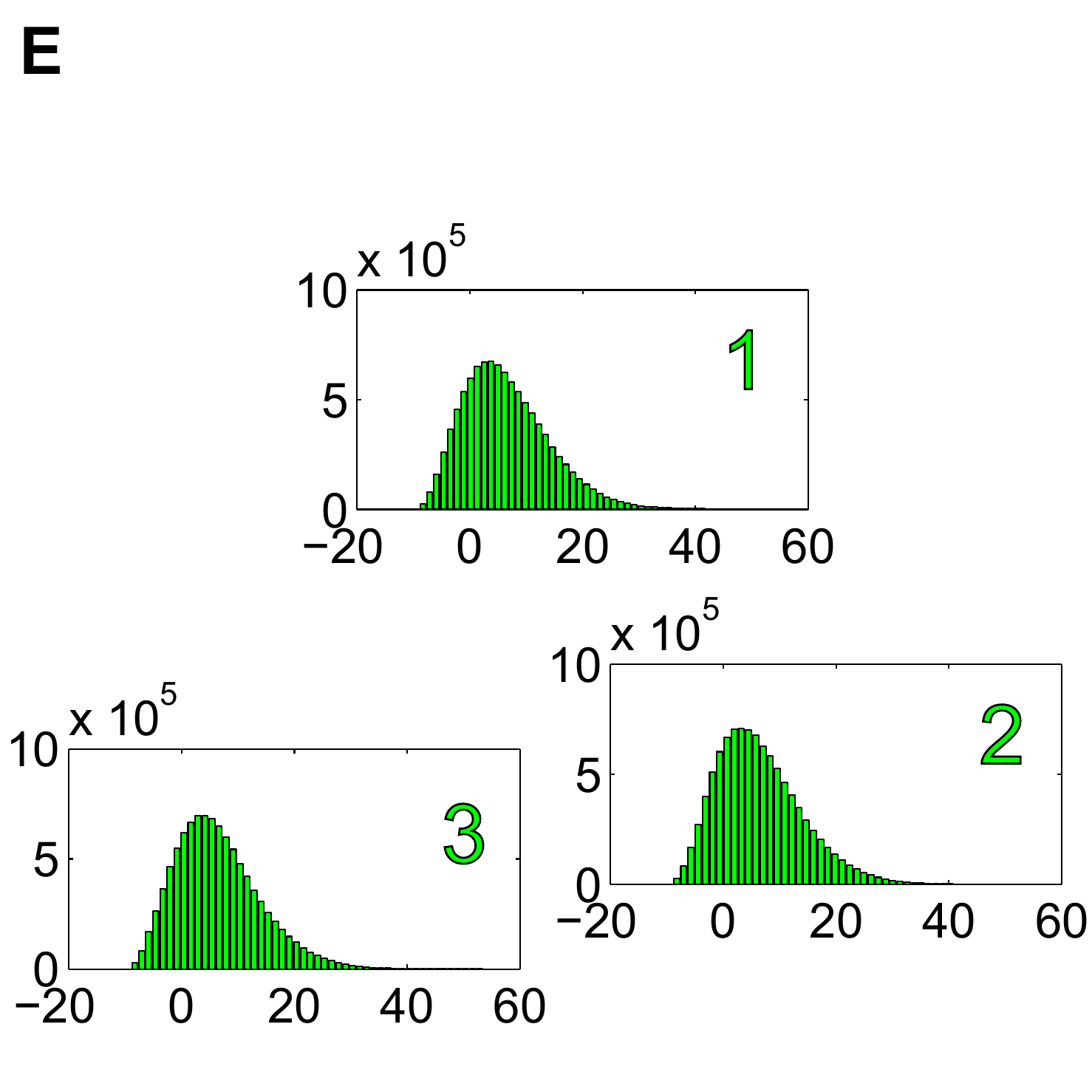}
\includegraphics[width=0.3\textwidth]{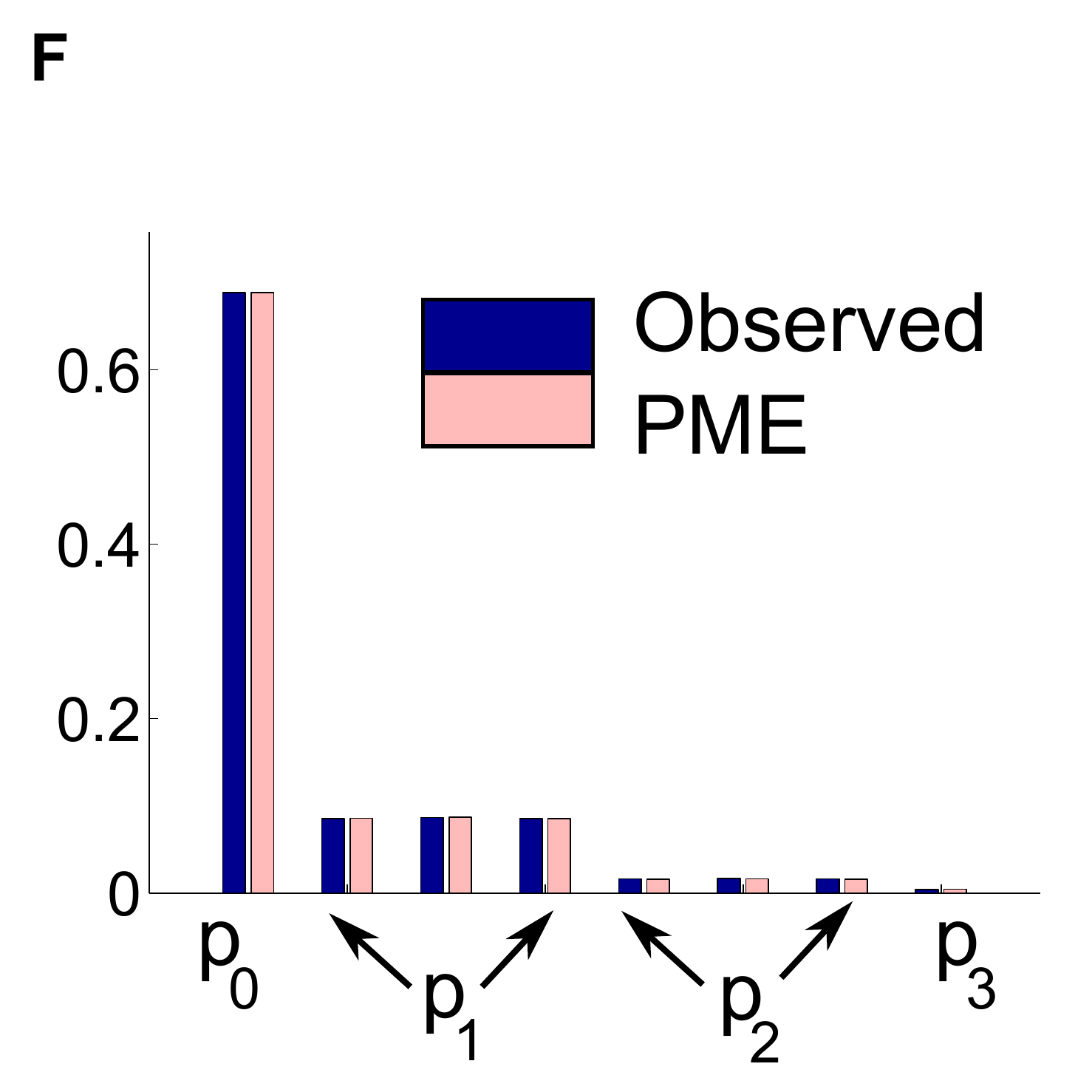}\\
\includegraphics[width=0.3\textwidth]{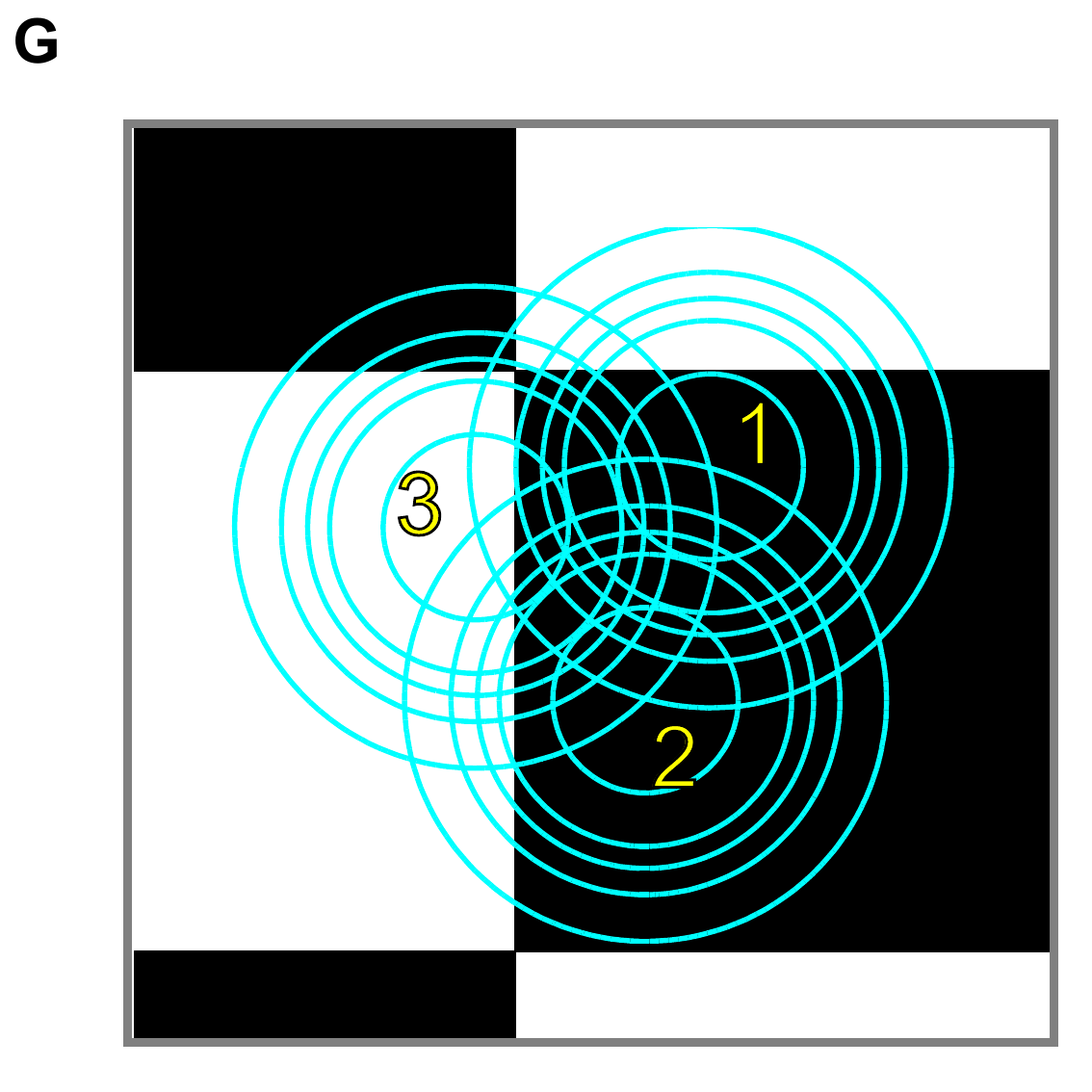}
\includegraphics[width=0.3\textwidth]{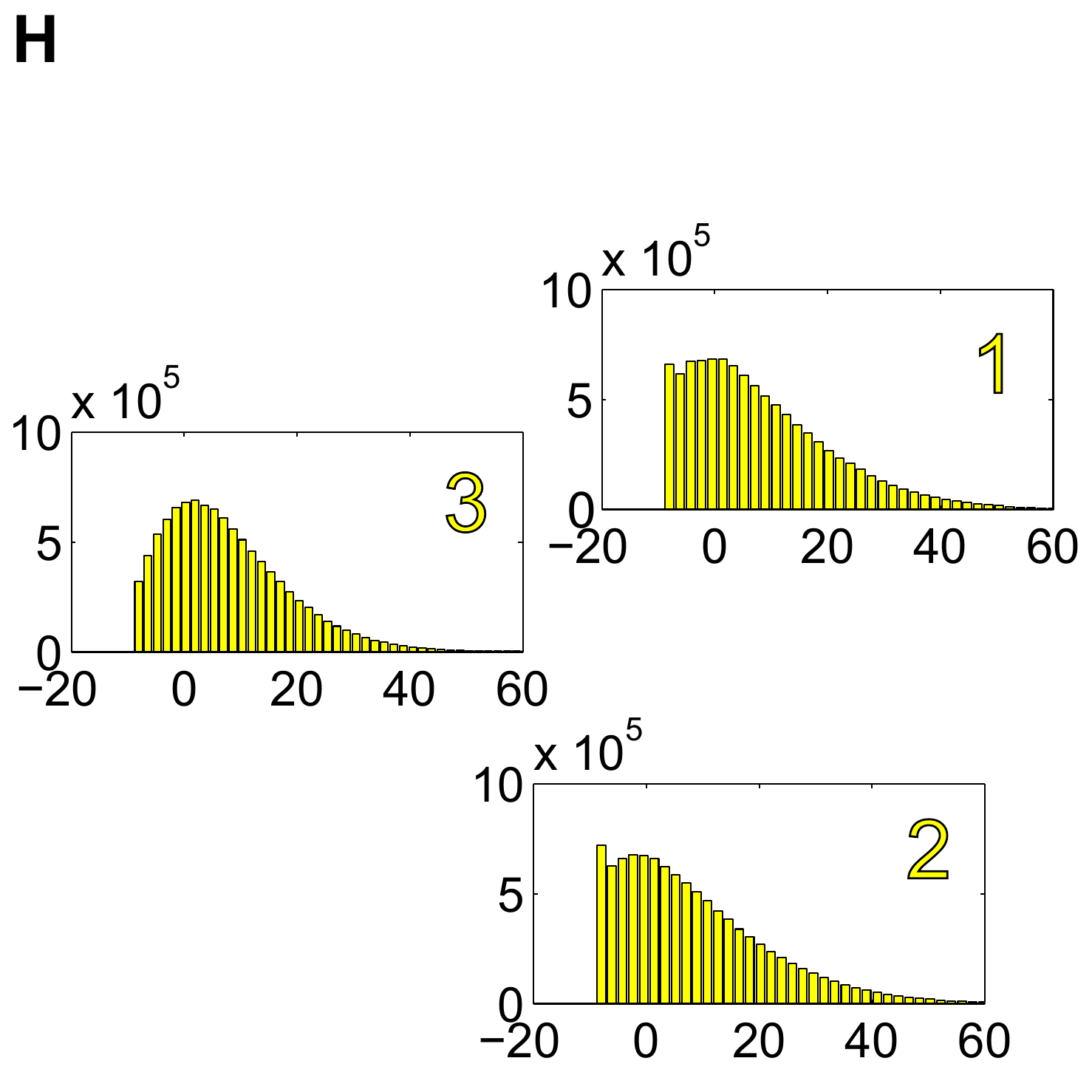}
\includegraphics[width=0.3\textwidth]{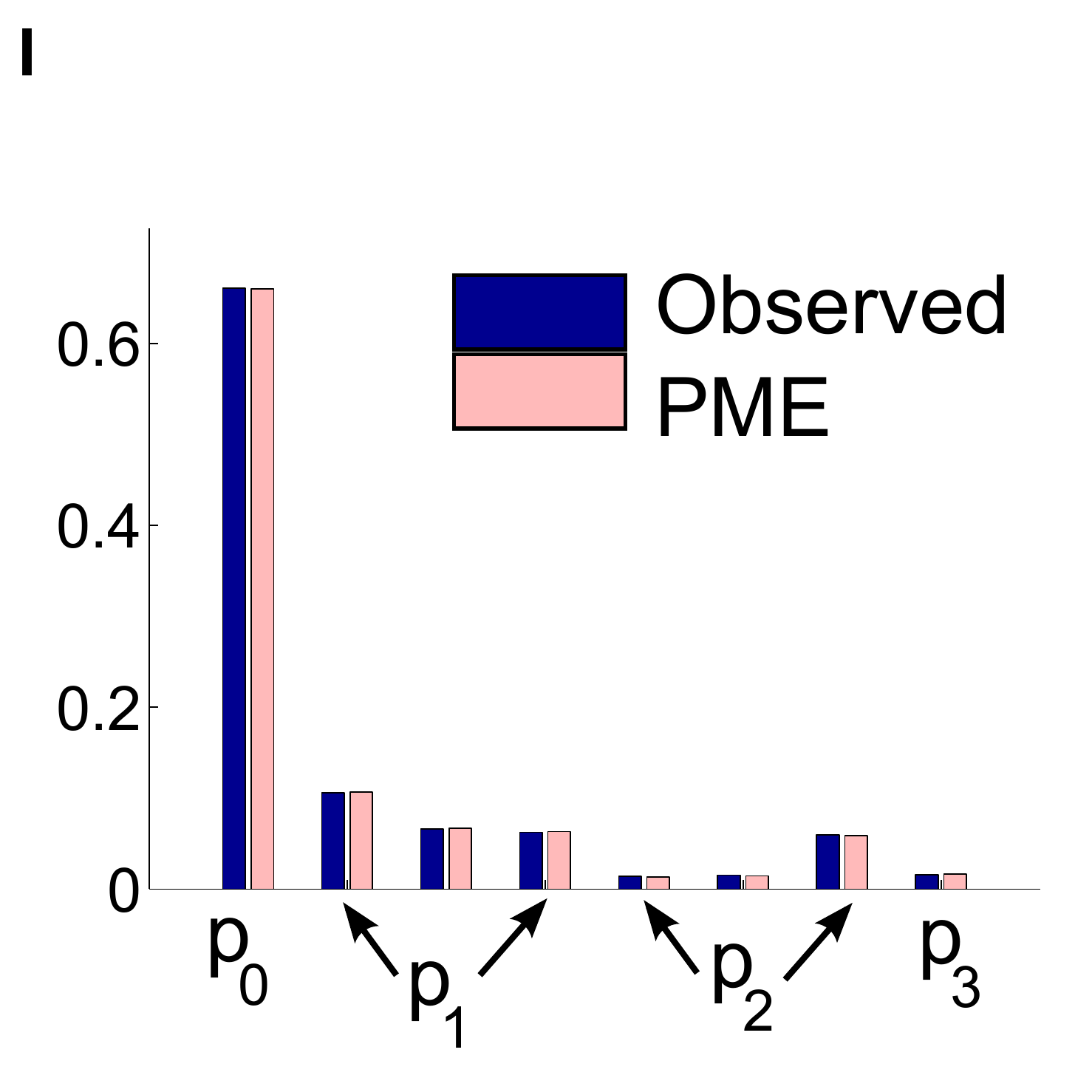}
%\caption{Mock up of Figure 8} \label{fig:stixel_fig}
\label{fig:stixel_pic_only}
\end{center}
\end{figure}

%\begin{figure}[p!]
%\begin{center}
%\begin{figure}[htbp]
%\begin{center}
%\includegraphics[width=5in]{../../figures/fig8_stixel.pdf}
%%\caption{Results for RGC simulations with light stimuli of varying spatial scale (``stixels").  With the exception of (A), we show data from the binary light distributions; results from the Gaussian case are similar.
%% (A) Average $D_{KL}(P,\tilde{P})$ as a function of stixel size.
%%Values were averaged over 5 stimulus positions, each with a different (random) stimulus rotation and translation; $512 \, \rm{\mu m}$ corresponds to full field stimuli.
%%(B,C)  Single (black outline) and double (cyan outline) spiking events; individual runs (dots) and averages (large circles). 
%%The black line indicates 
%%perfect homogeneity among cells (e.g. $P(1,0,0) = P(0,1,0) = P(0,0,1))$. Different colors indicate different stimulus positions. (D-F) Results for one stimulus position, with stixel size $256 \, \rm{\mu m}$. (D) Contour lines of the three receptive fields (at 0.5, 1, 1.5 and 2 SD; and at the zero contour line) superimposed on the stimulus checkerboard (for illustration, pictured in an alternating
%%black/white pattern). (E) Marginal distributions of the excitatory conductances, for each cell. (F) Spike pattern distribution; the three different probabilities labeled $p_1$ correspond to, e.g, $P(1,0,0)$, $P(0,1,0)$, and $P(0,0,1))$, demonstrating heterogenous responses among the RGCs. 
%%(G-I) As in (D-F), but for stixel size $60 \, \rm{\mu m}$.}
%\label{fig:stixel_pic_only}
%\end{center}
%\end{figure}

% caption only -----------------------------------
\setcounter{figure}{8}
\begin{figure}
\begin{center}
\caption{Results for RGC simulations with light stimuli of varying spatial scale (``stixels").  With the exception of (A), we show data from the binary light distributions; results from the Gaussian case are similar.
 (A) Average $D_{KL}(P,\tilde{P})$ as a function of stixel size.
Values were averaged over 5 stimulus positions, each with a different (random) stimulus rotation and translation; $512 \, \rm{\mu m}$ corresponds to full field stimuli.
(B,C)  Probability of singlet and doublet spiking events, under stimulation by movies of $256 \, \rm{\mu m}$ (B) and $60 \, \rm{\mu m}$ (C) stixels. 
Event probabilities are plotted in 3-space, with the $x$, $y$, and $z$ axes identifying the singlet (doublet) events $001$ ($011$),  $010$ ($101$), and $100$ ($110$) respectively. The black dashed line indicates 
perfect cell-to-cell homogeneity (e.g. $P(1,0,0) = P(0,1,0) = P(0,0,1))$. 
Both individual runs (dots) and averages over 20 runs (large circles) are shown, with averages outlined in black (singlet) and gray (doublet). Different colors indicate different stimulus positions. 
(D-F) Results for one stimulus position, with stixel size $60 \, \rm{\mu m}$. (D) Contour lines of the three receptive fields (at 0.5, 1, 1.5 and 2 SD; and at the zero contour line) superimposed on the stimulus checkerboard (for illustration, pictured in an alternating
black/white pattern). (E) Marginal distributions of the excitatory conductances, for each cell. (F) Spike pattern distribution; the three different probabilities labeled $p_1$ correspond to, e.g, $P(1,0,0)$, $P(0,1,0)$, and $P(0,0,1))$, demonstrating heterogenous responses among the RGCs. 
(G-I) As in (D-F), but for stixel size $256 \, \rm{\mu m}$.}
\label{fig:stixel}
\end{center}
\end{figure}

\begin{figure}[p!]
\begin{center}
\includegraphics[width=\textwidth]{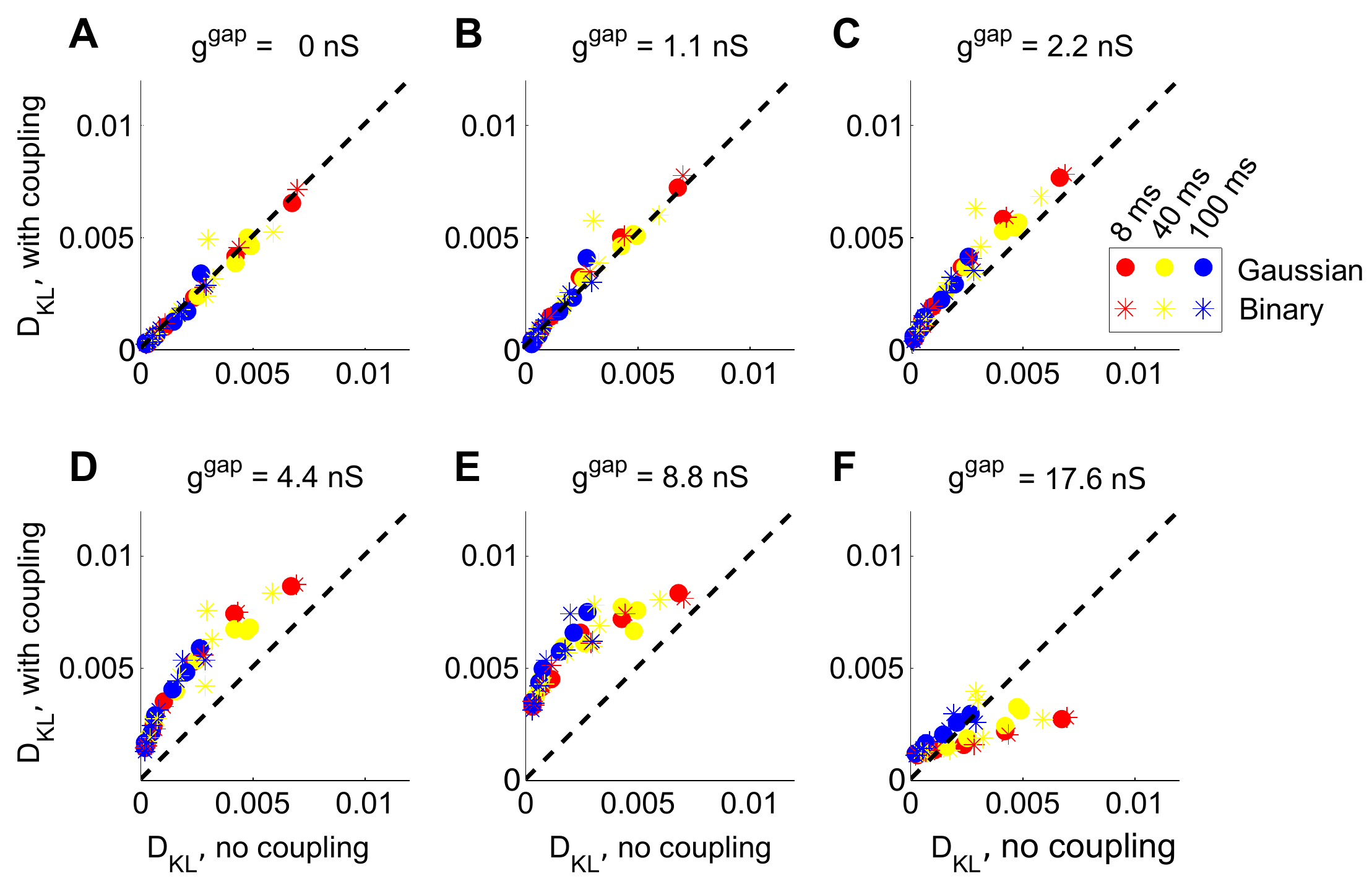}
\caption{The impact of recurrent coupling on RGC networks with full-field visual stimuli. 
The strength of gap junction connections was varied from a baseline level (relative magnitude $g=1$, or absolute magnitude $g^{\text{gap}} = 1.1$ nS) to an order of magnitude larger ($g=16$, or $g^{\text{gap}} = 17.6$ nS). (A) $D_{KL}(P,\tilde{P})$ obtained from two independent simulations without coupling, for each of 42 different stimulus ensembles (deviation of this data from the line $y=x$ thus gives a control for the statistical variability of the results in later panels). (B-F) $D_{KL}(P,\tilde{P})$ obtained with coupling, plotted versus the value obtained for the same stimulus ensemble without coupling, for each of 42 different stimulus ensembles.} \label{fig:recur_vs_orig}
\end{center}
\end{figure}

\begin{figure}[t!]
\begin{center}
\includegraphics[width=\textwidth]{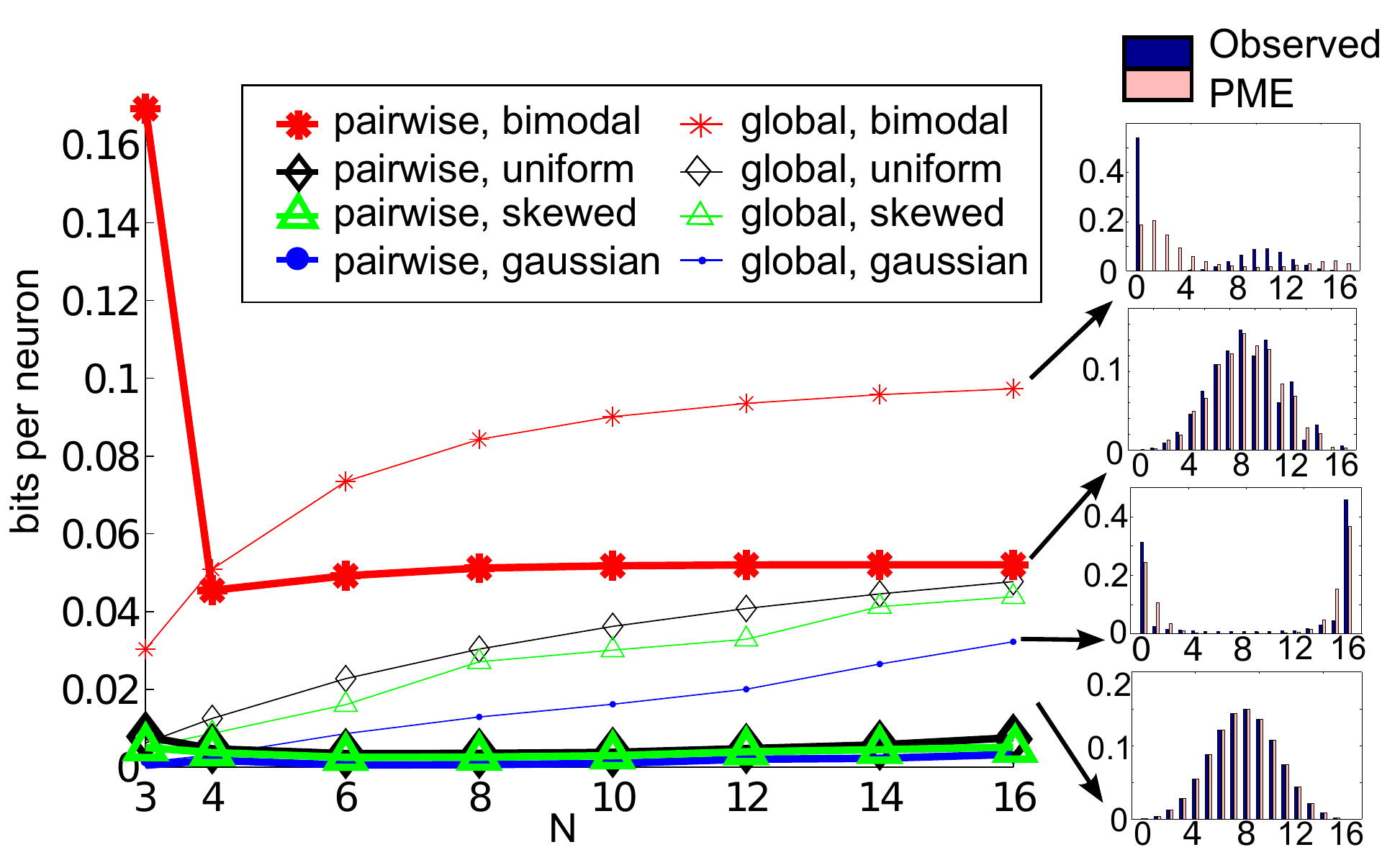}
\caption{Maximal deviation from PME fit for thresholding circuit model, forced by either global or local input against background noise, for increasing network size $N$.
For each $N$, possible input parameters are ranged as described in the Results:
over $c \in [0,1]$, $\sigma \in [0,4]$, and $\Theta \in [-1,3]$ (unimodal inputs), over $s \in [0,1]$ and $p \in [0,1]$ (bimodal inputs).  The sidebar shows sample distributions with maximal $D_{KL}(P,\tilde{P})$ for $N=16$; from top, global bimodal inputs, pairwise bimodal inputs, global Gaussian inputs, and pairwise Gaussian inputs.} 
\label{fig:max_DKL_Nmedium}
\end{center}
\end{figure}

\begin{figure}
\begin{center}
\includegraphics[width=\textwidth]{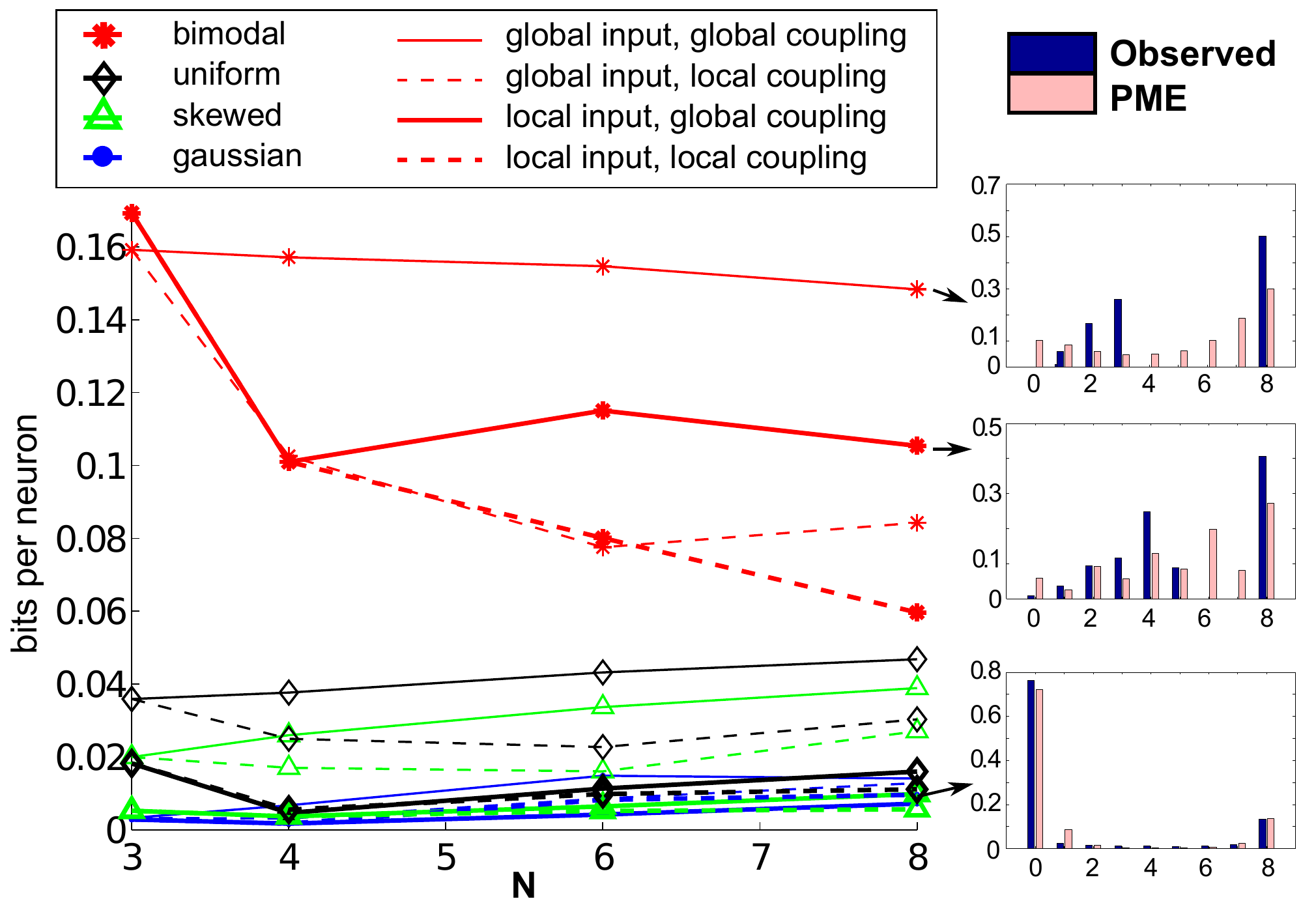}
\caption{Maximal deviation from PME fit for circuit forced by either global or local input against background noise
and either global or local recurrent coupling.
Plot shows the maximal $D_{KL}$ achieved over all parameters as $N$ increases.
For each $N$, possible input parameters are ranged described as in the Results;
over $c \in [0,1]$, $\sigma \in [0,4]$, and $\Theta \in [-1,3]$ (unimodal inputs), over $s \in [0,1]$, $p \in [0,1]$, and $\Theta \in [0.3,1.9]$ (bimodal inputs).  In addition, recurrent coupling strength $g$ is varied over $[0,2.4]$.  The sidebar shows sample distributions with maximal $D_{KL}(P,\tilde{P})$ for $N=8$; from top, global bimodal inputs with global coupling; pairwise bimodal inputs with global coupling; global Gaussian inputs with pairwise coupling.} 
\label{fig:max_DKL_Recur}
\end{center}
\end{figure}

\begin{figure}
\begin{center}
\includegraphics[width=5.0in]{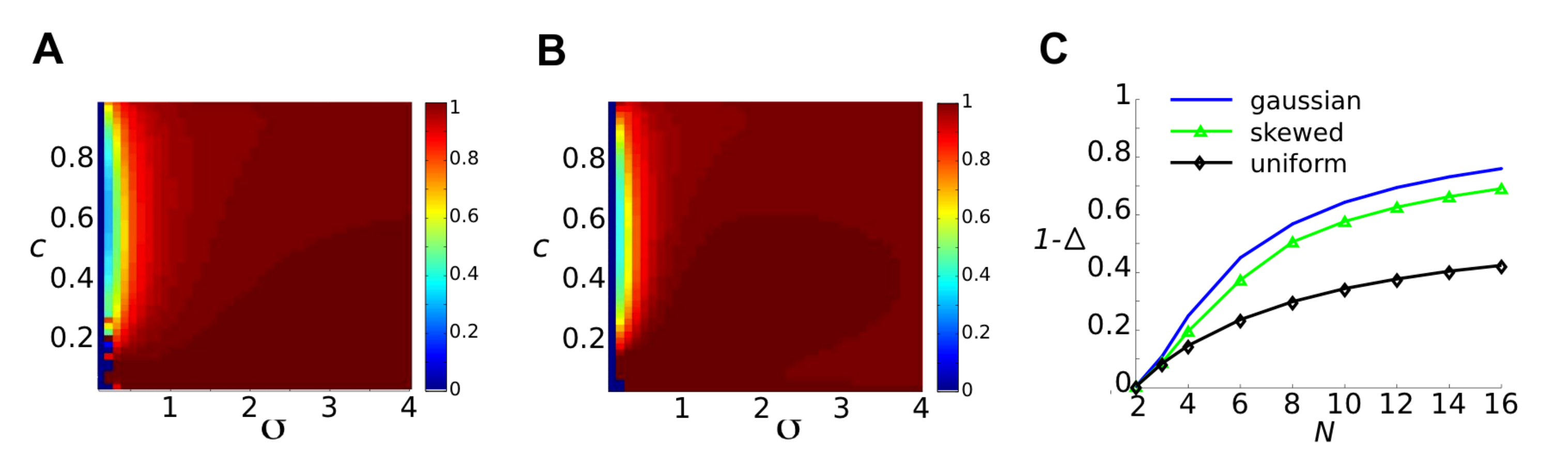}
\caption{ The fraction of multi-information captured by the PME model, $\Delta$, 
is low for some network parameters as population size increases. 
Here, we show data
obtained from feedforward networks of $N=12$ thresholding cells. For each choice of marginal input statistics, 
possible input parameters were varied over correlation coefficient $c \in [0,1]$ and input variance $\sigma \in [0,4]$.
The threshold was held at $\Theta = 1$. (A-B) $\Delta$ for (A) Gaussian and (B) skewed input, for $N=12$. (C) $1-\Delta$ vs. $N$ for Gaussian, skewed and uniform inputs at a fixed value of $c$ and $\sigma$ ($c = 0.56$; for Gaussian and skewed, $\sigma = 0.2$; for uniform, $\sigma = 0.6$). Because of the low firing rate, $1-\Delta$ must grow linearly with $N$ (see text).} \label{fig:outputstats_N12}
\end{center}
\end{figure}

\setcounter{figure}{0}
\makeatletter
\renewcommand{\thefigure}{S\@arabic\c@figure}

%---------------------------------------------------------
\begin{figure}[p!]
\begin{center}
\includegraphics[width=5in]{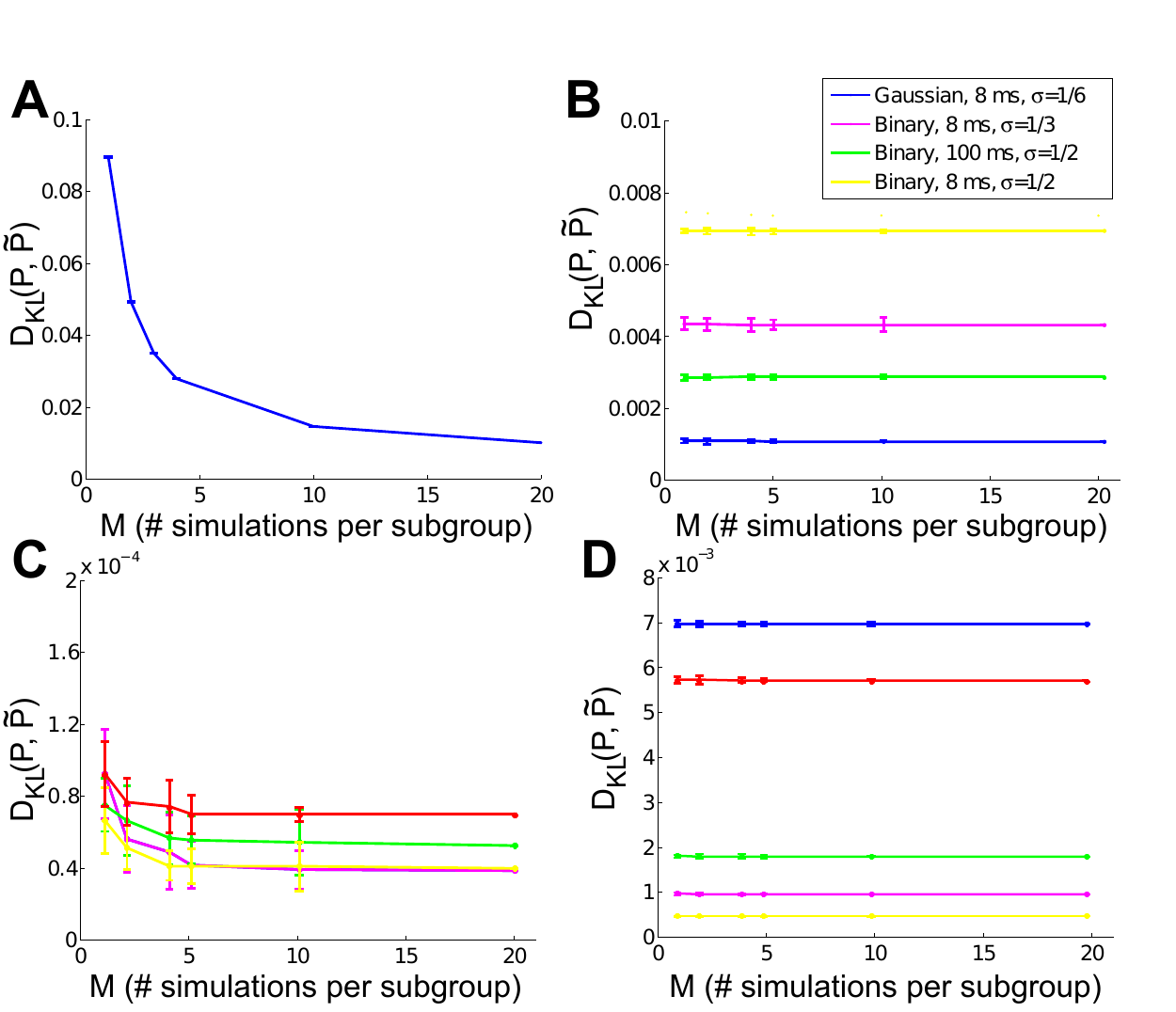}
\caption{Mean and estimated standard errors of $D_{KL}(P,\tilde{P})$, as a function of subgroup size.
The 20 simulations for each circuit condition (see Materials and Methods) were collected into subgroups of $M = 1$, 2, 4, 5, 10, and 20.
$M=20$ corresponds to the full simulation length --- $2 \times 10^6$ ms in (B), $2 \times 10^5$ ms in (C,D) --- reported in the text. 
As expected, bias decreases as the length of subgroup increases and asymptotes at  --- or before --- the full simulation length.
(A) $N=16$, Gaussian pairwise inputs, for the sum-and-threshold model.
(B) Full-field RGC simulations. (C) Spatially variable RGC simulations, binary stimulus, stixel size = $4 \, {\rm \mu m}$. Different colors signify different positions of stimulus relative to receptive field. (D) As in (C), but stixel size = $256 \, {\rm \mu m}$.} \label{fig:suppl}
\end{center}
\end{figure}

%---------------------------------------------------------
\begin{figure}[p!]
\begin{center}
\includegraphics[width=0.45\textwidth]{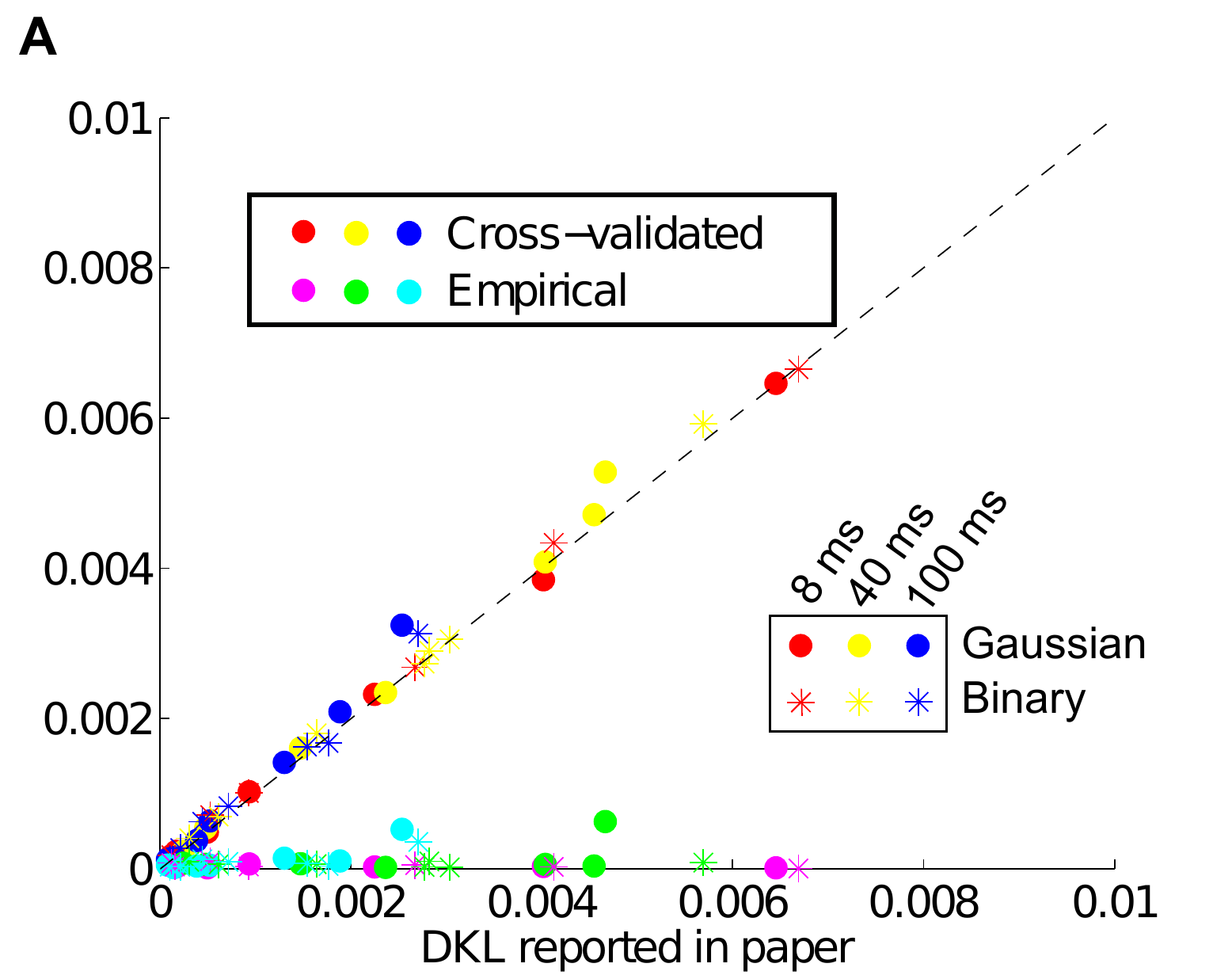}
\includegraphics[width=0.45\textwidth]{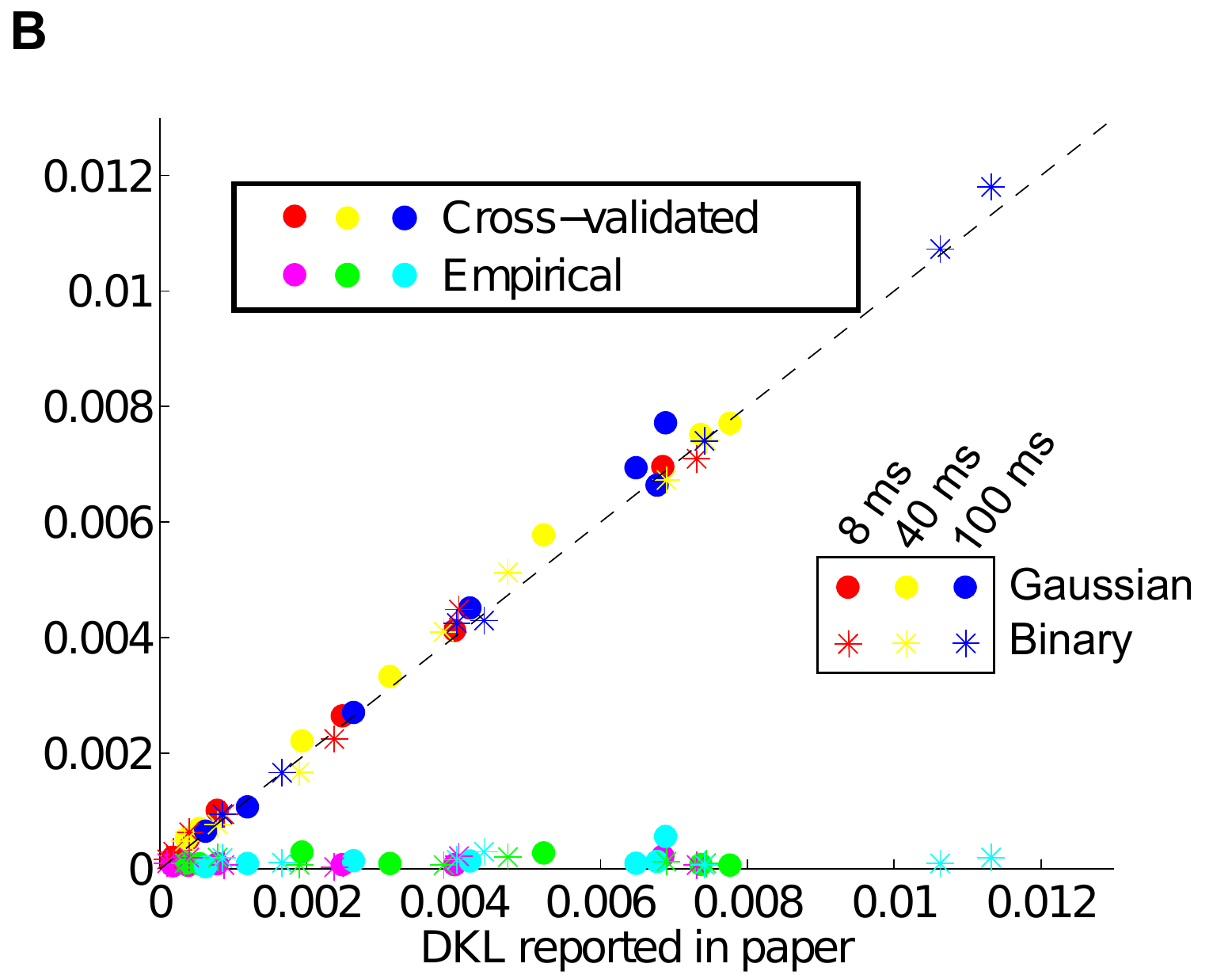}\\
\includegraphics[width=0.45\textwidth]{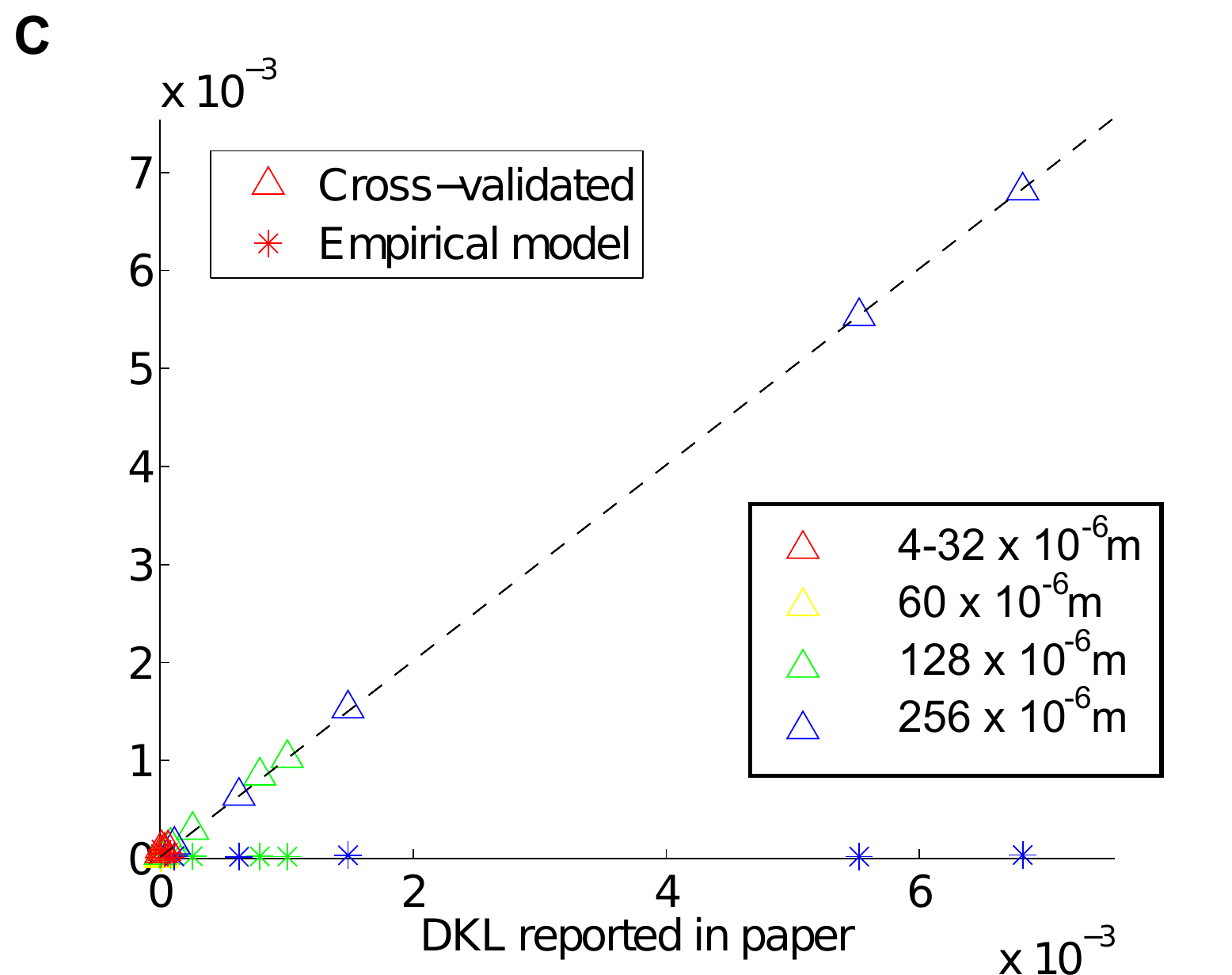}
\includegraphics[width=0.45\textwidth]{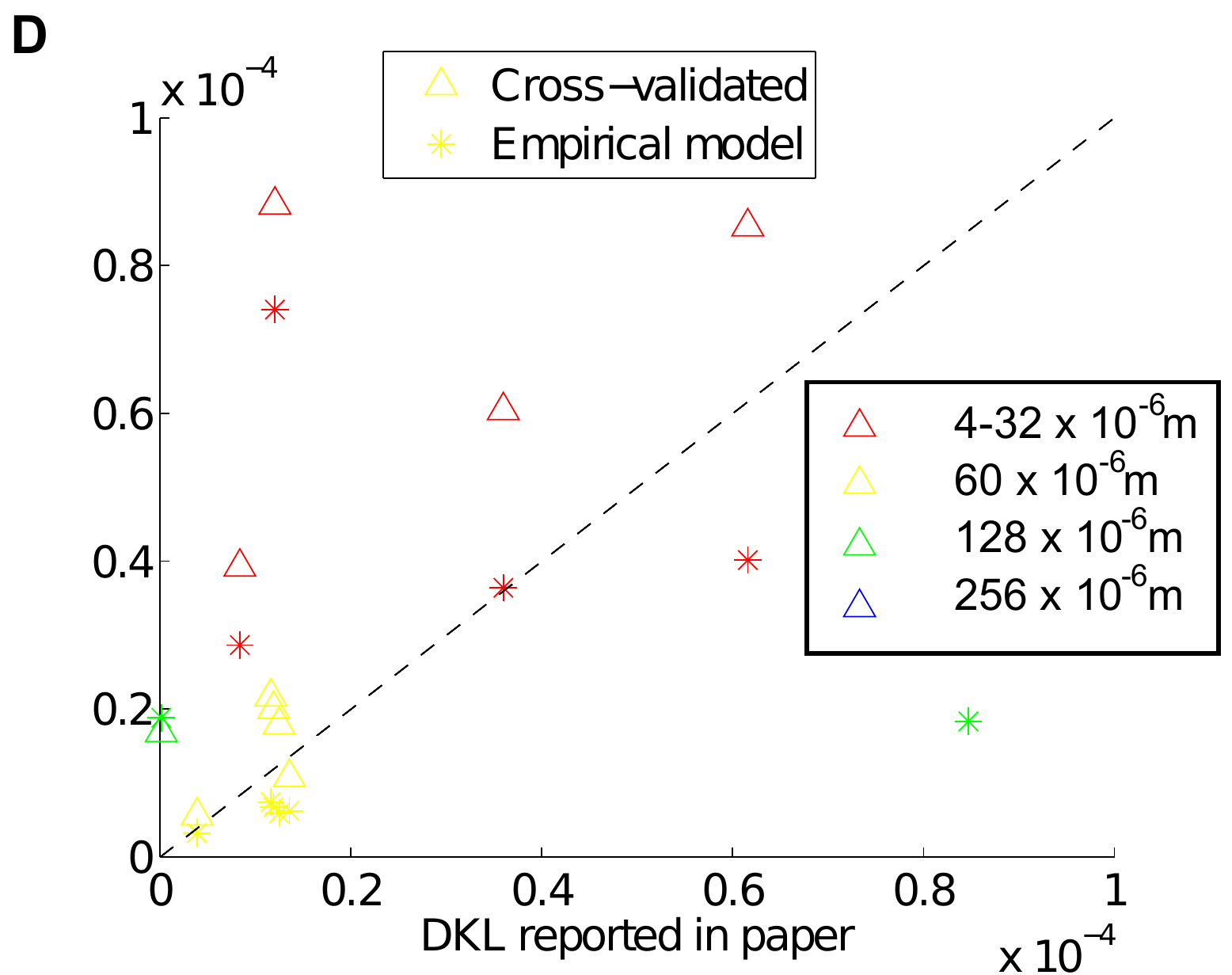}
\caption{Cross-validated and empirical values of $D_{KL}(P,\tilde{P})$, vs. values reported in the paper. (A) Full field, (B) Rectified, (C) Stixel simulations with binary inputs, (D) Same as  (C), zoomed into origin.} \label{fig:suppl_2}
\end{center}
\end{figure}

\end{document}